\documentclass[rmp,twocolumn,nofootinbib, letterpaper, noraggedfooter, showkeys]{revtex4}

\usepackage{amssymb,amsmath,amsfonts}
\usepackage{graphics,dcolumn,bm,fleqn,epic,eepic,float}
\usepackage{graphicx} % Include figure files
\usepackage{bm} % bold math
\usepackage{subfigure}
\usepackage{color} %  for colored comments
\begin{document}

\title{Quantifying firm-level economic systemic risk from nation-wide supply networks}

\author{Christian Diem$^{a,b}$, Andr\'as Borsos$^{a,c,d}$, Tobias Reisch$^{e,a}$, J\'anos Kert\'esz$^{d,a}$, Stefan Thurner$^{e,a,f}$}

\affiliation{$^a$Complexity Science Hub Vienna, Josefst\"adter Strasse 39, A-1080 Vienna, Austria}
\affiliation{$^b$Institute for Finance, Banking and Insurance,  Vienna University of Economics and Business, Welthandelsplatz 1, A-1080 Vienna, Austria}
\affiliation{$^c$Financial Systems Analysis, Central Bank of Hungary}
\affiliation{$^d$Department of Network and Data Science, Central European University,  Quellenstrasse 51, A-1080 Vienna, Austria}
\affiliation{$^e$Section for Science of Complex Systems, CeMSIIS, Medical University of Vienna, Spitalgasse 23, A-1090, Austria}
\affiliation{$^e$Santa Fe Institute, 1399 Hyde Park Road, Santa Fe, NM 85701, USA}

\email[E-mail: ]{diem@csh.ac.at}
\email[ ]{ stefan.thurner@muv.ac.at}

\date{\today}

\begin{abstract}
	Crises like COVID-19 or the Japanese earthquake in 2011 exposed the fragility of corporate supply networks. The production of goods and services is a highly interdependent process and can be severely impacted by the default of critical suppliers or customers. While knowing the impact of individual companies on national economies is a prerequisite for efficient risk management, the quantitative assessment of the involved economic systemic risks (ESR) is hitherto practically non-existent, mainly because of a lack of fine-grained data in combination with coherent methods.
	Based on a unique value added tax dataset we derive the detailed production network of an entire  country and present a novel approach for computing the ESR of all individual firms. We demonstrate that a tiny fraction (0.035\%) of companies has extraordinarily high systemic risk impacting about 23\% of the national economic production should any of them default. Firm size alone cannot explain the ESR of individual companies; their position in the production networks does matter substantially. If companies are ranked according to their economic systemic risk index (ESRI), firms with a rank above a characteristic value have very similar ESRI values, while for the rest the rank distribution of ESRI decays slowly as a power-law;  99.8\% of all companies have an impact on less than  1\% of the economy. We show that the assessment of ESR is impossible with aggregate data as used in traditional Input-Output Economics. We discuss how simple policies of introducing supply chain redundancies can reduce ESR of some extremely risky companies. 
	% max 250 ; now 250
\end{abstract}

\date{\today}


\maketitle

Increasing the efficiency of production processes and corporate supply chains has been a dominating economic paradigm of the past decades. 
Popular managerial concepts that exemplify this view are reflected in keywords such as supply chain management, lean production \citep{lamming1996squaring}, just-in-time delivery \citep{kannan2005just}, out- and global sourcing \citep{hummels2001nature, quelin2003bringing, trent2003understanding}, or supply base reduction \citep{trent1998purchasing, choi2006supply}. Efficiency gains are usually achieved by reducing inventory buffers, shorter lead times, supplier integration, or reducing the number of direct suppliers. Actions like these reduce production costs and increase profits. However, these actions also do have consequences in terms of resilience of the overall economy. It has been argued that increased levels of efficiency go hand in hand with a reduction of resilience \cite{choi2006supply, craighead2007severity}.

Supply chains of firms and consequently their production processes are highly interdependent. Supplier-buyer relations between companies lead to so-called production networks. The ongoing transformation of  production networks towards higher efficiency has made these networks more vulnerable to shocks  \cite{craighead2007severity}. On regional scales, hurricane Katrina or the Japanese earthquake in 2011 have shown the economic impacts that can arise due to subsequent cascading shock propagation along corporate supply chains \citep{hallegatte2008adaptive, carvalho2021supply, inoue2019firm}. The COVID-19 pandemic impressively revealed that not only overall economic activity can be affected by interruptions of supply chains, but they also may lead to shortages in basic supplies \citep{ivanov2020viability}, affecting people directly. This became apparent, for example, in food production \cite{bloomberg2021}, in vaccine supply  \cite{economist2021}, computer chips manufacturing, and car manufacturing \cite{Bushey2021, Miller2021}.
% Provide date of access for online sources.

The propagation of shocks through an economy is tightly related to classical input-output economics \cite{miller2009input}, where, however, only industry sectors are studied. For details and relations to this work, see Appendix  \ref{SI:IO}. The importance of supply chains at the firm level has been considered by economists, in particular the consequences of individual company failure on the overall economy \citep{bak1993aggregate, gabaix2011granular, acemoglu2012network}. The supply chain and production management literature studied firm level supply networks \citep{choi2001supply} and the spreading of disruptions along supply chains under various names, such as {\em supply chain resilience} \citep{craighead2007severity}, {\em snowball effect} \citep{swierczek2014impact}, the {\em ripple effect} \cite{ivanov2014ripple}, or {\em nexus suppliers} \citep{shao2018data}. Despite all these efforts, reliable and systematic estimates of firm level systemic risk for entire economies are hitherto not available. 

When compared to recent progress in the assessment of financial systemic risk in financial networks, the quantification of {\em economic systemic risk} (ESR) of individual companies in production networks is still in its infancy. From initial demonstrations of the relevance of network structure in the context of financial systemic risk \cite{allen2000financial, boss2004network, elsinger2006risk}, by now it is not only possible to assign systemic risk to individual players in financial networks \cite{battiston2012debtrank, thurner2013debtrank} but also to individual transactions \cite{poledna2016elimination}, and on multiple network layers \cite{poledna2015the}. These developments allow for novel policy paradigms for systemic risk regulation \cite{thurner2020macro}. To reliably assess systemic risks detailed and correct information on the underlying networks is essential \citep{battiston2016price}. For more information on financial systemic risk and its relation to the topic of the present paper, see Appendix \ref{SI:FSRI}.

\begin{figure}[tb]
	\centering
	\includegraphics[width=1 \columnwidth]{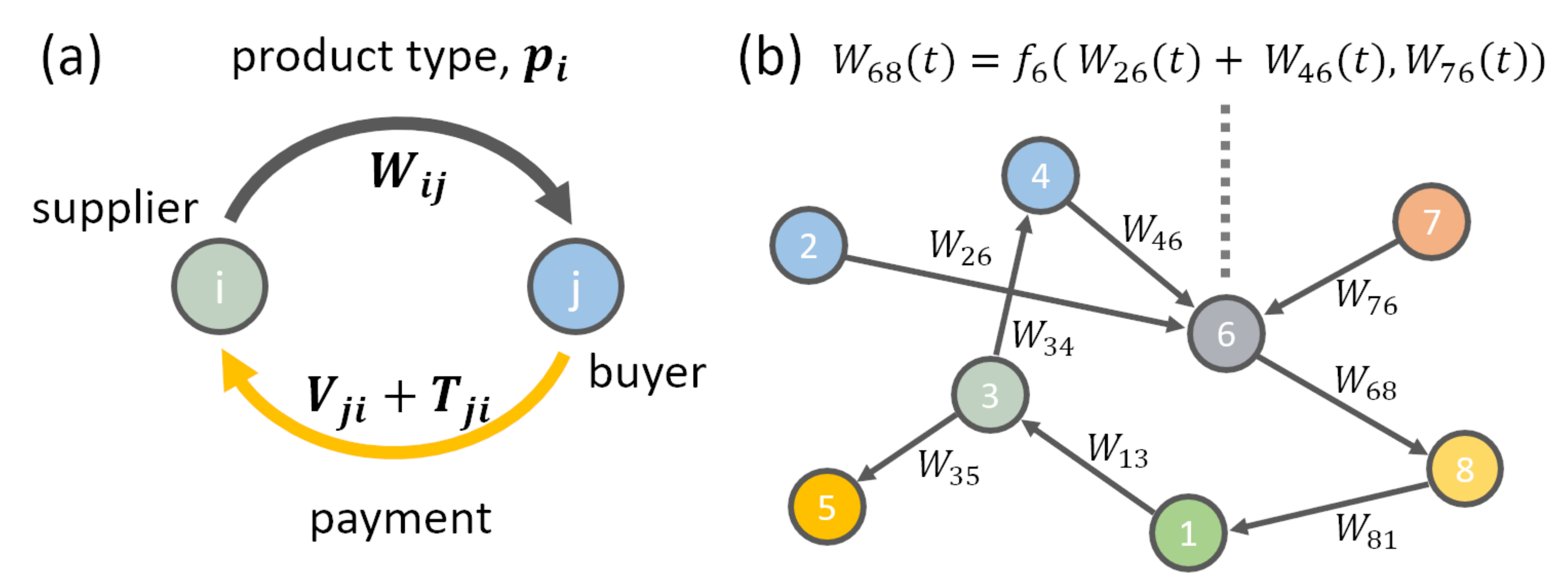} 
	\includegraphics[width=1 \columnwidth]{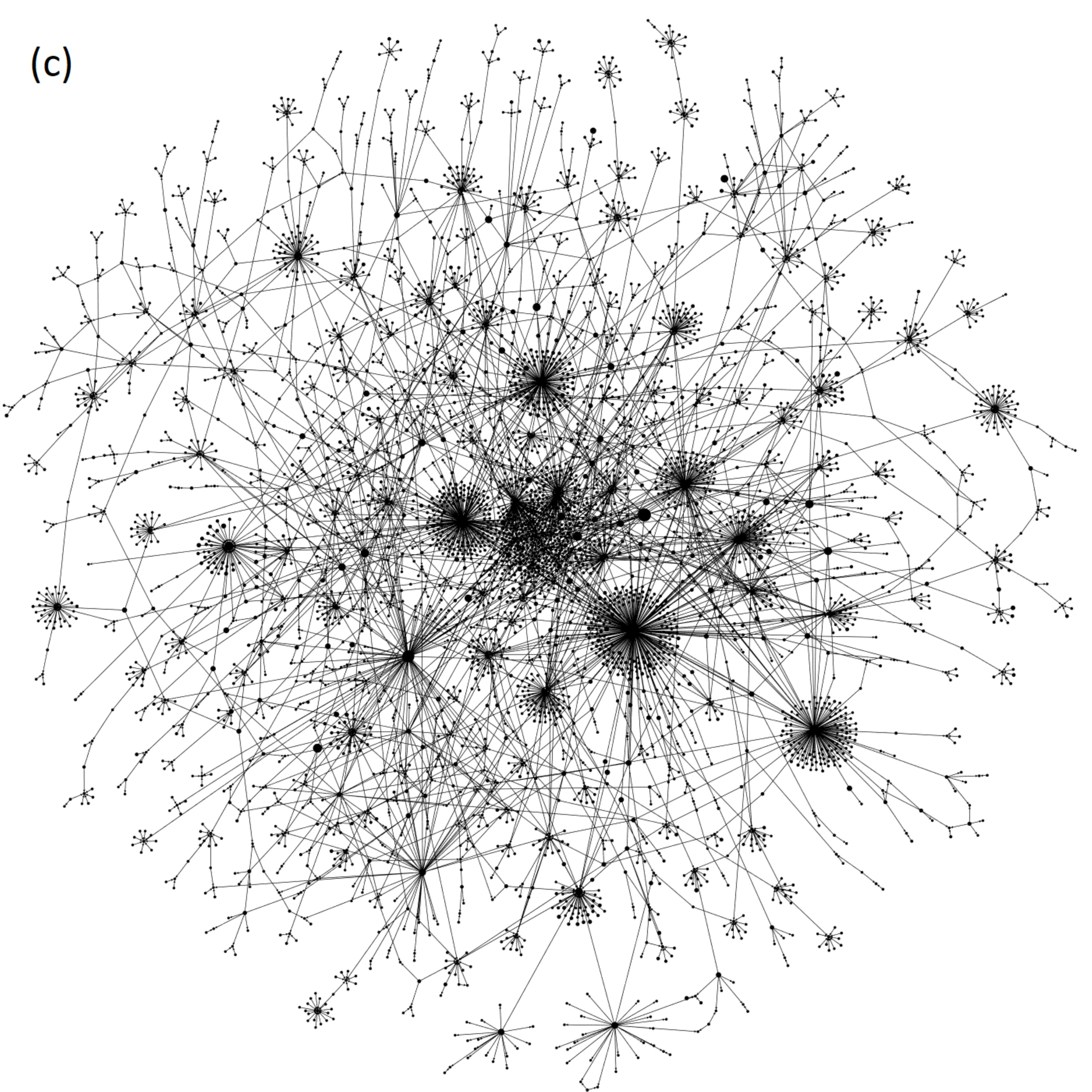}
	\caption{Reconstructing the production network. (a) Schematic supplier-buyer transaction between two companies, $i$ and $j$. Supplier $i$ produces product, $p_i$, and delivers a quantity of $W_{ij}$ to buyer $j$. The buyer makes a gross payment consisting of the net price, $V_{ji}$, and the value added tax, $T_{ji}$. 
		(b) Production network consisting of eight firms. Color represents their industry classification. Delivered goods and services are recorded in the weighted adjacency matrix, where the element $W_{ij}(t)$ records the volume  delivered from $i$ to $j$. The production function $f_6$ for firm 6 is shown as an example. It uses two types of input, $p_2 = p_4$ and $p_7$ from firms 2, 4, and 7 to produce the amount $W_{68}$ of product $p_6$. Firm 6 sells its entire output to firm 8.   
		(c) Section of the Hungarian production network with 4,070 nodes and 4,845 links. Node size corresponds to the total strength (proxy for company size). Many supply ``chains'' are seen to be connected to form a supply network.}
	\label{fig:transaction}
\end{figure}

It is impossible for decision makers to manage economic systemic risks proactively without understanding which companies pose exceptionally high risk to the entire economy in case of their (temporary) failure. It is therefore important to develop methodology to quantify these risks. Here we show how firm level supply transaction data can be used to compute an {\em economic systemic risk index} (ESRI) for {\em every} company within an entire country.

The most important ingredient for developing a meaningful ESRI is the underlying firm level production network, consisting of the supply relations between companies and their production processes. Country-wide production networks can be reconstructed from firm level value added tax (VAT) transaction data \citep{dhyne2015belgian, borsos2020unfolding}\footnote{The value added tax exists in all OECD countries, except for the USA \citep{oecd2020VAT}.}. The construction of the supply network is schematically depicted in FIG. \ref{fig:transaction} (a). For every sales transaction between a supplier $i$ and buyer $j$, the monetary value of the goods and services sold, $V_{ji}$, can be inferred, from the tax rate, $\tau$, and the tax amount paid, $T_{ji}= \tau V_{ji}$. We use this as an estimate for the volume, $W_{ij}$, of product type $p_i$, delivered from supplier $i$ to buyer $j$. Note that the proxy (volume = price $\times$ quantity) is a basic assumption in economics \citep{ miller2009input, lequiller2014understanding}. For notation, in-links to a node represent supply transactions (buying); out-links are sales transactions. We define the in-strength of node $i$ as the sum of all its in-links, $s_i^\text{in} = \sum_{j=1}^{n} W_{ji}$ (volume of purchased products), the out-strength is the sum of all out-links,  $s_i^\text{out} = \sum_{j=1}^{n} W_{ij}$ (sales). The (total) strength is defined as $s_i=s_i^\text{in}+s_i^\text{out}$ and serves as a proxy for firm-size. 

To assess the importance of the supply transaction, $W_{ij}$, between firms $i$ and $j$, it is essential to know which product type, $p_i$, is exchanged and how it is used in the production process of firm $j$. We use the term ``products'' for goods \emph{and} services. Economists typically resort to the simplifying assumption that every company produces one out of $m$ different products that is determined by the company's industry classification. We use an industry affiliation vector, $p$, that assigns one of $m$ industries to each firm $i$, $p_i \in \{1, 2, \dots, m \}$. We use the NACE (Statistical Classification of Economic Activities in the European Community~\cite{ramon2021nace}) classification scheme on the 4-digit level with $m=615$ categories. For more details on industry classifications, see Appendix \ref{S1:industry}. 

The production process of a company is commonly described with a \textit{production function}~\cite{carvalho2019production}, $f_i$, that determines the (maximal) amount of product $x_i$ (output) of type $p_i$ that firm $i$ can produce with a given amount of intermediate products, $\Pi_i=(\Pi_{i1}, \Pi_{i2}, \dots \Pi_{im})$ ($\Pi_{ik}$ is the amount of input of type $k$ of firm $i$), its employees (labour), $l_i$, and manufacturing equipment (capital), $c_i$. We use production functions that allow us to determine how much firm $i$ can still produce if a supplier $j$ fails to deliver its products of type $p_j=k$ to firm $i$ and hence reduces the availability of input $\Pi_{ik}$. This is illustrated with a small production network in FIG. \ref{fig:transaction} (b). In-links to a node (arrows pointing towards node) represent supply transactions (buying), out-links are sales transactions. By identifying the supply links belonging to the inputs in the production function, it can be seen how production depends on the current state of the network. This is illustrated for firm 6. It uses two types of inputs, product $a$ (supplied by firm 2 and 4) and product $b$ (supplied by firm 7), and produces an amount $x_6(t+1)=f_6\big(W_{26}(t)+W_{46}(t), W_{76}(t) \big)$ of product $c$. The delivery from 6 to 8, $W_{68}(t+1)$, depends on the in-links of 6, $W_{68}(t+1)=f_6\big(W_{26}(t)+W_{46}(t), W_{76} (t)\big)$. Should one of the in-links diminish or vanish, this impacts the output of 6. This illustrates a so-called \emph{downstream} shock propagation of a {\em supply shock}. Vice versa, if 8 decides to no longer buy from 6, the in-links $W_{26}(t),W_{46}(t), W_{76} (t)$ would  no longer be needed. This illustrates the \emph{upstream} propagation of a {\em demand shock}. 

The specific choice of the production function, $f_i$, determines the intensity of the downstream shock propagation. Frequently used production functions include the constant elasticity of substitution (CES) \citep{mcfadden1963constant, carvalho2021supply, moran2019may} and its special cases, the Cobb-Douglas  \citep{acemoglu2012network} and the Leontief \citep{inoue2019firm, pichler2020production} production functions, 
see Appendix \ref{SI:productionfunction} for details. Here we take a short term perspective and consider a {\em generalized Leontief production function} (GL) that accounts for the fact that even for companies with physical production processes, not all procured inputs are essential in the short term. Note that \cite{pichler2020production} provides a study on which inputs are essential for 56 industry sectors. For example, in tire production, rubber intermediates are essential inputs, while consulting services are not (short term). The GL treats non-essential inputs in a linear fashion, while essential inputs are treated in the non-linear Leontief way. We denote the set of essential inputs by $\mathcal{I}_i^\text{es}$ and  non-essential inputs by $\mathcal{I}_i^\text{ne}$, respectively. We define the GL as 
\begin{equation}
	x_i = \min\Bigg[
	\min_{k \in \mathcal{I}_i^\text{es}} \Big[ \frac{1}{\alpha_{ik}}\Pi_{ik}\Big], \: 
	\beta_{i} + \frac{1}{\alpha_i} \sum_{k \in  \mathcal{I}_i^\text{ne}} \Pi_{ik} \Bigg] \, ,
	\label{GLPF}
\end{equation}
where $\alpha_{ik}$ are technologically determined coefficients and $\beta_{i}$ is the production level that is possible without non-essential inputs $k \in \mathcal{I}_i^\text{ne}$;  $\alpha_{i}$ is chosen to interpolate between the full production level (with all inputs) and $\beta_{i}$. Both, $\alpha$ and $\beta$, are determined by $W$, $\mathcal{I}_i^\text{es}$ and $\mathcal{I}_i^\text{ne}$. Note, that we assume labour and capital to be fixed (in the short term) and omit them in the notation. The Leontief production function is a special case of Eq. (\ref{GLPF}) if all inputs are essential. The linear production function is the special case when all inputs are non-essential; see Appendix \ref{SI:productionfunction}.
An essential aspect that determines downstream shock propagation is how easily failing suppliers can be replaced. For our purposes we assume that companies with a low market share are easier to replace, than firms with large market shares; see Appendix \ref{SI:supplier_replaceability}.

For the empirical analysis we calibrate the GL in Eq. (\ref{GLPF}) to four scenarios based on the firms' NACE classifications. First, for a hypothetical purely linear production scenario (LIN), all firms have liner production functions, i.e. only non-essential inputs, or $ \mathcal{I}_i^\text{ne} = \{01,\dots, 99\}$). Second, in a purely Leontief scenario (LEO), all firms  have Leontief functions, i.e. only essential inputs, $ \mathcal{I}_i^\text{es} = \{01,\dots, 99\}$.  Third, in a mixed scenario (MIX) we assume all firms within NACE classes 01-45 (physical production) have only essential inputs $ \mathcal{I}_i^\text{es} = \{01,\dots, 99\} $ and all firms within NACE classes 46-99 (non-physical production) have only non-essential inputs $\mathcal{I}_i^\text{ne} = \{01,\dots, 99\} $. Fourth, in the generalized Leontief scenario (GL) we assume that for firms within NACE 01-45 the set of essential inputs consists of $\mathcal{I}_i^\text{es} = \{01,\dots, 45\}$ (supplied by physical producers) and the set of non-essential inputs consists of $\mathcal{I}_i^\text{ne} = \{46,\dots, 99\} $ (supplied by non-physical producers), while for firms within classes NACE 46-99 we assume they have only non-essential inputs  $ \mathcal{I}_i^\text{ne} = \{01,\dots, 99\}$. We assume that firms within NACE 01-45 have a physical production process while firms within NACE 46-99 predominantly trade or provide services. Note that LIN and LEO provide lower and upper bound scenarios for more realistic situations, where firms have distinct types of production functions as in the MIX and GL scenarios.

Given the details of a production network, $W$, in terms of weighted in-links, out-links, product classification $p_i$, and production functions $f$, the economic systemic risk index (ESRI) can finally be computed. It captures the consequences of both, the up- and downstream shock propagation following the default of a particular company, $i$ as outlined in Materials and Methods and Appendix \ref{SI:ESRI} and Appendix \ref{SI:derivation}. The quantity ${\rm ESRI}_i$ can be readily interpreted as the fraction of the production network's observed output that is likely to be affected if firm $i$ fails and its output and demand is not replaced by other firms. The value is computed by simulating the propagation of shocks with the recursive algorithm described in Appendix \ref{SI:propagation}.

\begin{figure*}[tb] 
	\centering
	\includegraphics[width=0.75 \columnwidth]{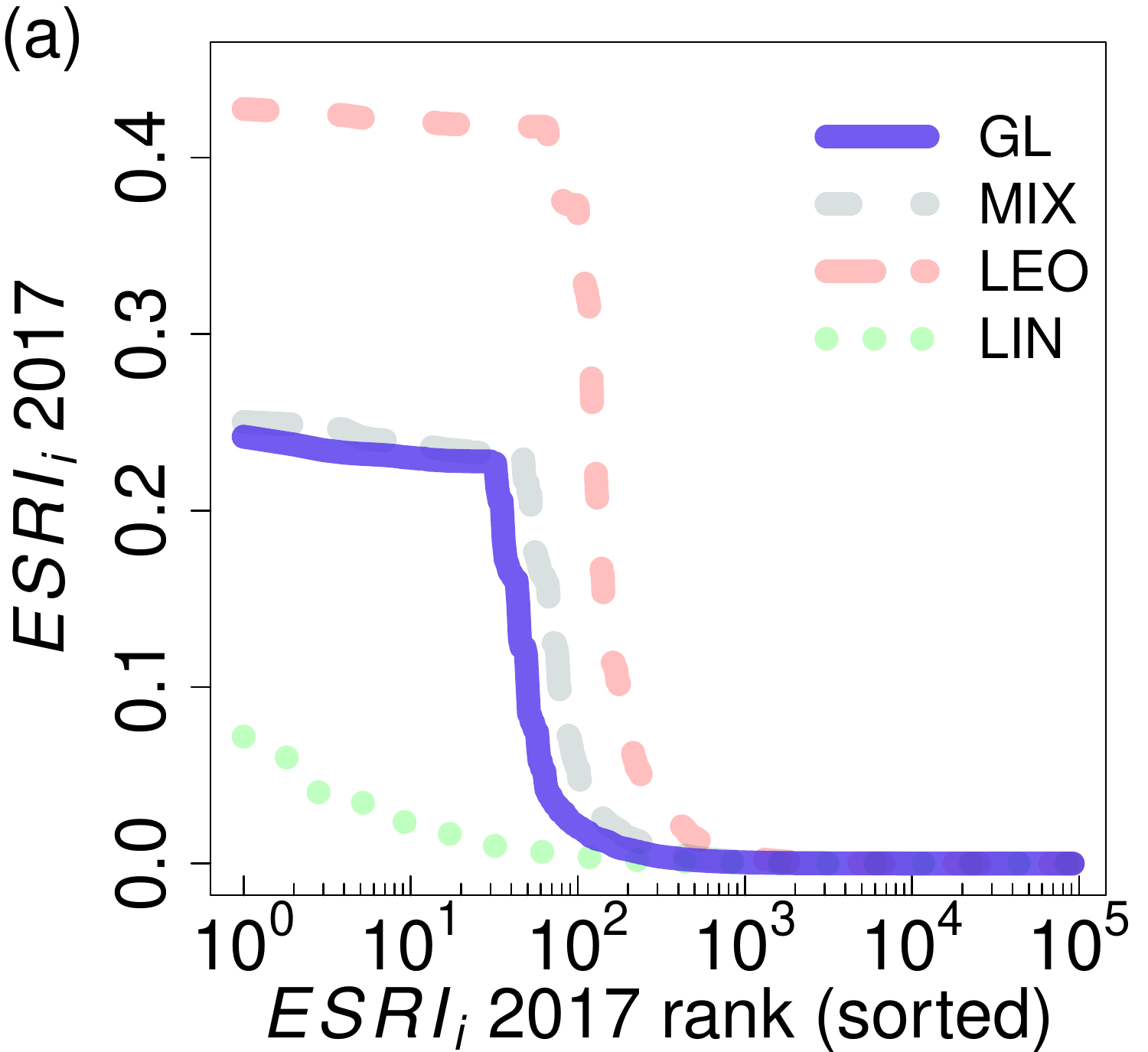} 
	\includegraphics[width=0.75 \columnwidth]{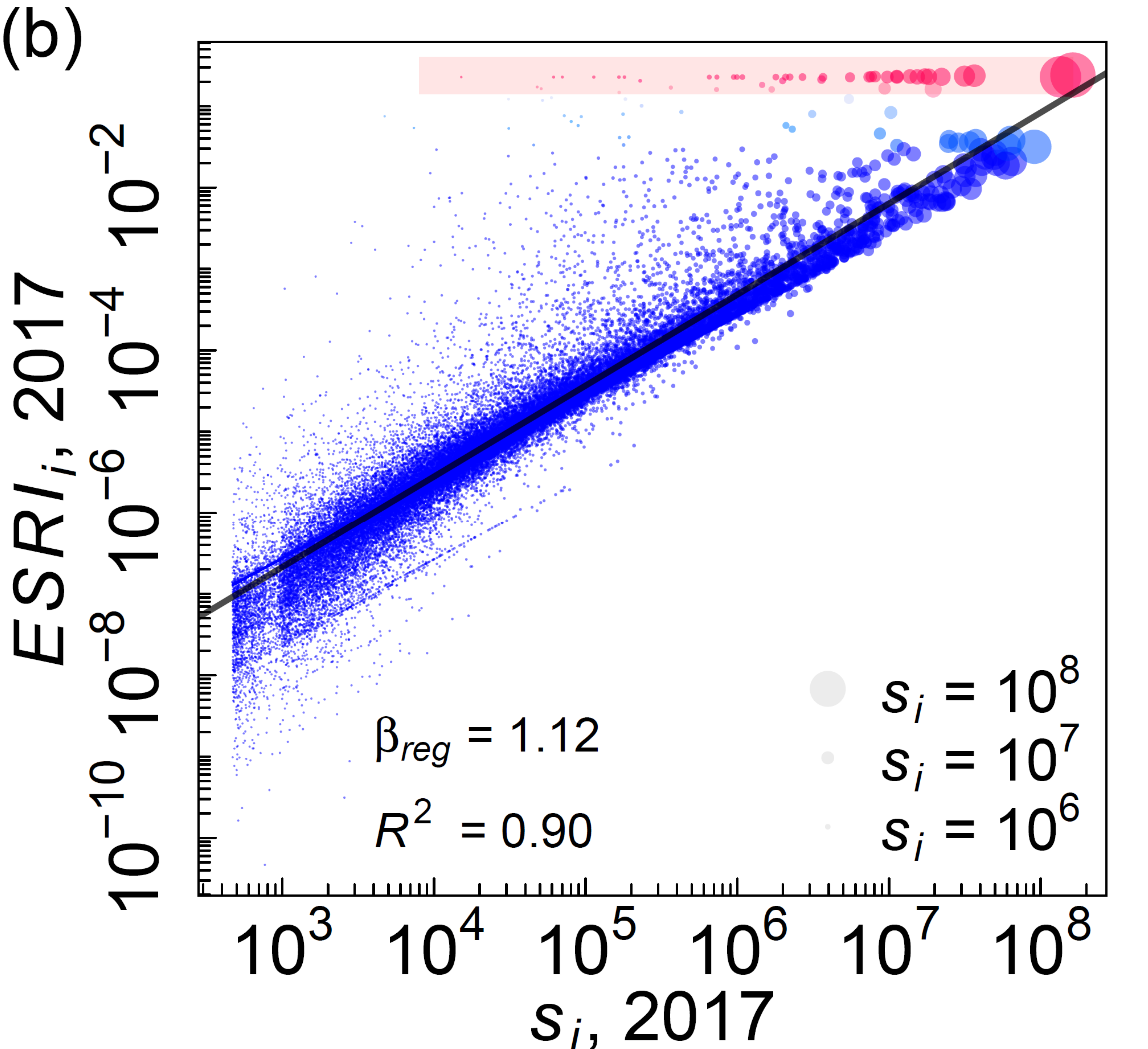} 
	\includegraphics[width=0.75 \columnwidth]{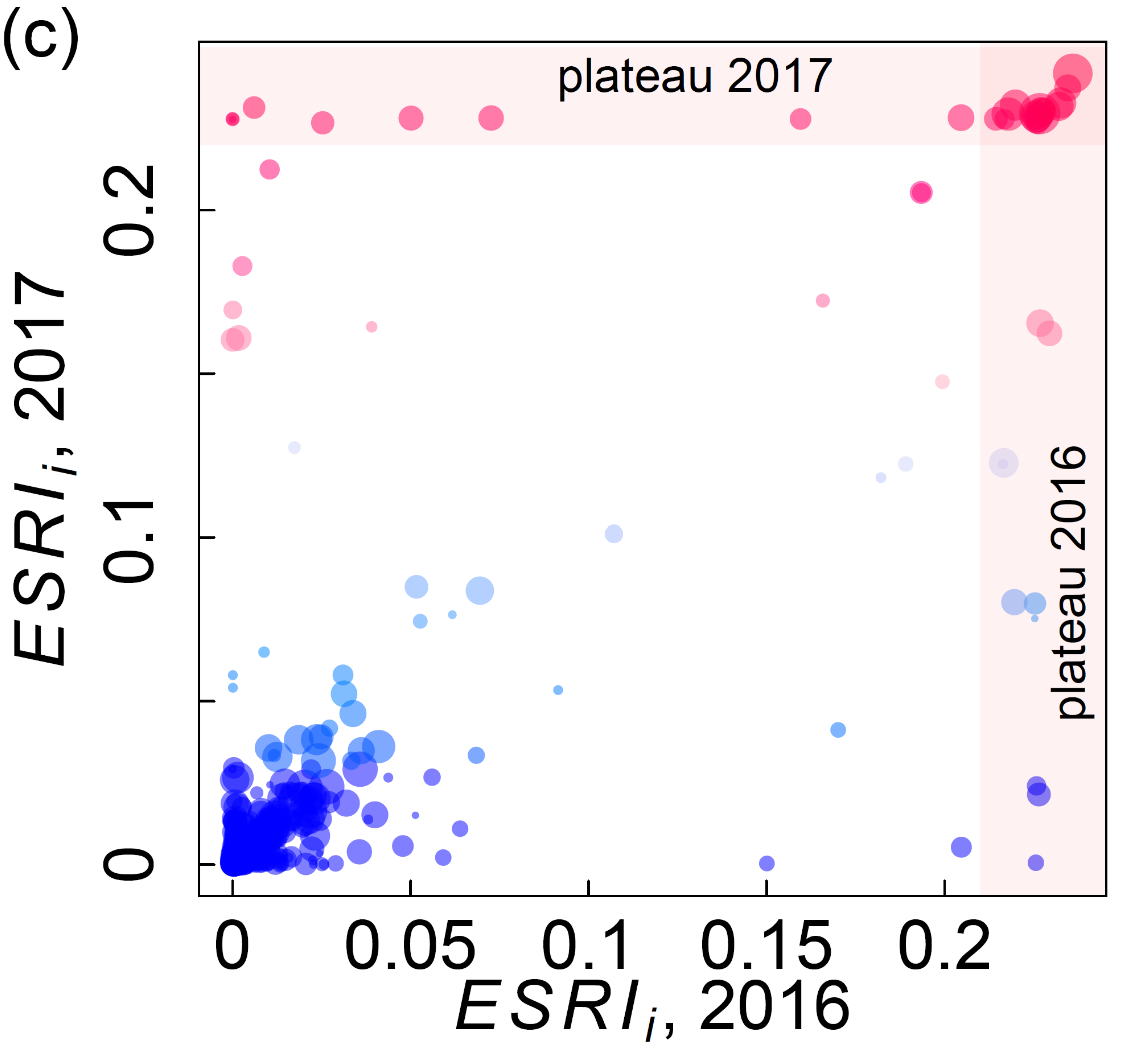} 
	\includegraphics[width=0.77 \columnwidth]{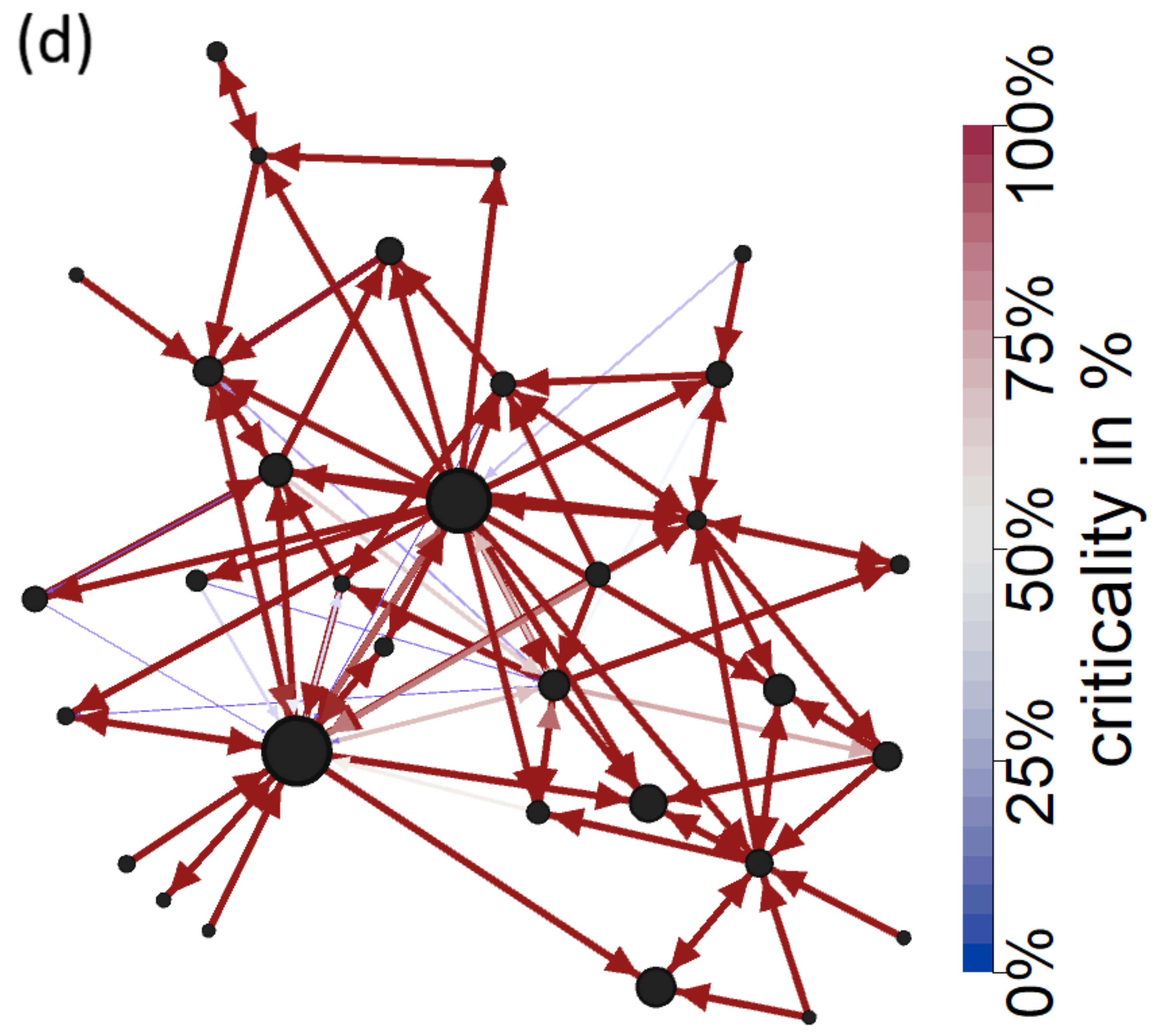} 
	\caption{Economic systemic risk of companies. (a) Economic systemic risk profile (distribution) ${\rm ESRI}_i$ of $n=91,595$ companies in linear-log scale for 2017. Distributions are rank-ordered, meaning the most risky company is to the very left. The blue line shows the result for the realistic GL scenario (production functions according to industry classification and classifying produced goods as essential or not). A plateau exists around an ESRI of 0.23, containing 32 firms for GL. There is a steep decline to ${\rm ESRI}_i \sim 0.05$ from rank 33 to 55. 165 firms have an ESRI larger than 0.01. For comparison the MIX scenario (light blue) is shown (production functions according to industry classification only).
		As the limiting cases we show the scenarios LIN (green), where all firms have linear production functions and LEO (red), where all firms have Leontief production functions. The tail of the ESRI profile decays as an approximate power-law. 
		(b) ESRI plotted against firm strength (firm-size) in log-log scale. Symbol size represents strength $s_i$, red symbols belong to the plateau, emphasised by the shaded area. We find large and small companies in the plateau, suggesting that very high ESRI is not determined by size, even though the correlation of ESRI and strength for the bulk of the companies is high.
		(c) Systemic risk in 2016 vs. 2017. Colors correspond to ESRI in 2017. Blue and red symbols indicate low and high values in 2017, respectively. Note the strong variability in the plateau firms. There is a significant correlation between ESRI in 2017 and 2016, that indicates relatively small temporal fluctuations for the bulk of the companies. 
		(d) Network of the 32 most systemically risky firms (plateau) in 2017. Node size is proportional to the square root of strength. Link colors correspond to the downstream 'criticality', i.e., the percentage of $j$'s production should $i$ stop functioning, $\Lambda_{ij}^d$. Red thick (blue, thin) links indicate very large (small) losses of production. Small companies predominately supply to large high-risk companies, thereby inheriting systemic risk. 
	}
	\label{fig:sr_profile}
\end{figure*}

\section*{Results}

The directed production network $W$ is reconstructed from fully anonymized VAT micro data of the Hungarian Central Bank. We consider all links, $W_{ij}$, where at least two recurring supply transactions occurred; see Appendix \ref{SI:data} and \cite{borsos2020unfolding}. Figure \ref{fig:transaction} (c) shows a section of 4,070 companies and 4,845 links of the empirically reconstructed Hungarian production network, $W_{ij}$. The section is obtained by sampling 1,500 random nodes and considering all their direct suppliers, yielding 6,113 nodes. Only the giant component (4,070 nodes) is shown. The entire network consists of $n=91,595$ companies; it is too large and dense to visualize its structure in a meaningful way. Already this small section reveals the fact that production by no means happens along independent supply {\em chains}, but on a tightly interwoven {\em complex network}. Merged chains that are quite extended are clearly visible. The network has an apparent core periphery structure.  For visualizations of other aspects of the production network, see Appendix \ref{SI:network_vis}. 

We next calculate the ${\rm ESRI}_i$ for the realistic GL and MIX scenarios as well as for the two limiting cases, LIN and LEO for all of the $91,595$ companies in the Hungarian VAT micro dataset in the year 2017 and for the $85,131$ firms in 2016. Rank-ordered distributions of the ESRI are shown in FIG. \ref{fig:sr_profile} (a) in linear-log scale for 2017. For 2016, see  FIG. \ref{fig:sr_profile2016} (a) in Appendix \ref{SI:2016results}. For the realistic scenario GL (blue), we find that 32 companies show extremely high levels of systemic risk, all being at a value of about $0.23$, meaning that about 23\% of the entire economy is affected should one of these companies fail and its supply and demand is not replaced. 63 and 165 firms have an ESRI larger than 0.05 and 0.01, respectively. For respective numbers for the other scenarios, see Table \ref{SI:table_large_obs2017} in Appendix \ref{SI:powerlaw}. The situation is similar for the MIX scenario, where 47 companies belong to the plateau with values around $0.23$. For the (unrealistic) reference case, LEO with {\em all} companies of Leontief type (red), for 66 firms we find much higher systemic risk levels of about $0.42$.

For ranks larger than a characteristic rank of 32, the pronounced plateau in the distribution is followed first by a steep decline and then by a slow decay of the ESRI values. The tail (without plateau and steep decline) can be fitted to a power-law with an exponent of roughly $\alpha^\text{GL}=0.67$ for GL,  $\alpha^\text{MIX}=0.63$ for MIX, and $\alpha^\text{LEO}=0.50$. For details of the fits, see Appendix \ref{SI:powerlaw},  FIG.  \ref{SI_fig:powerlaw} (a). The shape of the rank-ordered ESRI distribution (plateau and power-law tail) is similar to what was found in Fig 4. in \cite{moran2019may}. As expected, the reference case LIN (green) where {\em all} companies have linear production functions, generates substantially lower systemic risk levels than GL and MIX. LIN neither shows a plateau nor a power-law decay. The situation is very similar in 2016, see FIG.  \ref{SI_fig:powerlaw} (b) in Appendix \ref{SI:powerlaw}.

To better understand which companies are forming the plateau of extremely risky companies --- data protection regulations prevent us from showing company names or actual  turnover --- in FIG. \ref{fig:sr_profile} (b) we show ${\rm ESRI}_i$ as a function of the strength, $s_i$ (firm size within the network). Companies on the plateau belong to industry sectors like energy, manufacturing (electrical equipment, chemicals, computer and electronics, vehicles), or repair of machinery. Red color indicates the highly risky plateau companies; symbol size represents strength, $s_i$.
Clearly, the ESRI of plateau firms (located in the shaded area) is not changing with size; in the plateau we find large and small companies (note the range of strength of 4 orders of magnitude), suggesting that firm-size is not able to predict extreme ESRI values at all. For the bulk of companies we find a strong statistically significant correlation of log-ESRI and log-strength ($R^2=0.90$ and slope $\beta_{reg}=1.12$ in log-log regression). However, for individual companies strength is not a reliable predictor of ESRI, since the spread of the ESRI extends up to 4 orders of magnitude. A more detailed regression analysis, found in Appendix \ref{SI:regression_extended}, confirms that generally firm level quantities fail to explain ESRI that is a network-based measure. Note the relation to the Hulten theorem~\cite{Hulten1978Theorem}, stating ---in simplified terms--- that in an efficient economy the effect of a firm level shock on overall output is proportional to its revenue. Our results are in strong disagreement with this statement, adding further evidence for its limited practical validity \citep{baqaee2019macroeconomic}. Finally, we checked that all firms on the plateau are within NACE 01-45, meaning that their production functions are purely of generalized  Leontief type with a large share of essential inputs.

The reason for the formation of the plateau is twofold. First, about two thirds of its nodes form a strongly connected component based on highly critical supplier relationships (core of plateau). In this core the failure of one firm leads to the failure of the other members in the component, implying the default of anyone has the same consequences on the entire network (same ESRI). The second reason is that the other third of the plateau nodes are suppliers to the strongly connected component, i.e. their failure causes failure of the nodes in the strongly connected component. These peripheral nodes inherit high ESRI values by supplying (in)directly to inherently risky companies. Note that these two features are impossible to explain with only linear production functions (LIN). To illustrate this point, in FIG. \ref{fig:sr_profile} (d), we show the network spanned by the 32 plateau firms. Node size corresponds to the square root of strength,  $\sqrt{s_i}$; link colors correspond to the direct down-stream impact of the supplier on the buyer node (criticality), $\Lambda^\text{d}_{ij}$, (for the definition, see Appendix \ref{SI:propagation}). Red links indicate highly critical supplier-buyer relations; $\Lambda^\text{d}_{ij}=1$ means that 100\% of $j$'s production fails if $i$ fails. Blue links mark non-critical supplier-buyer relations; $\Lambda^\text{d}_{ij}=0.01$ means that $j$'s production reduces by 1\% if $i$ fails. It is visible that almost all links are critical for the production of the buyer. 

For systemic risk monitoring of companies it is crucial to track the change of their systemic riskiness (ESRI) over time. To estimate the extent of the typical yearly fluctuations, we compare the ESRI values for the years 2016 and 2017 in FIG. \ref{fig:sr_profile} (c). Size corresponds to firms' log-strength. Of the $68,254$ firms that appear in both years, the values of ESRI in 2017 and 2016 are correlated ($\rho ({\rm ESRI}^{16},{\rm ESRI}^{17})=0.78$,  $p=2.2\: 10^{-16}$). The extent of fluctuations is clearly seen in log-log representation in FIG. \ref{fig:sri_16_17_density} (a) in Appendix \ref{SI_annualchange}. For the logged variables we observe an even higher correlation ($\rho ({\rm log(ESRI)}^{16},{\rm log(ESRI)}^{17})=0.85$, $p=2.2\: 10^{-16}$). Figure \ref{fig:sri_16_17_density} (c) shows that relative changes in strength explain the relative changes in ESRI only partially.

In FIG. \ref{fig:sr_profile} (c) temporal changes of companies in the plateau between 2016 and 2017 are seen. The points in the horizontal shaded region are those companies in the plateau in 2017 that had (much) less ESR in 2016. The points in the vertical shaded area are those companies that are in the plateau in 2016, but not 2017. These reduced their ESRI. From the 32 plateau companies in 2016, 20 remain in the plateau (intersection of shaded regions). Of the 12 companies leaving the plateau from 2016 to 2017 some remain very risky (light red) others become less risky (blue). 12 companies enter the plateau in 2017 (also firms not contained in the 2016 dataset). 
Possible reasons for these strong fluctuations could be that core plateau firms added additional suppliers for the same input, that supply links were discontinued, or that market share dropped leading to higher supplier replaceability. From the data it is hard to decide how these factors contribute. 

\begin{figure*}[tb]
	\centering
	\includegraphics[width=0.75 \columnwidth]{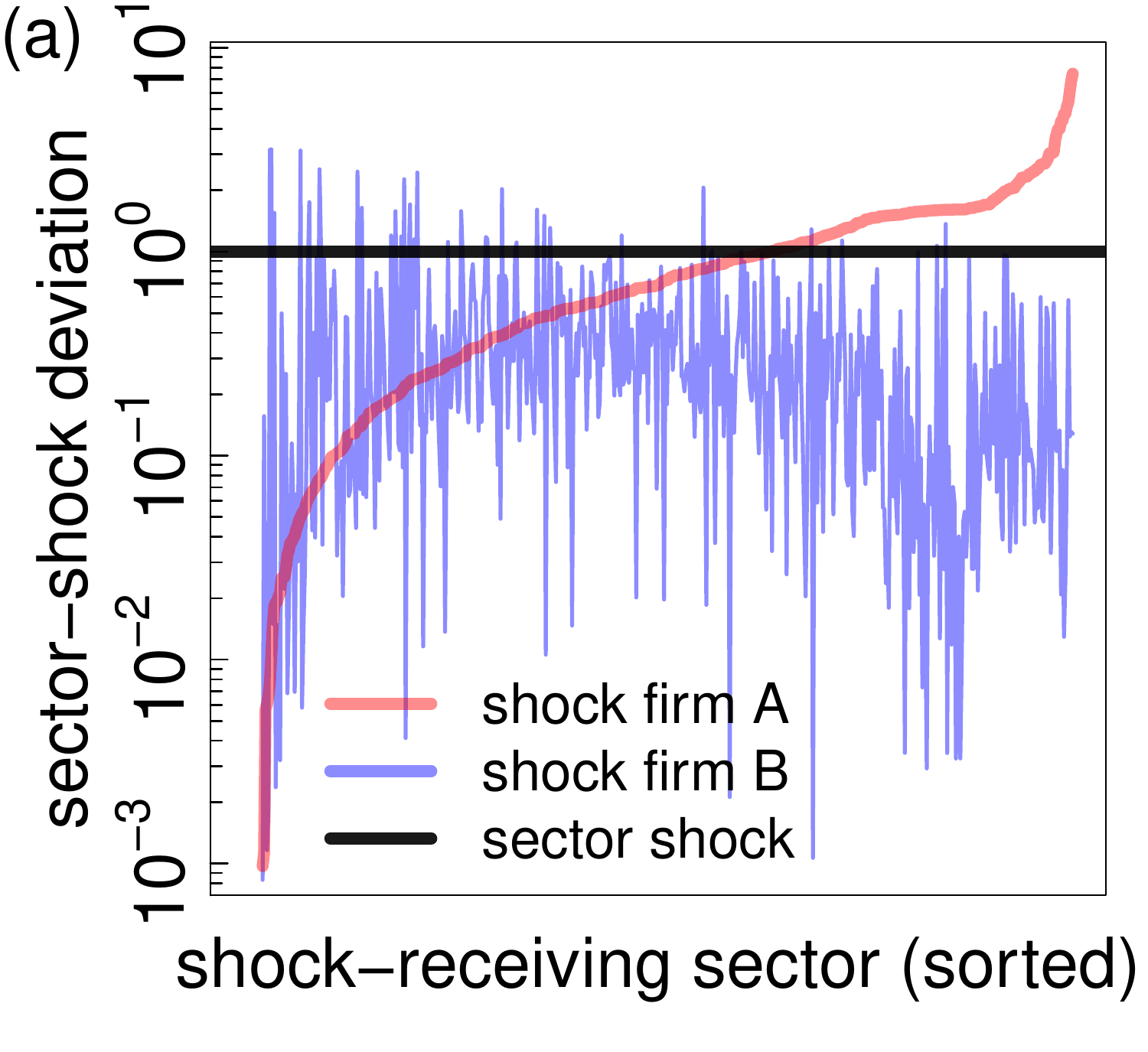}
	\includegraphics[width=0.75 \columnwidth]{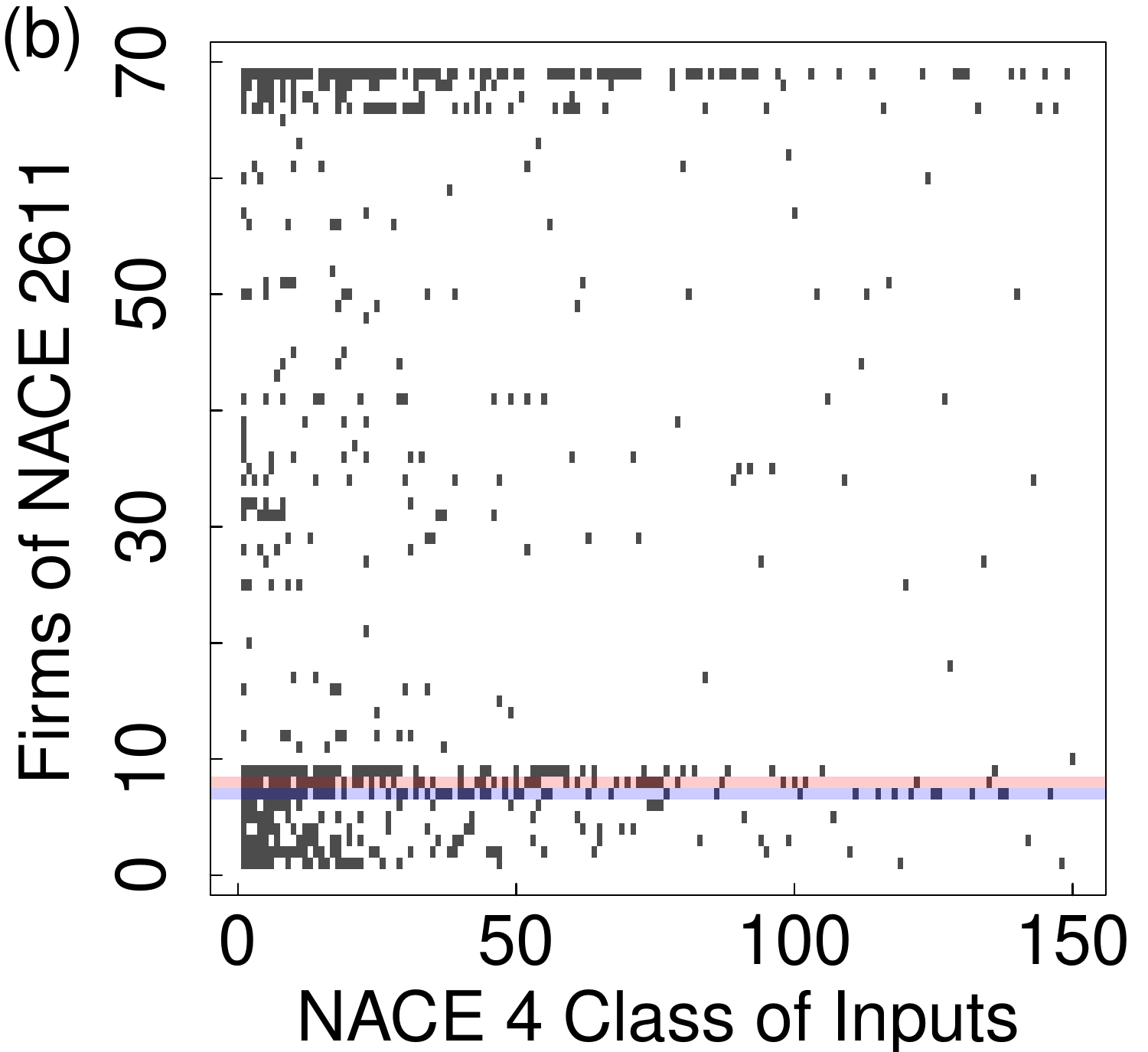}
	\caption{ 
		Importance of firm level analysis. 
		(a) Effect on all 568 NACE 4-digit industry sectors, following a general 18\% shock of the entire NACE sector 2611 (Manufacture of electronic components). The x-axis denotes the 568 shock receiving sectors, the relative deviation of received shocks from the 18\% homogeneous initial shock to sector 2611 is shown on the y-axis. The black line marks the reference scenario. The red line shows the relative deviation if firm A (in 2611) receives a 100\% shock; the blue line shows the situation for a 59\% shock to firm B. Both firm level shocks are equivalent in size to the 18\% sector shock (reference). The particular choice of the defaulting companies lead to drastically different results on how other economic sectors are affected.
		(b) Most ubiquitous inputs (at NACE 4 level, columns) for 69 firms (rows) in NACE class 2611. Clearly, even though all companies belong to the same sector, their input sector vectors are drastically different. Inputs of firm A (red, 50 distinct inputs) and firm B (blue, 49 distinct inputs) differ substantially (Jaccard index 0.25). Note that 18 firms have no input sectors (empty rows).
	}
	\label{fig:firm-level}
\end{figure*}

To demonstrate that it is necessary to use firm level data for any reasonable assessment of shock propagation in production networks, we compare it to a situation where shocks are studied on a more aggregated level. In FIG. \ref{fig:firm-level} (a) we show three different economic shock scenarios that affect the NACE class 2611 (Manufacture of electronic components). All three initial shocks have the same size on the sector level, but affect different companies within the sector. An {\em initial shock} is exogenously applied and triggers a spreading of shocks in the production network. The {\em received shock} is the shock each sector  receives as a result of the propagation of the initial shock. Shock size is measured as the fraction of the sector's overall strength, ($s^\text{2611}=\sum_{i|p_i = 2611}s_i$), affected by the initial shock. The first shock scenario serves as a reference scenario. It applies a homogeneous initial shock of 18\%  (reduced production) to all of the 69 firms in sector 2611. The received shock of all (568 NACE 4) sectors (shown on the x-axis) marks the reference response. The received shocks are shown on the y-axis (in log-scale) as a fraction of this reference response, and by construction, the received shock for the homogeneous scenario is 1 (black line). In the second scenario a 100\% shock (failure) is applied to a single firm, A, the received shocks in the different sectors (given as a fraction of the reference scenario) are shown as the red line. Sectors are sorted w.r.t. received shock sizes in this scenario (A). The third shock scenario applies a 59\% shock to firm B, the responses are shown by the blue line. Firm A is a plateau firm, firm B is not.  

Clearly, the specific choice of the defaulting company within the sector has a drastic  effect on how the different economic sectors are affected. Note that on the sector level the initial shocks are indistinguishable. In other words, \emph{homogeneous} sector shocks (aggregated view) yield a grossly false picture of the true shock propagation and thus systemic risks. As so often in networked dynamical systems: details do matter. If studying shocks on the sector level was sufficient, the three scenarios should be strongly correlated. However, the figure suggests that the relative deviations from the sector shock (blue and red line) are negatively correlated with -19\% ($p=5\cdot 10^{-6}$). In Appendix \ref{SI:firm_level_advantages} we show the actual received shocks (not their deviations from the reference scenario). 

One obvious reason for these large differences between scenarios A and B is made clear in FIG. \ref{fig:firm-level} (b), where we see for all companies in NACE sector 2611 the inputs at the NACE 4-digit level. Firms (rows) are sorted w.r.t. similarity of their inputs (columns). Even though all  companies belong to the same industry sector (NACE 2611) the sectors of their inputs vary substantially. The input sectors of Firm A (red shading) and firm B (blue shading) have only a small overlap, indicated by a Jaccard index of 0.25. Depending on which of two companies fails, different input sectors are affected.  The Jaccard index for the customer sectors is even smaller (0.13). For details on customer sectors, see Appendix \ref{SI:firm_level_advantages}. Among the 51 firms (that in NACE 2611 have NACE 4 classified inputs) 56\% have no pairwise input sector overlap (Jaccard index of zero) and 85\% have no pairwise customer sector overlap. This means that when choosing two firms randomly, the probability of having no common input (customer) sector is 56\% (85\%). Consequently, shocks to different firms within a sector must lead to different economic sectors being affected.

\section*{Discussion}

Based on the reconstruction of all relevant buyer-supplier relations between companies in an entire country from VAT data, we demonstrate that the real economy can not be viewed as a collection of separate supply chains, but is a tightly connected directed  network that has a strongly (weakly) connected component, containing 26\% (94\%) of all companies. The network allows us to develop and compute an index for the approximate systemic risk of every individual company to the economy, ESRI. We demonstrate explicitly that systemic risk based on aggregated sector level data yields a severely distorted picture.  

We find that the 32 top risky companies contribute to 45 \% of the entire systemic risk, the top 100 companies contribute to 74 \%. Only 165 companies have more than 1\% risk. The 32 top risky companies (0.035\% of all 91,595 companies) show extremely high systemic risk of about  23\%. Of those, only a fraction appears to be inherently risky (e.g., because of size or high market shares). The average systemic risk of a Hungarian company is $\overline{{\rm ESRI}_i} = 0.00018$ (mean). The median is $1.7 \cdot 10^{-6}$.

Approximately a third of the most risky companies constitute the periphery of the network of the plateau companies. These are small and inherit systemic risk from the core plateau companies because they are critical suppliers to the core. They would not show high ESRI values if they supplied to other --non-risky-- companies. This means that several of these smaller companies can be made less risky simply by increasing the number of suppliers of the specific good sold to the inherently risky companies.  By analyzing changes of systemic risk from 2016 to 2017, we find that ESRI is relatively stable for most companies. However, several smaller companies change their ESRI from marginal to extreme, and vice versa. The reasons for this might be the risk-inheritance of (non-inherently risky) companies that start (or stop) supplying to inherently risky ones, or changes in market shares that affect their replaceability. We confirmed that to a large extent systemic risk is not predictable with strength (firm-size). The position in the supply network matters much more than company-size, similarly to what has been observed in financial systemic risk in earlier studies \cite{battiston2012debtrank, markose2012systemic, thurner2013debtrank}. 

The presented study has limitations. The proposed measure is an economically motivated, straightforward  quantitative measure for systemic relevance of companies in a given production network. It captures up- and down-stream impacts independently. This doesn't cause distortions in tree-like supply chains, however, in networks with strongly connected components it does. In reality, the same firm can be affected by up- and down-stream shocks; these shocks can potentially interact. For example, the lack of a critical input (down-stream shock) can translate into an upstream shock for suppliers of complementary inputs. This second-order effect is not considered in the algorithm. Its magnitude needs to be estimated in future work and should be taken care of in more subtle algorithms. Nonetheless, ours is the first attempt to calculate company level ESRI for an entire national economy and the resulting index can be seen as a good first-order estimate for the systemic risk in production networks. 
We assume, somewhat unrealistically, that all customers and all suppliers are treated as being equally important (proportional rationing). Other rationing mechanisms, for example, prioritizing large customers and suppliers, could lead to modified ESRI estimates. For a detailed discussion of the effects of rationing mechanisms at the sector level, see \cite{pichler2021modeling}. Further, for the implementation of the algorithm we assumed a data-driven replaceability index for each firm that is based on its market share (fraction of output within its NACE 4 class). This implies that firms with high (low) market shares are  difficult (easy) to replace, see Appendix \ref{SI:supplier_replaceability}. Further improvement of the ESRI would be possible by modelling the substitutability of inputs and the replaceability of suppliers in more detail. However, this comes with the cost of more parameters to calibrate, more detailed data needs, and significantly higher computational effort.

For assigning production functions to companies, we used a simple matching based on NACE 2-digit categories. Although we think that this is a good first approximation, we emphasize that this can be a source of error. Given the current data availability, an exact and objective mapping of production functions to companies is not feasible. Large-scale firm level surveys providing information on how strongly firms are affected by the lack of specific inputs could be a useful first step clarifying this point.

We assume a static production network, which in reality, is a temporal network. With this simplification, seasonality effects in production and manufacturing are missed. Also, in the present implementation we don't consider competition between companies that would result in a dynamic restructuring of the network. Moreover, it is technically hard to interpret the time index $t$ in terms of how long shocks really need to spread and converge. For simplicity, we assume uniform spreading, i.e. all periods $t$ are of the same abstract length. Imports, exports, and production networks of other countries  are not considered due to lack of data. This is a limitation since Hungary is a small, open economy with significant exposures to shocks in the global economy. In principle the vulnerability to initial shocks from import and export relationships can be considered in our framework by using coarse grained sector level import-export information.

Despite these limitations, our findings have a number of policy implications. 

{\bf Monitoring.} The economic systemic risk index for individual companies, ESRI, allows countries to use VAT tax data to identify critical companies in the economy and their critical supply relations (via their marginal ESRI contributions, see \citep{poledna2016elimination}). Our findings indicate that only a few firms pose a substantial risk to the overall economy. These should be monitored closely if they produce goods of societal relevance. In particular, sharp increases in firms' systemic riskiness can be monitored over time. Several countries are currently implementing ``supply chain due diligence'' laws, for example, Germany \cite{Solomon2021} or the USA \citep{biden2021}. The possibilities for systemic risk monitoring as presented in this paper should be kept in mind, when designing new regulation. 

{\bf Systemic risk mitigation.} 
A straightforward way to increase resilience in the real economy is to introduce supply chain redundancies, where risk is reduced by altering the network structure and thereby reducing the probability for inherently systemically risky firms to suffer failures that result from a lack of inputs. We have shown that a simple network analysis allows us to identify the inherently risky firms. For those, contingency plans should be established to buy time for changing suppliers or to build alternative production capacities. The change of ESRI over time would allow countries to monitor if implemented policies really increase resilience levels of the economy. 

{\bf Avoiding risk concentration.}
In \cite{poledna2016elimination} it has been shown how incentive schemes can be designed to make financial networks safer without making them less efficient. A similar scheme could be devised for production networks by de-incentivizing critical supply relations. A simple scheme would be to have at least one back-up supplier for every critical product that is shipped to inherently risky firms. Like in regulations of the financial sector, risk concentration can be avoided by demanding that no customer should have more than a certain supply exposure (e.g. 10\% of total exposure). 

{\bf Inventory buffers.}
A natural possibility is to introduce mandatory inventory buffers for systemically risky companies, to ensure production in situations where a set of critical suppliers default.

{\bf Make economic systemic risk visible for firms.}
Today, companies usually manage their direct suppliers. Most companies do not know their higher-order dependencies in the production network \cite{geodis2017}. It is conceivable that it could be highly beneficial to many producing companies to obtain a better overview on their actual supply and customer risks. If countries provided a more global view on supply chains to identify critical situations, this could lead to deeper and more proactive systemic supply chain management, that would increase overall economic resilience by using market mechanisms. 

Policies and regulatory measures of this kind might run contrary to the hunt for efficiency that dominated the last decades, since supplier relations are time- and cost intensive.
More resilience does not pay off in the short term. It would be a fascinating question to see to what extent the economy could be made more resilient by not making it less efficient, i.e. to maximize the resilience-gain per link. For financial networks the potential of designing networks with lower systemic risk, but comparable economic function was recently highlighted in \cite{diem2020minimal, pichler2021systemic}. A related question is whether service-based economies are more resilient than production-based ones because services typically require less critical suppliers (due to their linear production functions). Instead of asking whether large economies are more stable than small ones \cite{moran2019may}, our methodology, given data from other countries, could contribute to the question, whether production-based economies or economies with high levels of input-redundancy are more resilient.

\section*{Material and Methods}

	The economic systemic risk index for firm $i$ is calculated as
	\begin{equation} \label{eq:matmethods_esri}
		{\rm ESRI}_{i} = \sum_{j=1}^{n} \frac{s^\text{out}_j }{\sum_{l=1}^n s^\text{out}_l }\big(1-h_j(T) \big) \quad ,
	\end{equation} 
	where $h_j(T) = \frac{x_j(T)}{x_j(0)}$ is the fraction of the output remaining of firm $j$ at time $T$, where $T$ is the convergence time, see Appendix \ref{SI:ESRI}. ESRI$_i$, is interpreted as the fraction of output that is lost in the entire production network in response to the failure of individual company $i$, given that its supply and demand can not be replaced. The vector $h(T)=\min(h^d(T),h^u(T))$ is the result of propagating the shock from firm $i$'s initial default downstream along the out-links by updating
	\begin{eqnarray}\label{eq:matmethods_downstream}
		x_l^{\text{d}}(t+1) & = &   f_j\Bigg(\sum_{j=1}^{n} W_{ji} h_j^\text{d}(t) \delta_{p_j,1},  \dots, \\ 
		& & \qquad \sum_{j=1}^{n} W_{ji}h_j^\text{d}(t) \delta_{p_j,m}\Bigg) \quad , \notag
	\end{eqnarray}
	and upstream along the in-links by updating 
	\begin{eqnarray}\label{eq:matmethods_upstream}
		x_l^{\text{u}}(t+1) & = &    \sum_{j=1}^{n} W_{lj}h_j^\text{u}(t) \quad. 
	\end{eqnarray} 
	For details of the calibration to the GL, see Appendix \ref{SI:ESRI}, Appendix \ref{SI:derivation}, for the implementation, see Appendix \ref{SI:propagation}.

% Bibliography

\bibliography{references}

\begin{thebibliography}{75}
\expandafter\ifx\csname natexlab\endcsname\relax\def\natexlab#1{#1}\fi
\expandafter\ifx\csname bibnamefont\endcsname\relax
  \def\bibnamefont#1{#1}\fi
\expandafter\ifx\csname bibfnamefont\endcsname\relax
  \def\bibfnamefont#1{#1}\fi
\expandafter\ifx\csname citenamefont\endcsname\relax
  \def\citenamefont#1{#1}\fi
\expandafter\ifx\csname url\endcsname\relax
  \def\url#1{\texttt{#1}}\fi
\expandafter\ifx\csname urlprefix\endcsname\relax\def\urlprefix{URL }\fi
\providecommand{\bibinfo}[2]{#2}
\providecommand{\eprint}[2][]{\url{#2}}

\bibitem[{\citenamefont{Acemoglu} \emph{et~al.}(2012)\citenamefont{Acemoglu,
  Carvalho, Ozdaglar, and Tahbaz-Salehi}}]{acemoglu2012network}
\bibinfo{author}{\bibnamefont{Acemoglu}, \bibfnamefont{D.}},
  \bibinfo{author}{\bibfnamefont{V.~M.} \bibnamefont{Carvalho}},
  \bibinfo{author}{\bibfnamefont{A.}~\bibnamefont{Ozdaglar}}, and
  \bibinfo{author}{\bibfnamefont{A.}~\bibnamefont{Tahbaz-Salehi}},
  \bibinfo{year}{2012}, \bibinfo{journal}{Econometrica}
  \textbf{\bibinfo{volume}{80}}(\bibinfo{number}{5}), \bibinfo{pages}{1977},
  \urlprefix\url{https://onlinelibrary.wiley.com/doi/abs/10.3982/ECTA9623}.

\bibitem[{\citenamefont{Allen and Gale}(2000)}]{allen2000financial}
\bibinfo{author}{\bibnamefont{Allen}, \bibfnamefont{F.}}, and
  \bibinfo{author}{\bibfnamefont{D.}~\bibnamefont{Gale}}, \bibinfo{year}{2000},
  \bibinfo{journal}{Journal of {P}olitical {E}conomy}
  \textbf{\bibinfo{volume}{108}}(\bibinfo{number}{1}), \bibinfo{pages}{1}.

\bibitem[{\citenamefont{Arinaminpathy}
  \emph{et~al.}(2012)\citenamefont{Arinaminpathy, Kapadia, and
  May}}]{arinaminpathy2012size}
\bibinfo{author}{\bibnamefont{Arinaminpathy}, \bibfnamefont{N.}},
  \bibinfo{author}{\bibfnamefont{S.}~\bibnamefont{Kapadia}}, and
  \bibinfo{author}{\bibfnamefont{R.~M.} \bibnamefont{May}},
  \bibinfo{year}{2012}, \bibinfo{journal}{Proceedings of the National Academy
  of Sciences} \textbf{\bibinfo{volume}{109}}(\bibinfo{number}{45}),
  \bibinfo{pages}{18338}, ISSN \bibinfo{issn}{0027-8424},
  \urlprefix\url{https://www.pnas.org/content/109/45/18338}.

\bibitem[{\citenamefont{Atalay} \emph{et~al.}(2011)\citenamefont{Atalay,
  Hortacsu, Roberts, and Syverson}}]{atalay2011network}
\bibinfo{author}{\bibnamefont{Atalay}, \bibfnamefont{E.}},
  \bibinfo{author}{\bibfnamefont{A.}~\bibnamefont{Hortacsu}},
  \bibinfo{author}{\bibfnamefont{J.}~\bibnamefont{Roberts}}, and
  \bibinfo{author}{\bibfnamefont{C.}~\bibnamefont{Syverson}},
  \bibinfo{year}{2011}, \bibinfo{journal}{Proceedings of the National Academy
  of Sciences} \textbf{\bibinfo{volume}{108}}(\bibinfo{number}{13}),
  \bibinfo{pages}{5199}, ISSN \bibinfo{issn}{0027-8424},
  \urlprefix\url{https://www.pnas.org/content/108/13/5199}.

\bibitem[{\citenamefont{Bak} \emph{et~al.}(1993)\citenamefont{Bak, Chen,
  Scheinkman, and Woodford}}]{bak1993aggregate}
\bibinfo{author}{\bibnamefont{Bak}, \bibfnamefont{P.}},
  \bibinfo{author}{\bibfnamefont{K.}~\bibnamefont{Chen}},
  \bibinfo{author}{\bibfnamefont{J.}~\bibnamefont{Scheinkman}}, and
  \bibinfo{author}{\bibfnamefont{M.}~\bibnamefont{Woodford}},
  \bibinfo{year}{1993}, \bibinfo{journal}{Ricerche Economiche}
  \textbf{\bibinfo{volume}{47}}(\bibinfo{number}{1}), \bibinfo{pages}{3}, ISSN
  \bibinfo{issn}{0035-5054},
  \urlprefix\url{https://www.sciencedirect.com/science/article/pii/003550549390023V}.

\bibitem[{\citenamefont{Baqaee and Farhi}(2019)}]{baqaee2019macroeconomic}
\bibinfo{author}{\bibnamefont{Baqaee}, \bibfnamefont{D.~R.}}, and
  \bibinfo{author}{\bibfnamefont{E.}~\bibnamefont{Farhi}},
  \bibinfo{year}{2019}, \bibinfo{journal}{Econometrica}
  \textbf{\bibinfo{volume}{87}}(\bibinfo{number}{4}), \bibinfo{pages}{1155},
  \urlprefix\url{https://onlinelibrary.wiley.com/doi/abs/10.3982/ECTA15202}.

\bibitem[{\citenamefont{Bardoscia} \emph{et~al.}(2015)\citenamefont{Bardoscia,
  Battiston, Caccioli, and Caldarelli}}]{bardoscia2015debtrank}
\bibinfo{author}{\bibnamefont{Bardoscia}, \bibfnamefont{M.}},
  \bibinfo{author}{\bibfnamefont{S.}~\bibnamefont{Battiston}},
  \bibinfo{author}{\bibfnamefont{F.}~\bibnamefont{Caccioli}}, and
  \bibinfo{author}{\bibfnamefont{G.}~\bibnamefont{Caldarelli}},
  \bibinfo{year}{2015}, \bibinfo{journal}{{PLOS ONE}}
  \textbf{\bibinfo{volume}{10}}(\bibinfo{number}{7}),
  \bibinfo{pages}{e0130406}.

\bibitem[{\citenamefont{Barrot and Sauvagnat}(2016)}]{barrot2016input}
\bibinfo{author}{\bibnamefont{Barrot}, \bibfnamefont{J.-N.}}, and
  \bibinfo{author}{\bibfnamefont{J.}~\bibnamefont{Sauvagnat}},
  \bibinfo{year}{2016}, \bibinfo{journal}{The Quarterly Journal of Economics}
  \textbf{\bibinfo{volume}{131}}(\bibinfo{number}{3}), \bibinfo{pages}{1543},
  ISSN \bibinfo{issn}{0033-5533},
  \urlprefix\url{https://doi.org/10.1093/qje/qjw018}.

\bibitem[{\citenamefont{Battiston} \emph{et~al.}(2016)\citenamefont{Battiston,
  Caldarelli, May, Roukny, and Stiglitz}}]{battiston2016price}
\bibinfo{author}{\bibnamefont{Battiston}, \bibfnamefont{S.}},
  \bibinfo{author}{\bibfnamefont{G.}~\bibnamefont{Caldarelli}},
  \bibinfo{author}{\bibfnamefont{R.~M.} \bibnamefont{May}},
  \bibinfo{author}{\bibfnamefont{T.}~\bibnamefont{Roukny}}, and
  \bibinfo{author}{\bibfnamefont{J.~E.} \bibnamefont{Stiglitz}},
  \bibinfo{year}{2016}, \bibinfo{journal}{Proceedings of the National Academy
  of Sciences} \textbf{\bibinfo{volume}{113}}(\bibinfo{number}{36}),
  \bibinfo{pages}{10031}, ISSN \bibinfo{issn}{0027-8424},
  \urlprefix\url{https://www.pnas.org/content/113/36/10031}.

\bibitem[{\citenamefont{Battiston} \emph{et~al.}(2012)\citenamefont{Battiston,
  Puliga, Kaushik, Tasca, and Caldarelli}}]{battiston2012debtrank}
\bibinfo{author}{\bibnamefont{Battiston}, \bibfnamefont{S.}},
  \bibinfo{author}{\bibfnamefont{M.}~\bibnamefont{Puliga}},
  \bibinfo{author}{\bibfnamefont{R.}~\bibnamefont{Kaushik}},
  \bibinfo{author}{\bibfnamefont{P.}~\bibnamefont{Tasca}}, and
  \bibinfo{author}{\bibfnamefont{G.}~\bibnamefont{Caldarelli}},
  \bibinfo{year}{2012}, \bibinfo{journal}{Scientific Reports}
  \textbf{\bibinfo{volume}{2}}(\bibinfo{number}{541}).

\bibitem[{\citenamefont{Beale} \emph{et~al.}(2011)\citenamefont{Beale, Rand,
  Battey, Croxson, May, and Nowak}}]{beale2011individual}
\bibinfo{author}{\bibnamefont{Beale}, \bibfnamefont{N.}},
  \bibinfo{author}{\bibfnamefont{D.~G.} \bibnamefont{Rand}},
  \bibinfo{author}{\bibfnamefont{H.}~\bibnamefont{Battey}},
  \bibinfo{author}{\bibfnamefont{K.}~\bibnamefont{Croxson}},
  \bibinfo{author}{\bibfnamefont{R.~M.} \bibnamefont{May}}, and
  \bibinfo{author}{\bibfnamefont{M.~A.} \bibnamefont{Nowak}},
  \bibinfo{year}{2011}, \bibinfo{journal}{Proceedings of the National Academy
  of Sciences} \textbf{\bibinfo{volume}{108}}(\bibinfo{number}{31}),
  \bibinfo{pages}{12647}, ISSN \bibinfo{issn}{0027-8424},
  \urlprefix\url{https://www.pnas.org/content/108/31/12647}.

\bibitem[{\citenamefont{Biden}(2021)}]{biden2021}
\bibinfo{author}{\bibnamefont{Biden}, \bibfnamefont{J.~R.}},
  \bibinfo{year}{2021}, \bibinfo{title}{Executive order on america’s supply
  chains},
  \urlprefix\url{https://www.whitehouse.gov/briefing-room/presidential-actions/2021/02/24/executive-order-on-americas-supply-chains/}.

\bibitem[{\citenamefont{Bloomberg}(2020)}]{bloomberg2021}
\bibinfo{author}{\bibnamefont{Bloomberg}}, \bibinfo{year}{2020},
  \bibinfo{title}{Meat-{S}hortage {R}isk {C}limbs with 25\% of {U.S.} {P}ork
  {C}apacity {O}ffline},
  \urlprefix\url{https://www.bloomberg.com/news/articles/2020-04-22/tyson-foods-to-indefinitely-suspend-waterloo/operations-k9bbgnr9}.

\bibitem[{\citenamefont{Borsos and Stancsics}(2020)}]{borsos2020unfolding}
\bibinfo{author}{\bibnamefont{Borsos}, \bibfnamefont{A.}}, and
  \bibinfo{author}{\bibfnamefont{M.}~\bibnamefont{Stancsics}},
  \bibinfo{year}{2020}, \emph{\bibinfo{title}{Unfolding the hidden structure of
  the {H}ungarian multi-layer firm network}}, \bibinfo{type}{Technical Report},
  \bibinfo{institution}{Magyar Nemzeti Bank (Central Bank of Hungary)}.

\bibitem[{\citenamefont{Boss}
  \emph{et~al.}(2004{\natexlab{a}})\citenamefont{Boss, Elsinger, Summer, and
  4}}]{boss2004network}
\bibinfo{author}{\bibnamefont{Boss}, \bibfnamefont{M.}},
  \bibinfo{author}{\bibfnamefont{H.}~\bibnamefont{Elsinger}},
  \bibinfo{author}{\bibfnamefont{M.}~\bibnamefont{Summer}}, and
  \bibinfo{author}{\bibfnamefont{S.~T.} \bibnamefont{4}},
  \bibinfo{year}{2004}{\natexlab{a}}, \bibinfo{journal}{Quantitative Finance}
  \textbf{\bibinfo{volume}{4}}(\bibinfo{number}{6}), \bibinfo{pages}{677},
  \urlprefix\url{https://www.tandfonline.com/doi/abs/10.1080/14697680400020325}.

\bibitem[{\citenamefont{Boss}
  \emph{et~al.}(2004{\natexlab{b}})\citenamefont{Boss, Summer, and
  Thurner}}]{boss2004contagion}
\bibinfo{author}{\bibnamefont{Boss}, \bibfnamefont{M.}},
  \bibinfo{author}{\bibfnamefont{M.}~\bibnamefont{Summer}}, and
  \bibinfo{author}{\bibfnamefont{S.}~\bibnamefont{Thurner}},
  \bibinfo{year}{2004}{\natexlab{b}}, in
  \emph{\bibinfo{booktitle}{International Conference on Computational Science}}
  (\bibinfo{organization}{Springer}), pp. \bibinfo{pages}{1070--1077}.

\bibitem[{\citenamefont{Carvalho}(2014)}]{carvalho2014micro}
\bibinfo{author}{\bibnamefont{Carvalho}, \bibfnamefont{V.~M.}},
  \bibinfo{year}{2014}, \bibinfo{journal}{Journal of Economic Perspectives}
  \textbf{\bibinfo{volume}{28}}(\bibinfo{number}{4}), \bibinfo{pages}{23},
  \urlprefix\url{https://www.aeaweb.org/articles?id=10.1257/jep.28.4.23}.

\bibitem[{\citenamefont{Carvalho} \emph{et~al.}(2020)\citenamefont{Carvalho,
  Nirei, Saito, and Tahbaz-Salehi}}]{carvalho2021supply}
\bibinfo{author}{\bibnamefont{Carvalho}, \bibfnamefont{V.~M.}},
  \bibinfo{author}{\bibfnamefont{M.}~\bibnamefont{Nirei}},
  \bibinfo{author}{\bibfnamefont{Y.~U.} \bibnamefont{Saito}}, and
  \bibinfo{author}{\bibfnamefont{A.}~\bibnamefont{Tahbaz-Salehi}},
  \bibinfo{year}{2020}, \bibinfo{journal}{The Quarterly Journal of Economics}
  \textbf{\bibinfo{volume}{136}}(\bibinfo{number}{2}), \bibinfo{pages}{1255},
  ISSN \bibinfo{issn}{0033-5533},
  \urlprefix\url{https://doi.org/10.1093/qje/qjaa044}.

\bibitem[{\citenamefont{Carvalho and
  Tahbaz-Salehi}(2019)}]{carvalho2019production}
\bibinfo{author}{\bibnamefont{Carvalho}, \bibfnamefont{V.~M.}}, and
  \bibinfo{author}{\bibfnamefont{A.}~\bibnamefont{Tahbaz-Salehi}},
  \bibinfo{year}{2019}, \bibinfo{journal}{Annual Review of Economics}
  \textbf{\bibinfo{volume}{11}}(\bibinfo{number}{1}), \bibinfo{pages}{635},
  \urlprefix\url{https://doi.org/10.1146/annurev-economics-080218-030212}.

\bibitem[{\citenamefont{Choi} \emph{et~al.}(2001)\citenamefont{Choi, Dooley,
  and Rungtusanatham}}]{choi2001supply}
\bibinfo{author}{\bibnamefont{Choi}, \bibfnamefont{T.~Y.}},
  \bibinfo{author}{\bibfnamefont{K.~J.} \bibnamefont{Dooley}}, and
  \bibinfo{author}{\bibfnamefont{M.}~\bibnamefont{Rungtusanatham}},
  \bibinfo{year}{2001}, \bibinfo{journal}{Journal of Operations Management}
  \textbf{\bibinfo{volume}{19}}(\bibinfo{number}{3}), \bibinfo{pages}{351},
  ISSN \bibinfo{issn}{0272-6963},
  \urlprefix\url{https://www.sciencedirect.com/science/article/pii/S0272696300000681}.

\bibitem[{\citenamefont{Choi and Krause}(2006)}]{choi2006supply}
\bibinfo{author}{\bibnamefont{Choi}, \bibfnamefont{T.~Y.}}, and
  \bibinfo{author}{\bibfnamefont{D.~R.} \bibnamefont{Krause}},
  \bibinfo{year}{2006}, \bibinfo{journal}{Journal of Operations Management}
  \textbf{\bibinfo{volume}{24}}(\bibinfo{number}{5}), \bibinfo{pages}{637},
  ISSN \bibinfo{issn}{0272-6963},
  \urlprefix\url{https://www.sciencedirect.com/science/article/pii/S0272696305001233}.

\bibitem[{\citenamefont{Clauset} \emph{et~al.}(2009)\citenamefont{Clauset,
  Shalizi, and Newman}}]{clauset2009power}
\bibinfo{author}{\bibnamefont{Clauset}, \bibfnamefont{A.}},
  \bibinfo{author}{\bibfnamefont{C.~R.} \bibnamefont{Shalizi}}, and
  \bibinfo{author}{\bibfnamefont{M.~E.~J.} \bibnamefont{Newman}},
  \bibinfo{year}{2009}, \bibinfo{journal}{SIAM Review}
  \textbf{\bibinfo{volume}{51}}(\bibinfo{number}{4}), \bibinfo{pages}{661},
  \urlprefix\url{https://doi.org/10.1137/070710111}.

\bibitem[{\citenamefont{Colon} \emph{et~al.}(2021)\citenamefont{Colon,
  Hallegatte, and Rozenberg}}]{colon2021criticality}
\bibinfo{author}{\bibnamefont{Colon}, \bibfnamefont{C.}},
  \bibinfo{author}{\bibfnamefont{S.}~\bibnamefont{Hallegatte}}, and
  \bibinfo{author}{\bibfnamefont{J.}~\bibnamefont{Rozenberg}},
  \bibinfo{year}{2021}, \bibinfo{journal}{Nature Sustainability}
  \textbf{\bibinfo{volume}{4}}(\bibinfo{number}{3}), \bibinfo{pages}{209}.

\bibitem[{\citenamefont{Cont} \emph{et~al.}(2010)\citenamefont{Cont, Moussa,
  and Santos}}]{cont2010network}
\bibinfo{author}{\bibnamefont{Cont}, \bibfnamefont{R.}},
  \bibinfo{author}{\bibfnamefont{A.}~\bibnamefont{Moussa}}, and
  \bibinfo{author}{\bibfnamefont{E.}~\bibnamefont{Santos}},
  \bibinfo{year}{2010}, \bibinfo{journal}{SSRN, doi:10.2139/ssrn.1733528} .

\bibitem[{\citenamefont{Craighead} \emph{et~al.}(2007)\citenamefont{Craighead,
  Blackhurst, Rungtusanatham, and Handfield}}]{craighead2007severity}
\bibinfo{author}{\bibnamefont{Craighead}, \bibfnamefont{C.~W.}},
  \bibinfo{author}{\bibfnamefont{J.}~\bibnamefont{Blackhurst}},
  \bibinfo{author}{\bibfnamefont{M.~J.} \bibnamefont{Rungtusanatham}}, and
  \bibinfo{author}{\bibfnamefont{R.~B.} \bibnamefont{Handfield}},
  \bibinfo{year}{2007}, \bibinfo{journal}{Decision Sciences}
  \textbf{\bibinfo{volume}{38}}(\bibinfo{number}{1}), \bibinfo{pages}{131},
  \urlprefix\url{https://onlinelibrary.wiley.com/doi/abs/10.1111/j.1540-5915.2007.00151.x}.

\bibitem[{\citenamefont{Dhyne} \emph{et~al.}(2015)\citenamefont{Dhyne,
  Magerman, and Rub{\'\i}nov{\'a}}}]{dhyne2015belgian}
\bibinfo{author}{\bibnamefont{Dhyne}, \bibfnamefont{E.}},
  \bibinfo{author}{\bibfnamefont{G.}~\bibnamefont{Magerman}}, and
  \bibinfo{author}{\bibfnamefont{S.}~\bibnamefont{Rub{\'\i}nov{\'a}}},
  \bibinfo{year}{2015}, \emph{\bibinfo{title}{The {B}elgian production network
  2002-2012}}, \bibinfo{type}{Technical Report}, \bibinfo{institution}{NBB
  Working Paper}.

\bibitem[{\citenamefont{Diem} \emph{et~al.}(2020)\citenamefont{Diem, Pichler,
  and Thurner}}]{diem2020minimal}
\bibinfo{author}{\bibnamefont{Diem}, \bibfnamefont{C.}},
  \bibinfo{author}{\bibfnamefont{A.}~\bibnamefont{Pichler}}, and
  \bibinfo{author}{\bibfnamefont{S.}~\bibnamefont{Thurner}},
  \bibinfo{year}{2020}, \bibinfo{journal}{Journal of Economic Dynamics and
  Control} \textbf{\bibinfo{volume}{116}}, \bibinfo{pages}{103900}, ISSN
  \bibinfo{issn}{0165-1889},
  \urlprefix\url{https://www.sciencedirect.com/science/article/pii/S0165188920300683}.

\bibitem[{\citenamefont{Douglas}(1976)}]{douglas1976cobb}
\bibinfo{author}{\bibnamefont{Douglas}, \bibfnamefont{P.~H.}},
  \bibinfo{year}{1976}, \bibinfo{journal}{Journal of Political Economy}
  \textbf{\bibinfo{volume}{84}}(\bibinfo{number}{5}), \bibinfo{pages}{903},
  \urlprefix\url{https://doi.org/10.1086/260489}.

\bibitem[{\citenamefont{Eisenberg and Noe}(2001)}]{eisenberg2001systemic}
\bibinfo{author}{\bibnamefont{Eisenberg}, \bibfnamefont{L.}}, and
  \bibinfo{author}{\bibfnamefont{T.~H.} \bibnamefont{Noe}},
  \bibinfo{year}{2001}, \bibinfo{journal}{Management Science}
  \textbf{\bibinfo{volume}{47}}(\bibinfo{number}{2}), \bibinfo{pages}{236},
  \urlprefix\url{https://doi.org/10.1287/mnsc.47.2.236.9835}.

\bibitem[{\citenamefont{Elsinger} \emph{et~al.}(2006)\citenamefont{Elsinger,
  Lehar, and Summer}}]{elsinger2006risk}
\bibinfo{author}{\bibnamefont{Elsinger}, \bibfnamefont{H.}},
  \bibinfo{author}{\bibfnamefont{A.}~\bibnamefont{Lehar}}, and
  \bibinfo{author}{\bibfnamefont{M.}~\bibnamefont{Summer}},
  \bibinfo{year}{2006}, \bibinfo{journal}{Management Science}
  \textbf{\bibinfo{volume}{52}}(\bibinfo{number}{9}), \bibinfo{pages}{1301},
  \urlprefix\url{https://doi.org/10.1287/mnsc.1060.0531}.

\bibitem[{\citenamefont{EUROSTAT}(2021)}]{ramon2021nace}
\bibinfo{author}{\bibnamefont{EUROSTAT}}, \bibinfo{year}{2021},
  \bibinfo{title}{Your companion guide to international statistical
  classifications. section iv - description of the main economic
  classifications},
  \urlprefix\url{https://ec.europa.eu/eurostat/ramon/miscellaneous/index.cfm?TargetUrl=DSP_GENINFO_CLASS_4}.

\bibitem[{\citenamefont{{Financial Times}}(2021{\natexlab{a}})}]{Miller2021}
\bibinfo{author}{\bibnamefont{{Financial Times}}},
  \bibinfo{year}{2021}{\natexlab{a}}, \bibinfo{title}{Chip shortage forces audi
  to delay production},
  \urlprefix\url{https://www.ft.com/content/8cb74f6a-3859-4b13-885e-0e79a4d5de1a}.

\bibitem[{\citenamefont{{Financial Times}}(2021{\natexlab{b}})}]{Bushey2021}
\bibinfo{author}{\bibnamefont{{Financial Times}}},
  \bibinfo{year}{2021}{\natexlab{b}}, \bibinfo{title}{Ford says chip shortage
  could knock \$2.5bn from earnings},
  \urlprefix\url{https://www.ft.com/content/a0d5ca20-2559-4d47-9106-cf50eaf97720}.

\bibitem[{\citenamefont{{Financial Times}}(2021{\natexlab{c}})}]{Solomon2021}
\bibinfo{author}{\bibnamefont{{Financial Times}}},
  \bibinfo{year}{2021}{\natexlab{c}}, \bibinfo{title}{German cabinet backs law
  to protect human rights in global supply chain},
  \urlprefix\url{https://www.ft.com/content/2b969d2c-ad1e-48c2-b318-2dd20bb1662f}.

\bibitem[{\citenamefont{Freixas} \emph{et~al.}(2000)\citenamefont{Freixas,
  Parigi, and Rochet}}]{freixas1998systemic}
\bibinfo{author}{\bibnamefont{Freixas}, \bibfnamefont{X.}},
  \bibinfo{author}{\bibfnamefont{B.~M.} \bibnamefont{Parigi}}, and
  \bibinfo{author}{\bibfnamefont{J.~C.} \bibnamefont{Rochet}},
  \bibinfo{year}{2000}, \bibinfo{journal}{Journal of Money, Credit and Banking}
  \textbf{\bibinfo{volume}{32}}(\bibinfo{number}{3}), \bibinfo{pages}{611}.

\bibitem[{\citenamefont{Furfine}(2003)}]{furfine2003interbank}
\bibinfo{author}{\bibnamefont{Furfine}, \bibfnamefont{C.~H.}},
  \bibinfo{year}{2003}, \bibinfo{journal}{Journal of Money, Credit and Banking}
  \textbf{\bibinfo{volume}{35}}(\bibinfo{number}{1}), \bibinfo{pages}{111},
  ISSN \bibinfo{issn}{00222879, 15384616},
  \urlprefix\url{http://www.jstor.org/stable/3649847}.

\bibitem[{\citenamefont{Gabaix}(2011)}]{gabaix2011granular}
\bibinfo{author}{\bibnamefont{Gabaix}, \bibfnamefont{X.}},
  \bibinfo{year}{2011}, \bibinfo{journal}{Econometrica}
  \textbf{\bibinfo{volume}{79}}(\bibinfo{number}{3}), \bibinfo{pages}{733},
  \urlprefix\url{https://onlinelibrary.wiley.com/doi/abs/10.3982/ECTA8769}.

\bibitem[{\citenamefont{Gai and Kapadia}(2010)}]{gai2010contagion}
\bibinfo{author}{\bibnamefont{Gai}, \bibfnamefont{P.}}, and
  \bibinfo{author}{\bibfnamefont{S.}~\bibnamefont{Kapadia}},
  \bibinfo{year}{2010}, \bibinfo{journal}{Proceedings of the Royal Society A:
  Mathematical, Physical and Engineering Sciences}
  \textbf{\bibinfo{volume}{466}}(\bibinfo{number}{2120}),
  \bibinfo{pages}{2401}.

\bibitem[{\citenamefont{GEODIS}(2017)}]{geodis2017}
\bibinfo{author}{\bibnamefont{GEODIS}}, \bibinfo{year}{2017},
  \emph{\bibinfo{title}{{SUPPLY CHAIN WORLDWIDE SURVEY}}},
  \bibinfo{type}{Technical Report}, \bibinfo{institution}{GEODIS},
  \urlprefix\url{https://geodis.com/sites/default/files/2019-03/170509_GEODIS_WHITE-PAPER.PDF}.

\bibitem[{\citenamefont{Giannetti} \emph{et~al.}(2011)\citenamefont{Giannetti,
  Burkart, and Ellingsen}}]{giannetti2011you}
\bibinfo{author}{\bibnamefont{Giannetti}, \bibfnamefont{M.}},
  \bibinfo{author}{\bibfnamefont{M.}~\bibnamefont{Burkart}}, and
  \bibinfo{author}{\bibfnamefont{T.}~\bibnamefont{Ellingsen}},
  \bibinfo{year}{2011}, \bibinfo{journal}{The Review of Financial Studies}
  \textbf{\bibinfo{volume}{24}}(\bibinfo{number}{4}), \bibinfo{pages}{1261},
  ISSN \bibinfo{issn}{0893-9454},
  \urlprefix\url{https://doi.org/10.1093/rfs/hhn096}.

\bibitem[{\citenamefont{Hallegatte}(2008)}]{hallegatte2008adaptive}
\bibinfo{author}{\bibnamefont{Hallegatte}, \bibfnamefont{S.}},
  \bibinfo{year}{2008}, \bibinfo{journal}{Risk Analysis: An International
  Journal} \textbf{\bibinfo{volume}{28}}(\bibinfo{number}{3}),
  \bibinfo{pages}{779},
  \urlprefix\url{https://onlinelibrary.wiley.com/doi/abs/10.1111/j.1539-6924.2008.01046.x}.

\bibitem[{\citenamefont{Hanel} \emph{et~al.}(2017)\citenamefont{Hanel,
  Corominas-Murtra, Liu, and Thurner}}]{hanel2017fitting}
\bibinfo{author}{\bibnamefont{Hanel}, \bibfnamefont{R.}},
  \bibinfo{author}{\bibfnamefont{B.}~\bibnamefont{Corominas-Murtra}},
  \bibinfo{author}{\bibfnamefont{B.}~\bibnamefont{Liu}}, and
  \bibinfo{author}{\bibfnamefont{S.}~\bibnamefont{Thurner}},
  \bibinfo{year}{2017}, \bibinfo{journal}{PloS one}
  \textbf{\bibinfo{volume}{12}}(\bibinfo{number}{2}),
  \bibinfo{pages}{e0170920}.

\bibitem[{\citenamefont{Hulten}(1978)}]{Hulten1978Theorem}
\bibinfo{author}{\bibnamefont{Hulten}, \bibfnamefont{C.~R.}},
  \bibinfo{year}{1978}, \bibinfo{journal}{The Review of Economic Studies}
  \textbf{\bibinfo{volume}{45}}(\bibinfo{number}{3}), \bibinfo{pages}{511},
  ISSN \bibinfo{issn}{00346527, 1467937X},
  \urlprefix\url{http://www.jstor.org/stable/2297252}.

\bibitem[{\citenamefont{Hummels} \emph{et~al.}(2001)\citenamefont{Hummels,
  Ishii, and Yi}}]{hummels2001nature}
\bibinfo{author}{\bibnamefont{Hummels}, \bibfnamefont{D.}},
  \bibinfo{author}{\bibfnamefont{J.}~\bibnamefont{Ishii}}, and
  \bibinfo{author}{\bibfnamefont{K.-M.} \bibnamefont{Yi}},
  \bibinfo{year}{2001}, \bibinfo{journal}{Journal of International Economics}
  \textbf{\bibinfo{volume}{54}}(\bibinfo{number}{1}), \bibinfo{pages}{75}, ISSN
  \bibinfo{issn}{0022-1996},
  \urlprefix\url{https://www.sciencedirect.com/science/article/pii/S0022199600000933}.

\bibitem[{\citenamefont{Inoue and Todo}(2019)}]{inoue2019firm}
\bibinfo{author}{\bibnamefont{Inoue}, \bibfnamefont{H.}}, and
  \bibinfo{author}{\bibfnamefont{Y.}~\bibnamefont{Todo}}, \bibinfo{year}{2019},
  \bibinfo{journal}{Nature Sustainability}
  \textbf{\bibinfo{volume}{2}}(\bibinfo{number}{9}), \bibinfo{pages}{841}.

\bibitem[{\citenamefont{Ivanov and Dolgui}(2020)}]{ivanov2020viability}
\bibinfo{author}{\bibnamefont{Ivanov}, \bibfnamefont{D.}}, and
  \bibinfo{author}{\bibfnamefont{A.}~\bibnamefont{Dolgui}},
  \bibinfo{year}{2020}, \bibinfo{journal}{International Journal of Production
  Research} \textbf{\bibinfo{volume}{58}}(\bibinfo{number}{10}),
  \bibinfo{pages}{2904},
  \urlprefix\url{https://doi.org/10.1080/00207543.2020.1750727}.

\bibitem[{\citenamefont{Ivanov} \emph{et~al.}(2014)\citenamefont{Ivanov,
  Sokolov, and Dolgui}}]{ivanov2014ripple}
\bibinfo{author}{\bibnamefont{Ivanov}, \bibfnamefont{D.}},
  \bibinfo{author}{\bibfnamefont{B.}~\bibnamefont{Sokolov}}, and
  \bibinfo{author}{\bibfnamefont{A.}~\bibnamefont{Dolgui}},
  \bibinfo{year}{2014}, \bibinfo{journal}{International Journal of Production
  Research} \textbf{\bibinfo{volume}{52}}(\bibinfo{number}{7}),
  \bibinfo{pages}{2154},
  \urlprefix\url{https://doi.org/10.1080/00207543.2013.858836}.

\bibitem[{\citenamefont{Kannan and Tan}(2005)}]{kannan2005just}
\bibinfo{author}{\bibnamefont{Kannan}, \bibfnamefont{V.~R.}}, and
  \bibinfo{author}{\bibfnamefont{K.~C.} \bibnamefont{Tan}},
  \bibinfo{year}{2005}, \bibinfo{journal}{Omega}
  \textbf{\bibinfo{volume}{33}}(\bibinfo{number}{2}), \bibinfo{pages}{153},
  ISSN \bibinfo{issn}{0305-0483},
  \urlprefix\url{https://www.sciencedirect.com/science/article/pii/S030504830400060X}.

\bibitem[{\citenamefont{Lamming}(1996)}]{lamming1996squaring}
\bibinfo{author}{\bibnamefont{Lamming}, \bibfnamefont{R.}},
  \bibinfo{year}{1996}, \bibinfo{journal}{International Journal of Operations
  \& Production Management} \textbf{\bibinfo{volume}{16}}(\bibinfo{number}{2}),
  \bibinfo{pages}{183},
  \urlprefix\url{https://doi.org/10.1108/01443579610109910}.

\bibitem[{\citenamefont{Leontief}(1928)}]{Leontieff1928}
\bibinfo{author}{\bibnamefont{Leontief}, \bibfnamefont{W.}},
  \bibinfo{year}{1928}, \bibinfo{journal}{{Archiv fur Sozialwissenschaft und
  Sozialpolitik}} \textbf{\bibinfo{volume}{60}}, \bibinfo{pages}{577}.

\bibitem[{\citenamefont{Leontief}(1991)}]{leontief1991economy}
\bibinfo{author}{\bibnamefont{Leontief}, \bibfnamefont{W.}},
  \bibinfo{year}{1991}, \bibinfo{journal}{Structural Change and Economic
  Dynamics} \textbf{\bibinfo{volume}{2}}(\bibinfo{number}{1}),
  \bibinfo{pages}{181}, ISSN \bibinfo{issn}{0954-349X},
  \urlprefix\url{https://www.sciencedirect.com/science/article/pii/0954349X9190012H}.

\bibitem[{\citenamefont{Lequiller and
  Blades}(2014)}]{lequiller2014understanding}
\bibinfo{author}{\bibnamefont{Lequiller}, \bibfnamefont{F.}}, and
  \bibinfo{author}{\bibfnamefont{D.}~\bibnamefont{Blades}},
  \bibinfo{year}{2014}, \emph{\bibinfo{title}{Understanding National Accounts}}
  (\bibinfo{publisher}{OECD Paris}),
  \urlprefix\url{https://www.oecd-ilibrary.org/content/publication/9789264214637-en}.

\bibitem[{\citenamefont{Magerman} \emph{et~al.}(2016)\citenamefont{Magerman,
  De~Bruyne, Dhyne, and Van~Hove}}]{magerman2016heterogeneous}
\bibinfo{author}{\bibnamefont{Magerman}, \bibfnamefont{G.}},
  \bibinfo{author}{\bibfnamefont{K.}~\bibnamefont{De~Bruyne}},
  \bibinfo{author}{\bibfnamefont{E.}~\bibnamefont{Dhyne}}, and
  \bibinfo{author}{\bibfnamefont{J.}~\bibnamefont{Van~Hove}},
  \bibinfo{year}{2016}, \emph{\bibinfo{title}{Heterogeneous firms and the micro
  origins of aggregate fluctuations}}, \bibinfo{type}{Technical Report},
  \bibinfo{institution}{NBB Working Paper}.

\bibitem[{\citenamefont{Markose} \emph{et~al.}(2012)\citenamefont{Markose,
  Giansante, and Shaghaghi}}]{markose2012systemic}
\bibinfo{author}{\bibnamefont{Markose}, \bibfnamefont{S.}},
  \bibinfo{author}{\bibfnamefont{S.}~\bibnamefont{Giansante}}, and
  \bibinfo{author}{\bibfnamefont{A.~R.} \bibnamefont{Shaghaghi}},
  \bibinfo{year}{2012}, \bibinfo{journal}{Journal of Economic Behavior \&
  Organization} \textbf{\bibinfo{volume}{83}}(\bibinfo{number}{3}),
  \bibinfo{pages}{627}, ISSN \bibinfo{issn}{0167-2681},
  \urlprefix\url{https://www.sciencedirect.com/science/article/pii/S0167268112001254}.

\bibitem[{\citenamefont{McFadden}(1963)}]{mcfadden1963constant}
\bibinfo{author}{\bibnamefont{McFadden}, \bibfnamefont{D.}},
  \bibinfo{year}{1963}, \bibinfo{journal}{The Review of Economic Studies}
  \textbf{\bibinfo{volume}{30}}(\bibinfo{number}{2}), \bibinfo{pages}{73}, ISSN
  \bibinfo{issn}{00346527, 1467937X},
  \urlprefix\url{http://www.jstor.org/stable/2295804}.

\bibitem[{\citenamefont{Miller and Blair}(2009)}]{miller2009input}
\bibinfo{author}{\bibnamefont{Miller}, \bibfnamefont{R.~E.}}, and
  \bibinfo{author}{\bibfnamefont{P.~D.} \bibnamefont{Blair}},
  \bibinfo{year}{2009}, \emph{\bibinfo{title}{Input-{O}utput {A}nalysis:
  {F}oundations and {E}xtensions}} (\bibinfo{publisher}{Cambridge University
  Press}).

\bibitem[{\citenamefont{Moran and Bouchaud}(2019)}]{moran2019may}
\bibinfo{author}{\bibnamefont{Moran}, \bibfnamefont{J.}}, and
  \bibinfo{author}{\bibfnamefont{J.-P.} \bibnamefont{Bouchaud}},
  \bibinfo{year}{2019}, \bibinfo{journal}{Phys. Rev. E}
  \textbf{\bibinfo{volume}{100}}, \bibinfo{pages}{032307},
  \urlprefix\url{https://link.aps.org/doi/10.1103/PhysRevE.100.032307}.

\bibitem[{\citenamefont{Nier} \emph{et~al.}(2007)\citenamefont{Nier, Yang,
  Yorulmazer, and Alentorn}}]{nier2007network}
\bibinfo{author}{\bibnamefont{Nier}, \bibfnamefont{E.}},
  \bibinfo{author}{\bibfnamefont{J.}~\bibnamefont{Yang}},
  \bibinfo{author}{\bibfnamefont{T.}~\bibnamefont{Yorulmazer}}, and
  \bibinfo{author}{\bibfnamefont{A.}~\bibnamefont{Alentorn}},
  \bibinfo{year}{2007}, \bibinfo{journal}{Journal of Economic Dynamics and
  Control} \textbf{\bibinfo{volume}{31}}(\bibinfo{number}{6}),
  \bibinfo{pages}{2033}.

\bibitem[{\citenamefont{OECD}(2020)}]{oecd2020VAT}
\bibinfo{author}{\bibnamefont{OECD}}, \bibinfo{year}{2020},
  \emph{\bibinfo{title}{Consumption Tax Trends 2020}} (\bibinfo{publisher}{OECD
  Paris}),
  \urlprefix\url{https://www.oecd-ilibrary.org/content/publication/152def2d-en}.

\bibitem[{\citenamefont{Pichler and Farmer}(2021)}]{pichler2021modeling}
\bibinfo{author}{\bibnamefont{Pichler}, \bibfnamefont{A.}}, and
  \bibinfo{author}{\bibfnamefont{J.~D.} \bibnamefont{Farmer}},
  \bibinfo{year}{2021}, \bibinfo{title}{Modeling simultaneous supply and demand
  shocks in input-output networks}.

\bibitem[{\citenamefont{Pichler} \emph{et~al.}(2020)\citenamefont{Pichler,
  Pangallo, del Rio-Chanona, Lafond, and Farmer}}]{pichler2020production}
\bibinfo{author}{\bibnamefont{Pichler}, \bibfnamefont{A.}},
  \bibinfo{author}{\bibfnamefont{M.}~\bibnamefont{Pangallo}},
  \bibinfo{author}{\bibfnamefont{R.~M.} \bibnamefont{del Rio-Chanona}},
  \bibinfo{author}{\bibfnamefont{F.}~\bibnamefont{Lafond}}, and
  \bibinfo{author}{\bibfnamefont{J.~D.} \bibnamefont{Farmer}},
  \bibinfo{year}{2020}, \bibinfo{title}{Production {N}etworks and {E}pidemic
  {S}preading: {H}ow to {R}estart the {UK} {E}conomy?}

\bibitem[{\citenamefont{Pichler} \emph{et~al.}(2021)\citenamefont{Pichler,
  Poledna, and Thurner}}]{pichler2021systemic}
\bibinfo{author}{\bibnamefont{Pichler}, \bibfnamefont{A.}},
  \bibinfo{author}{\bibfnamefont{S.}~\bibnamefont{Poledna}}, and
  \bibinfo{author}{\bibfnamefont{S.}~\bibnamefont{Thurner}},
  \bibinfo{year}{2021}, \bibinfo{journal}{Journal of Financial Stability}
  \textbf{\bibinfo{volume}{52}}, \bibinfo{pages}{100809}, ISSN
  \bibinfo{issn}{1572-3089},
  \urlprefix\url{https://www.sciencedirect.com/science/article/pii/S1572308920301121}.

\bibitem[{\citenamefont{Poledna} \emph{et~al.}(2015)\citenamefont{Poledna,
  Molina-Borboa, Martínez-Jaramillo, {van der Leij}, and
  Thurner}}]{poledna2015the}
\bibinfo{author}{\bibnamefont{Poledna}, \bibfnamefont{S.}},
  \bibinfo{author}{\bibfnamefont{J.~L.} \bibnamefont{Molina-Borboa}},
  \bibinfo{author}{\bibfnamefont{S.}~\bibnamefont{Martínez-Jaramillo}},
  \bibinfo{author}{\bibfnamefont{M.}~\bibnamefont{{van der Leij}}}, and
  \bibinfo{author}{\bibfnamefont{S.}~\bibnamefont{Thurner}},
  \bibinfo{year}{2015}, \bibinfo{journal}{Journal of Financial Stability}
  \textbf{\bibinfo{volume}{20}}, \bibinfo{pages}{70}, ISSN
  \bibinfo{issn}{1572-3089},
  \urlprefix\url{https://www.sciencedirect.com/science/article/pii/S1572308915000856}.

\bibitem[{\citenamefont{Poledna and Thurner}(2016)}]{poledna2016elimination}
\bibinfo{author}{\bibnamefont{Poledna}, \bibfnamefont{S.}}, and
  \bibinfo{author}{\bibfnamefont{S.}~\bibnamefont{Thurner}},
  \bibinfo{year}{2016}, \bibinfo{journal}{Quantitative Finance}
  \textbf{\bibinfo{volume}{16}}(\bibinfo{number}{10}), \bibinfo{pages}{1599}.

\bibitem[{\citenamefont{Quélin and Duhamel}(2003)}]{quelin2003bringing}
\bibinfo{author}{\bibnamefont{Quélin}, \bibfnamefont{B.}}, and
  \bibinfo{author}{\bibfnamefont{F.}~\bibnamefont{Duhamel}},
  \bibinfo{year}{2003}, \bibinfo{journal}{European Management Journal}
  \textbf{\bibinfo{volume}{21}}(\bibinfo{number}{5}), \bibinfo{pages}{647},
  ISSN \bibinfo{issn}{0263-2373},
  \urlprefix\url{https://www.sciencedirect.com/science/article/pii/S0263237303001130}.

\bibitem[{\citenamefont{Rauch}(1999)}]{rauch1999networks}
\bibinfo{author}{\bibnamefont{Rauch}, \bibfnamefont{J.~E.}},
  \bibinfo{year}{1999}, \bibinfo{journal}{Journal of International Economics}
  \textbf{\bibinfo{volume}{48}}(\bibinfo{number}{1}), \bibinfo{pages}{7}, ISSN
  \bibinfo{issn}{0022-1996},
  \urlprefix\url{https://www.sciencedirect.com/science/article/pii/S0022199698000099}.

\bibitem[{\citenamefont{Shao} \emph{et~al.}(2018)\citenamefont{Shao, Shi, Choi,
  and Chae}}]{shao2018data}
\bibinfo{author}{\bibnamefont{Shao}, \bibfnamefont{B.~B.}},
  \bibinfo{author}{\bibfnamefont{Z.~M.} \bibnamefont{Shi}},
  \bibinfo{author}{\bibfnamefont{T.~Y.} \bibnamefont{Choi}}, and
  \bibinfo{author}{\bibfnamefont{S.}~\bibnamefont{Chae}}, \bibinfo{year}{2018},
  \bibinfo{journal}{Decision Support Systems} \textbf{\bibinfo{volume}{114}},
  \bibinfo{pages}{37}, ISSN \bibinfo{issn}{0167-9236},
  \urlprefix\url{https://www.sciencedirect.com/science/article/pii/S0167923618301374}.

\bibitem[{\citenamefont{{The Economist}}(2021)}]{economist2021}
\bibinfo{author}{\bibnamefont{{The Economist}}}, \bibinfo{year}{2021},
  \bibinfo{title}{How vaccines are made, and why it is hard},
  \urlprefix\url{https://www.economist.com/science-and-technology/2021/02/06/how-vaccines-are-made-and-why-it-is-hard}.

\bibitem[{\citenamefont{Thurner}(forthcoming)}]{thurner2020macro}
\bibinfo{author}{\bibnamefont{Thurner}, \bibfnamefont{S.}},
  \bibinfo{year}{forthcoming}, \bibinfo{title}{A {C}omplex {S}ystems
  {P}erspective on {M}acroprudential {R}egulation}.

\bibitem[{\citenamefont{Thurner and Poledna}(2013)}]{thurner2013debtrank}
\bibinfo{author}{\bibnamefont{Thurner}, \bibfnamefont{S.}}, and
  \bibinfo{author}{\bibfnamefont{S.}~\bibnamefont{Poledna}},
  \bibinfo{year}{2013}, \bibinfo{journal}{Scientific Reports}
  \textbf{\bibinfo{volume}{3}}, \bibinfo{pages}{1888}.

\bibitem[{\citenamefont{Trent and Monczka}(1998)}]{trent1998purchasing}
\bibinfo{author}{\bibnamefont{Trent}, \bibfnamefont{R.~J.}}, and
  \bibinfo{author}{\bibfnamefont{R.~M.} \bibnamefont{Monczka}},
  \bibinfo{year}{1998}, \bibinfo{journal}{International Journal of Purchasing
  and Materials Management} \textbf{\bibinfo{volume}{34}}(\bibinfo{number}{3}),
  \bibinfo{pages}{2},
  \urlprefix\url{https://onlinelibrary.wiley.com/doi/abs/10.1111/j.1745-493X.1998.tb00296.x}.

\bibitem[{\citenamefont{Trent and Monczka}(2003)}]{trent2003understanding}
\bibinfo{author}{\bibnamefont{Trent}, \bibfnamefont{R.~J.}}, and
  \bibinfo{author}{\bibfnamefont{R.~M.} \bibnamefont{Monczka}},
  \bibinfo{year}{2003}, \bibinfo{journal}{International Journal of Physical
  Distribution \& Logistics Management}
  \textbf{\bibinfo{volume}{33}}(\bibinfo{number}{7}), \bibinfo{pages}{607}.

\bibitem[{\citenamefont{Varian}(2014)}]{varian2014intermediate}
\bibinfo{author}{\bibnamefont{Varian}, \bibfnamefont{H.~R.}},
  \bibinfo{year}{2014}, \emph{\bibinfo{title}{Intermediate Microeconomics: A
  Modern Approach: Ninth International Student Edition}}
  (\bibinfo{publisher}{WW Norton \& Company}).

\bibitem[{\citenamefont{Wu}(2016)}]{wu2016shock}
\bibinfo{author}{\bibnamefont{Wu}, \bibfnamefont{D.}}, \bibinfo{year}{2016},
  \bibinfo{title}{Essays on the interface between finance and technology}.

\bibitem[{\citenamefont{Świerczek}(2014)}]{swierczek2014impact}
\bibinfo{author}{\bibnamefont{Świerczek}, \bibfnamefont{A.}},
  \bibinfo{year}{2014}, \bibinfo{journal}{International Journal of Production
  Economics} \textbf{\bibinfo{volume}{157}}, \bibinfo{pages}{89}, ISSN
  \bibinfo{issn}{0925-5273}, \bibinfo{note}{the International Society for
  Inventory Research, 2012},
  \urlprefix\url{https://www.sciencedirect.com/science/article/pii/S0925527313003654}.

\end{thebibliography}

%%%%%%%%%%%%%%%%%%%%%%%%%%%%%

\begin{acknowledgments}
This work was supported in part by 
the OeNB Hochschuljubil\"aumsfund P17795
the Austrian Science Promotion Agency FFG under 857136, 
the Austrian Science Fund FWF under P29252,
H2020 SoBigData-PlusPlus grant agreement ID 871042.   
\end{acknowledgments}

\clearpage
 
\onecolumngrid
\appendix

\section{Relation to I-O analysis}\label{SI:IO}

On the industry level studies on the propagation of shocks in production networks exist at least since the famous Leontief Input-Output analysis  \cite{Leontieff1928, leontief1991economy, miller2009input}. 
In \cite{bak1993aggregate} it is shown theoretically that small shocks to individual sectors can have effects on aggregate output. Refs. \cite{gabaix2011granular,acemoglu2012network,carvalho2014micro} confirm this finding.  \cite{hallegatte2008adaptive} assess higher order shocks from the Hurricane Katrina disaster. More recently, \cite{pichler2020production} investigated how demand and supply constraints on the sector level affect GDP. In \cite{colon2021criticality} it is modeled how shocks spread through the Tanzanian supply and transport network. The sector level analysis produced  valuable insights, but it has certain limitations due to the nature of the underlying data. Only recently high-quality large-scale firm level data has become available that allows for new insights.

The methodology presented in this paper addresses three usual shortcomings of sector level analyses. First, the data shows that even within fine grained industry classifications (NACE 4) firms tend to have considerably heterogeneous input sectors and customer sectors. In fact when comparing the pairwise input (customer) sectors of firms we see that for all firms in NACE 2611 (Manufacture of electronic components) 56\% (85\%) have no overlap of input (customer) sectors even though they belong to the same NACE 4 category, see FIG. \ref{fig:firm-level} in the main text and Appendix \ref{SI:firm_level_advantages}. Second, this intra-sector heterogeneity can lead to inaccurate results when using them for assessing shock propagation in production networks. Especially if the initial crises scenario does not affect all firms within a sector to the same extent (for example in the current COVID-19 crises). Our proposed framework takes this into account and yields different cascades for shocks which would appear to be the same at industry level, but are actually distributed heterogeneously among firms within an industry. FIG. \ref{fig:firm-level} (a) shows in a simple example how two cascades that are the same on the industry level lead to very different impacts on other firms and industry sectors. Third, in contrast to sector level models, each firm in our approach has a specific production function based on its industry classification that is calibrated to the observed individual input vector of the respective firm; see Appendix \ref{SI:productionfunction} for more information on the calibration of production functions. This matters especially since we show in FIG. \ref{fig:firm-level} (b) that firm input vectors, even in the fine grained industry classifications, vary on the sector level. The same is shown for customer sectors, see Appendix \ref{SI:firm_level_advantages}, FIG. \ref{fig:firm-level_SI} (b). 

Since, data has become available also firm level analysis has been performed. Ref. \cite{atalay2011network} uses a generative model for firm level production networks that  matches degree distributions better than scale-free frameworks. Belgian VAT data has been used to study productivity shocks to individual firms and their effects on aggregate output with a computeable equilibrium model \cite{magerman2016heterogeneous}. The theoretical study \cite{moran2019may} shows that shocks propagate widely as soon as production function have a ``Leontief'' component, but not when they are of pure Cobb-Douglas type. Probably the closest study to ours, \cite{inoue2019firm}, investigates how firm level shock propagation in response to an initial shock --the great earthquake in Japan in 2011-- based on an estimate of the Japanese production network. However, they focus on effects on aggregate output and do not compute firm level systemic risk.

\section{Relations to financial systemic risk}\label{SI:FSRI}

In the area of financial networks systemic risk has been extensively studied for about two decades \cite{freixas1998systemic, boss2004network}. The importance of being able to measure systemic risk of single firms (banks, insurance, funds) became apparent in the 2008 financial crises and the European government debt crises 2012. Initial systemic risk assessment methodologies have been shown by, for example, \citep{eisenberg2001systemic, furfine2003interbank, boss2004contagion} they have been refined to macroprudential stresstesting models that can assess the effects of adverse macro-economic scenarios  \citep{elsinger2006risk} and to measure the impact single banks have on the entire network  \citep{cont2010network, battiston2012debtrank}. A wide range of studies revealed various properties of financial networks that improve the understanding of how systemic risk emerges, for example, the role of network topology \citep{nier2007network, gai2010contagion}, the role of diversification of external assets \cite{beale2011individual}, or the role played by large banks \cite{arinaminpathy2012size}. These aspects are crucial for the management of systemic risk. The design of regulatory policies to address the adverse economic and societal effects of systemic risk has strongly benefited from these academic contributions. However, there are important differences between financial and production networks and how stress is spreading there, that require adaptions to how systemic risk  and the spreading of shocks is defined and modelled. For example, in production networks in- and out-links affect heterogeneous production process in contrast to stock quantities like equity or liquidity buffers in financial networks. Further, production networks tend to be orders of magnitude larger than banking networks. Here we extent the ideas of measuring systemic risk in financial networks \citep{cont2010network, battiston2012debtrank, bardoscia2015debtrank} to a more general framework of companies in a production network at a national scale. To estimate it we use micro level VAT data of Hungary \citep{borsos2020unfolding}.

\section{Industry classification schemes} \label{S1:industry}

Around the world there are millions of different products produced, all having small distinctive features. However, to allow for a meaningful statistical and economic analysis these products are usually grouped and categorized into product classes according to some common features. Examples are the Cooperative Patent Classification (CPC) and the EU's classification of products by activity (CPA). They contain 2647 and 3142 classes, respectively. The second kind of classification aims at grouping companies that produce these product groups based on common activities. Industry classifications have various levels of granularity, for example the International Standard Classification of All Economic Activities, ISIC, has 88 categories at the 2-digit and 419 at the 4-digit level; the Statistical Classification of Economic Activities in the European Community NACE has 88 categories at the 2-digit and 615 4 digit level; the North American Industry Classification System, NAIC, features 1057 categories at the 6-digit level. Note that the CPA and NACE classifications can be mapped onto each other. A concise overview can be found in \cite{ramon2021nace}.

In practice however, these classifications are not linked to the single supply transactions,  $W_{ij}$. As outlined in the introduction of the main-text economists typically resort to the assumption that each company $i$ produces one of $m$ possible different products, identified by the firm's industry classification. In mathematical notation we use the industry affiliation vector, $p$, to identify the industry affiliation $p_i \in \{1, 2, \dots, m \}$ for each firm $i$ in its entries. On the modelling side, this simplifying assumptions amounts to reducing a firm level input-output matrix, with each column determining the necessary inputs for each product firm $i$ produces, to a single input vector. In mathematical terms this simplification reduces the output vector $x_{i1},x_{i2},\dots,x_{im}$ of company $i$ to a scalar output $x_i$ of type $p_i$. Consequentially, the input matrix $(\Pi^i$ ---with element $\Pi^i_{kl}$ determining the amount of input, $k$, used to produce output $x_{il}$--- reduces to an input vector, $(\Pi_{i1}, \Pi_{i2}, \dots \Pi_{im})$ and the production function which in reality is a map $f: \mathbb{R}_+^m \rightarrow \mathbb{R}_+^m$ reduces to $f: \mathbb{R}_+^m \rightarrow \mathbb{R}_+$. In short, a multi-layer network is aggregated to a single network layer, where each buyer-supplier relation, $W^1_{ij}(t), W^2_{ij}(t), \dots W^m_{ij}(t)$, is simplified to a single transaction, $W_{ij}(t)$, of product, $p_i$.

The reduction of a distinct input and customer vector for each product a firm produces to a single vector for each firm leads to a blur of the up- and downstream contagion. This is illustrated with the case where firm $i$ produces two different products, $A_i$ and $B_i$, requiring each two types of inputs. $A_i$ requires $\alpha$ and $\gamma$, $B_i$ requires $\beta$ and $\gamma$. If either the suppliers of input $\alpha$ or input $\beta$ fail to deliver only the production of one of the two products $A_i$ or $B_i$ is adversely affected. Consequently,  a customer that only buys $A_i$ is affected only in case input $\alpha$ is becoming scant, while another customer who only buys $B_i$ is affected in the case where input $\beta$ is not available. In the aggregated picture where there is only one abstract product of type $p_i$, both customers are affected by a loss of input $\alpha$ or $\beta$. Note in the case of Leontief production functions, if either input $\alpha$ or $\beta$ is not available, this would affect  the entire production of firm $i$, even though in reality it would only affect the production of either $A_i$ or $B_i$. Similarly, for upstream contagion, if only one of the two customers stops buying, just the suppliers of either $\alpha$ or $\beta$ are affected but not both. The presented framework can be easily extended to account for this as soon as appropriate data becomes available. 

\begin{figure*}[t]
	\centering
	\includegraphics[width=1 \columnwidth]{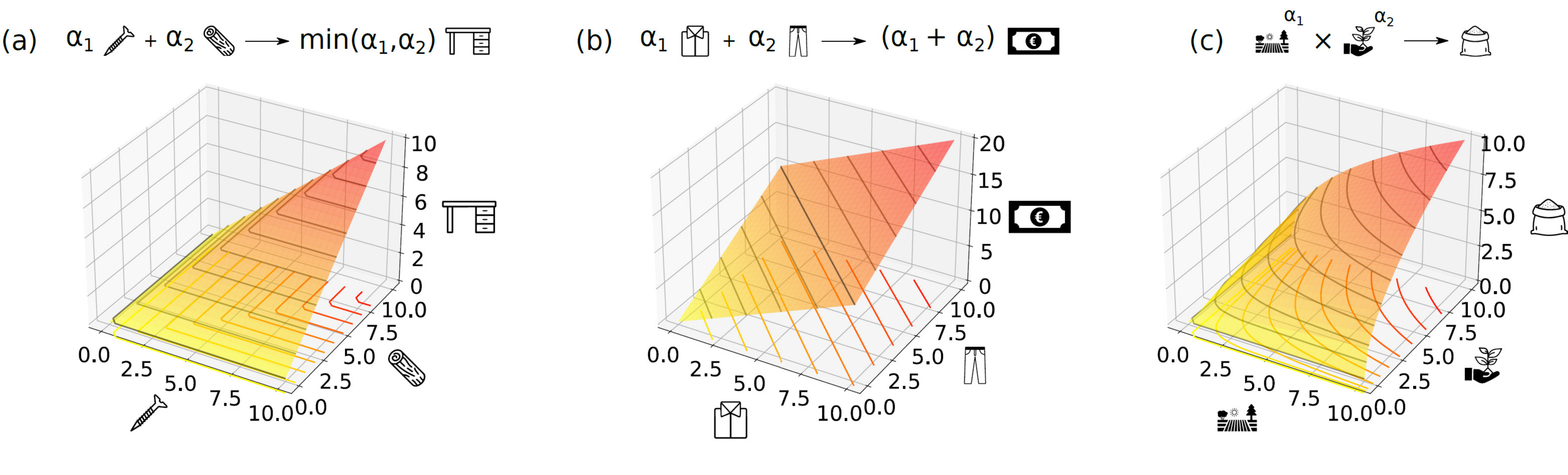}
	\caption{Three different types of production function.
		(a) The Leontief production function, $x_i = \min\Big( \frac{\Pi_{i1}}{\alpha_{i1}}, \frac{\Pi_{i2}}{{\alpha_{i2}}}\Big)$, is constrained by the least available input. We illustrate this with a carpenter building tables. She requires an exact amount of wood and nails for every table. 
		(b) Linear production function of two inputs, $x_i = \frac{ \Pi_{i1} }{\alpha_{i1}}+ \frac{\Pi_{i2}}{\alpha_{i2} }$. For a linear production function the output is proportional to a linear combination of the inputs, here illustrated by a retail company, selling shirts and trousers. If shirts are unavailable as an input, the revenue generated by trousers is not affected.
		(c) Cobb-Douglas production function, $x_i = \beta_i \Pi_{i1}^{\alpha_{i1}}\cdot\Pi_{i2}^{\alpha_{i2}} $, with $\alpha_1 + \alpha_2 = 1$. The Cobb-Douglas production function can be interpreted as an intermediate case, allowing for {\em partial substitution}. Exemplified by the production of crops, the amount of land used for the crops can be reduced if e.g. more fertilizer is invested. However, with the complete lack of either production factors, any production is impossible.
	}	
	\label{SIfig:production_fct}
\end{figure*}

\section{Production functions} \label{SI:productionfunction}

The mere existence of a buyer-supplier connection does not say much about how their production processes depend on each other, i.e. how the failure of one affects the other. This exposure depends primarily on the type of goods and services delivered and how these specific intermediate products (inputs) are used in combination with each other. Naturally, some inputs are more critical for specific production processes than others. We need to know the type of goods and services that are delivered and an approximations for the firms' production processes to model these exposures. As outlined in the main text, production processes of firms in the economics literature are commonly modelled with production functions. The standard choices are the constant elasticity of substitution production function (CES) \citep{mcfadden1963constant} and its special cases, the Cobb-Douglas and the Leontief production function. Note the linear production function is also a special case of CES. 

\textbf{Intuition.}
In FIG. \ref{SIfig:production_fct} we show a schematic visualisation of the three special cases. 
Figure \ref{SIfig:production_fct} (a) shows a Leontief production function for firm $i$, with two types of input, $\Pi_{i1}$ (screws) on the x-axis,  $\Pi_{i2}$ (wood) on the y-axis, and the resulting production level, $x_i$ (tables) on the z-axis,
$$x_i = \min\Big( \frac{1}{\alpha_{i1}}\Pi_{i1}, \frac{1}{{\alpha_{i2}}}\Pi_{i2}\Big) \quad . $$
The amount of tables produced depends on the availability of both products according to a fixed production recipe, $ \alpha_{i1}$ screws and $ \alpha_{i2}$ cubic meters of wood. Since, we only know the volumes (= price $\times$ quantity), the interpretation changes such that $ \alpha_{i1}$ is the money spent on screws as a fraction of the table's value and $ \alpha_{i2}$ is the money spent on wood as a fraction of the table's value. In this case, if 50\% less of one of the inputs is available, only 50\% of the tables are produced. In industrial processes the production recipes encoded in the $\alpha_{ik}$ are commonly used for production planning and management.

Figure \ref{SIfig:production_fct} (b) shows a linear production function of firm $i$ with two different types of inputs, $\Pi_{i1}$ (trousers) on the x-axis and $\Pi_{i2}$ (jackets) on the y-axis and the resulting production level, $x_i$ (sales of trousers and jackets) on the z-axis, 
$$x_i = \frac{1}{\alpha_{i1}} \Pi_{i1} + \frac{1}{\alpha_{i2} }\Pi_{i2} \quad .$$
It is clear that ``production'' in this case is possible with only one of the inputs. The coefficients $\alpha_{i1}$ and $\alpha_{i2}$ determine how important each of the inputs is for the output. The ratio, $\alpha_{i1}/\alpha_{i2}$, determines how much of input $1$ needs to be added to substitute the one unit loss of input $2$ to keep the same level of production. In our example the inputs are equally important. The linear production function can  be  used to reasonably describe a trades business or services. In the case of a trade business the coefficients $\alpha_{ik}$ correspond to the \emph{markup} firm $i$ charges when re-selling input $k$ to customers. Similarly, for the delivery of services, inputs are not used jointly as intermediate inputs to produce a new physical product, but to support the creation process. Illustrative examples are a hairdresser, running out of shampoo can still cut hair, a consulting firm running out of office supplies that still can produce powerpoint slides, a software developing company facing delays of new computers that can still produce new code on the old machines. Even though the processes are not becoming impossible to execute, they become less efficient or reduce in quality, thus leading to effectively less output. Text books usually state that linear production functions are of limited or no practical use \citep{varian2014intermediate}. However, it is obviously a sensible approximation for the above mentioned situations. Clearly, it fails to describe the production process of physical goods where inputs are inherently combined and transformed to become an output of a new type. 

Figure \ref{SIfig:production_fct} (c) shows a Cobb-Douglas production function for firm $i$ with two different types of input, $\Pi_{i1}$ on the x-axis, $\Pi_{i2}$ on the y-axis, and the resulting production level, $x_i$ on the z-axis, 
$$x_i = \beta_i \Pi_{i1}^{\alpha_{i1}}\cdot\Pi_{i2}^{\alpha_{i2}} \quad . $$
It is clear that ``production'' is more efficient if both inputs are available, because the fewer of input one is available, the more of input two is needed to compensate for this loss and keeping the production level constant. In the extreme, an infinite amount of input one would be needed to completely substitute two. This is of course not possible in reality, but  indicates that production with one input alone is not possible. In its original formulation it was not intended to describe production based on different intermediate inputs, but on labour ($l_i$) and capital ($c_i$), $x_i=b\cdot  l_i^{\alpha_i}\cdot c_i^{1-\alpha_i}$ \cite{douglas1976cobb}. Note that if the $\log$ is taken on both sides of the equation, this yields a linear production function. The interpretation is that if $\Pi_{i1}$ grows (decreases) by 1\%, output will grow (decrease) by ${\alpha_{i1}}$ \%, at the given production level. On the sector level $\alpha_{ik}$ can be set to the technical coefficients, derived from input-output tables \citep{miller2009input, carvalho2019production}, i.e., $\alpha_{ik}=\Pi_{ik}/x_i$. This is equally possible on the firm level. \\

\textbf{Interpretation in the context of shock spreading.}
Usually the production function specifies how much goods and services (output), $x_i$, of type $p_i \in \{1,2, \dots,m \}$ firm $i$ can produce with a given amount of $m$ different intermediate products (inputs) $(\Pi_{i1}, \Pi_{i2}, \dots \Pi_{im}) \in \mathbb{R}_+^m$, manufacturing equipment (capital) $c_i$, and its employees (labour), $l_i$. 
The production functions are usually used in a long-term perspective, where different inputs (traditionally labour and capital) can be substituted, by for example, using more machines and less labour. In practical terms this implies a change of a firm's production process. Such changes are usually not possible in the short-term and often need at least a few months (lead times for buying new machinery, designing the new business processes, hiring differently qualified employees, train employees on new machines or software, etc.). This is especially true for complicated production processes of complex goods. Thus, for short-term spreading of shocks caused by, for example, the failure of a supplier, we assume that the production processes are constant and we neglect capital and labour within the production function. Similarly, we assume that intermediate inputs can not be substituted in the short-term by simply buying more of a different input. In the long-term, however, certainly, missing inputs can be substituted by other inputs when the production process is sufficiently adapted. 
Consequently, for the analysis of shock propagation in the production network, we use the production functions to determine how much a firm can still produce in the short-term if a certain input is not available. As discussed in Appendix \ref{SI:supplier_replaceability}, the assumption of replacing a failing supplier by another that can deliver a comparable input, has a strong impact of how shocks are spreading. \\

\textbf{Leontief-, linear-, and generalized Leontief production functions.}
We now show the exact specifications of the production functions used to model the spreading of up- and downstream shocks in the economic systemic risk index, ESRI. Note that the calibrations are made for an observed production network, $W$, and an industry affiliation vector, $p$. These two objects determine the output volume of every firm and the volume of respective inputs purchased in the observed production network. We specifically use the linear production function for trades businesses and services companies, and a modified Leontief production function for firms with a physical production processes, to allow for a different treatment of inputs from {\em physical} production processes and from services. 

We start from observed outputs and inputs of firms determined by the supply network, $W$, where $W_{ij}$ denotes the volume (value) supplied from firm $i$ to firm $j$ in the respective observation period. The input, $k$, of firm $i$ is mapped to its suppliers by,
\begin{equation}
	\Pi_{jk}(t)=\sum_{i=1}^{n} W_{ij}(t)\delta_{p_i,k} \quad ,
\end{equation}
where  $\delta_{a,b}$ is the Kronecker delta that is equal to $1$, if $a=b$ and $0$,  otherwise. The output, $x_i = \sum_{l=1}W_{il}$, is the sum over all supply transactions to other firms, $l$. 

First, we consider the Leontief production function for company $i$ as 
\begin{eqnarray} \label{eq:def_leontief}
	x_i  &  = & \min\Big[ \frac{1}{\alpha_{i1}} \sum_{j=1}^{n}  W_{ji} \delta_{1, p_{j}} ,  \frac{1}{\alpha_{i2}} \sum_{j=1}^{n}  W_{ji} \delta_{2, p_{j}} ,  \dots ,  \frac{1}{\alpha_{im}} \sum_{j=1}^{n}  W_{ji} \delta_{m, p_{j}} \Big]  \quad .
\end{eqnarray}
Note that this formulation implies that all suppliers $j$ of $i$ supplying the same input, $p_j=k$ (industry classification) are treated as {\em perfect substitutes}. The corresponding parameters $\alpha_k$ for company $i$ are the {\em technical coefficients}, 
\begin{equation}\label{alphadef}
	\alpha_{ik} = \frac{ \sum_{j=1}^{n}  W_{ji} \delta_{p_{j},k} }{ \sum_{l=1}^{n}  W_{il} } \quad .
\end{equation}
$\alpha_{ik}$ determines the fraction of output that firm $i$ needs to  spend on input $k$ to produce this respective output. 

Second, we consider the simplified linear production function  
\begin{equation} \label{eq:def_linear}
	x_i = \frac{1}{\alpha_i} \sum_{j=1}^{n} W_{ji} \quad,
\end{equation}
with 
\begin{equation} \label{def_alpha_linear_pf}
	\alpha_i = \frac{\sum_{j=1}^{n} W_{ji}}{\sum_{l=1}^{n} W_{il}} \quad.
\end{equation}
It is clear  that Eq. (\ref{eq:def_linear}) is a special case of Eq. (\ref{eq:def_leontief}) when all product types are the same. For shock propagation this implies that the loss of an input affects the output proportional to the value of the inputs. Note that the linear production function has a natural interpretation for a trades business, where $1/\alpha_i-1$ corresponds to the definition of the \emph{markup} ($Markup = (Sale\; Price - Cost)/Cost$). In our 2017 dataset, trades businesses (NACE classes 46 and 47) make up for 17\% of firms. 
\footnote{In practice distributors apply a different markup for different product groups. If sufficient data were available this can be simply corrected by applying a different markup for each product group. The general linear production function is
	\begin{equation} \label{def_linear_pf_general}
		x_i  =  \frac{1}{\alpha_{i1}} \sum_{j=1}^{n}  W_{ji} \delta_{1, p_{j}} +  \frac{1}{\alpha_{i2}} \sum_{j=1}^{n}  W_{ji} \delta_{2, p_{j}} + \dots +  \frac{1}{\alpha_{im}} \sum_{j=1}^{n}  W_{ji} \delta_{m, p_{j}})  \quad .
	\end{equation} 
	Then, for trades businesses the parameters $\alpha_{ik}$ would be chosen such that $1/\alpha_{ik}-1$ corresponds to the markup of product group $k$ for company $i$. }.  

Third, we consider a generalized Leontief production function (GL) where only essential inputs are treated in the Leontief sense and non-essential inputs are treated in the linear sense. We denote the set, $\mathcal{I}_i^\text{es}$, that contains all essential input types $k$, entering production in a Leontief way, and the set $\mathcal{I}_i^\text{ne}$ that contains all non-essential input types $k$ entering in a linear way. Then, we define the modified Leontief production function as
\begin{eqnarray} \label{eq:def_modified_leontief_pf}
	x_i & = & \min \Bigg[
	\min_{k \in \mathcal{I}_i^\text{es}} \Bigg( \frac{1}{\alpha_{ik}}\sum_{j=1}^{n}  W_{ji} \delta_{p_{j},k}\Bigg), \:\beta_{i} + \frac{1}{\alpha_i} \sum_{k \in  \mathcal{I}_i^\text{ne}}  \sum_{j=1}^{n}  W_{ji} \delta_{p_{j},k} \Bigg]
\end{eqnarray}

The parameter $\beta_i$ is defined as the production level that is attainable with only essential inputs $k \in \mathcal{I}_i^\text{es}$, i.e.,
\begin{equation}
	\beta_i = \Bigg(\sum_{l=1}^{n} W_{il} \Bigg) \frac{ \sum_{k \in  \mathcal{I}_i^\text{es}}  \sum_{j=1}^{n}  W_{ji} \delta_{p_{j},k} }{  \sum_{j=1}^{n}  W_{ji} \delta_{p_{j},k} } \quad. 
\end{equation}
As for the linear production function, the parameter $\alpha_i$ is the fraction of output spent on the value of inputs. It ensures that a lack of non-essential inputs interpolates between the full production level $x_i$ and $\beta_{i}$. It is clear that both the Leontief and the linear production function are special cases when either all inputs are essential or non-essential. Note that here non-essential means that production is not stopped when the input is lacking, but there is still an impact on output, i.e. the input is non-essential yet relevant. This formulation can be easily extended to cover the additional case that inputs are non-essential and non-relevant. In that case a third group of inputs needs to be specified and those inputs simply do not affect the production function. We leave this for future research.

\section{Replaceability of suppliers} \label{SI:supplier_replaceability}

On shorter time horizons the mode of production can only be changed in relatively subtle ways, e.g., by using a given input from a different supplier and even that is sometimes not possible if goods are not highly standardized. However, major modifications to the production process itself are much harder, especially for sophisticated products. Detailed modelling of how different companies can be replaced by others (as suppliers) would require the detailed knowledge of inventory levels, production capacities of suppliers, and detailed product information (e.g. if two suppliers in the same industry can produce the same good). With such additional data available, a consistent modelling of supplier replacement could be implemented with a dynamic rewiring of the network, i.e., companies looking for new suppliers with idle production capacities or inventory to replace a failed supplier. However, it is difficult to model this behaviour in a realistic way, without many additional parameters. Such extended modelling efforts would be computationally significantly more expensive. 

Only a few strategies in the firm level supply chain contagion literature tackle this challenge, however, generally all attempts suffer from limited data availability. Ref. \cite{inoue2019firm} does consider replaceability only between existing suppliers. \cite{wu2016shock} uses a different strategy and creates a measure of replaceability based on the weight of a given supplier in its costumers' production-related cost. This measure assumes that if a supplier is an important part of a firm's production, then it is harder to find a substitute for it. Another, frequently used solution, is to distinguish between standardized goods (goods with a clear reference price listed in trade publications) and differentiated goods (goods with multidimensional characteristics) based on \cite{rauch1999networks}. Although this distinction can be used as a categorical variable in econometric estimations, e.g., \cite{giannetti2011you} it is not informative regarding the extent of replaceability even for standardized products. Ref. \cite{barrot2016input} uses two other proxies to measure the specificity of suppliers: the level of R\&D expenditures and the number of patents held by a firm. Unfortunately, these pieces of information are only relevant for a tiny fraction of companies and not at all applicable to the entire network of Hungarian firms.

We propose a different strategy that employs a straightforward, data driven way to construct a short-term supplier replaceability index based on intra-industry market shares. The basic intuition is that a supplier, having a small market share within its industry, on average should be relatively easy to replace by a small increase of the production of its competitors to cover the additional demand for their products. However, a supplier producing a considerable share of the goods in a given industry is more difficult to replace, as it is unlikely that its competitors can increase their production immediately. Having a large market share within an industry category in a country does of course not necessarily mean that a firm is large.
It also needs to be taken into account that potential alternative suppliers in the given industry might also have experienced shocks during the contagion processes in the model. To make this approach more realistic, we take the deterioration in their production capacity into account. Here, it is important to distinguish  between the two different sources that could cause reductions in the production level of these potential alternative suppliers. On the one hand, one should account for downstream shocks, i.e. shocks coming from the suppliers, which is a truly limiting disruption in their production. On the other hand, one should \emph{disregard} the upstream shocks experienced by them, because demand shocks coming from customers are not actual restraints on their production (at least from the point of view of replacing reduced output of their competitors). We can also account for the fact that in the case of a system-wide crisis it might not be possible to find alternative suppliers; see Eq. (\ref{eq:replaceability_factor}) for the mathematical formulation of the supplier replaceability. 

The interpretation of the replaceability of suppliers in our approach can be illustrated by the following example. Assume that a supplier with a 10\% market share within its industry (after considering also its competitors' states) is responsible for 50\% of the input required by one of its customers in a given product category. If the production level of this supplier drops to 80\%, then the production of the customer will decrease by 1\% $(= 10\% \times 50\% \times (1-80\%))$.
If we disregarded the possibility of replacing this supplier, the corresponding decrease in the firm's output would be 10\%. This way, we are able to replace missing supplies w.r.t. the market conditions; i.e. our replaceability factor reflects not only the fact that the given input in this example can be bought from the remaining 90\% of the market, but also acknowledge that this replacement might not be possible entirely, or it entails some costs. In future research the interaction of upstream and downstream shocks needs to be taken into account also for modelling the replaceability. This is important because, if some producers receive upstream shocks, they would have an additional capacity to supply other firms with this, which themselves might be suffering a downstream shock. If these idle supply and idle demand is matched, this would result in less contagion.

\section{Economic systemic risk index (ESRI)}\label{SI:ESRI}

We quantify the systemic risk of a (temporally) failing firm  to the overall production network -- and hence economy and society (e.g. through  employment problems) --  by the hypothetical direct and \textit{indirect} up- and downstream effects of this initial failure on the remaining production network. The basic philosophy of our systemic risk measure --- considering the  up- and downstream dependencies of firms --- is captured in the following recursion scheme. First, we define the observed production network (for a given year) as our initial state, $W(0)$, and calibrate the chosen production functions to this state $t=0$, as shown in Appendix  \ref{SI:productionfunction}. Second, we recursively simulate how the production levels of the other firms in the network are affected in response to the initial shock (failure of a firm). Third, we observe the production levels of all firms after the recursion reaches a new stable state (shocks stop propagating). We define the economic systemic risk index, ESRI$_i$, of company $i$ as the fraction of the production networks' overall output that is (temporally) impeded in response to the firm's (temporal) failure. 
The downstream recursion is derived from linking the amount firm $i$ can produce at $t+1$ with the available inputs at $t$,  
\begin{equation}
	x_i^{d}(t+1)=f_i\big(\Pi_{i1}(t), \Pi_{i2}(t),\dots,\Pi_{im}(t) \big) \quad ,
\end{equation}
to the production network, $W(t)$. The amount $\Pi_{ik}(t)$ firm $i$ uses from input type $k$ in period $t$ can be mapped to its in-links (suppliers), $W_{ij}(t)$ by  $\Pi_{ik}(t)=\sum_{j=1}^{n} W_{ji}(t)\delta_{p_j,k}$, where $\delta_{a,b}$ is the Kronecker delta. Similarly, the upstream recursion is derived from the dependence of firm $i$'s output on the available demand for its products. The demand is mapped to the out-links (buyers), $W_{il}$, via  
\begin{equation}
	x_i^{\text{u}}(t+1) = \sum_{l=1}^{n} W_{il}(t) \quad .
\end{equation}
The in-links of node $i$, $W_{ji}(t)$,  themselves depend on the production $x^\text{d}_j$ of suppliers $j \in \{1,2,\dots,n \}$ and subsequently on their in-links (supply). The out-links, $W_{il}$, depend on the buyers $l \in \{1,2,\dots,n \}$ production $x^\text{u}_l$ and subsequently their own out-links (demand). Consequently,  the production of each firm recursively depends on the other firms' production. To simulate the impacts of firm $j$'s default, we assume an initially stable state at $t=0$, corresponding to the observed production network, $W(0)$. The initial output levels are given by the initial state $W(0)$ and $x^\text{d}(0)=x^\text{u}(0)=x(0)$ and $x_i(0)=\sum_{l=1}^{n} W_{il}(0)$. 
We apply an initial shock to the network by assuming the (temporary) failure of firm $j$ at $t=1$. We set $x_j^\text{d}(1)=x_j^\text{u}(1)=0$, and consequently, the out-links $W_{ji}(1)=0$, for all $i$ and the in-links $W_{lj}(1)=0$, for all $l$.
We can simulate the shock propagation by updating the production levels of all other firms. We continue updating the production levels iteratively until production levels of firms do not change anymore. 

The relative production level at time $t$ depending on the received up- and downstream shocks is given by  
\begin{equation}
	h_i^\text{d}(t)= \frac{x^\text{d}_{i}(t)}{x^\text{d}_{i}(0)} \qquad {\rm and} \qquad  h_i^\text{u}(t)= \frac{x^\text{u}_{i}(t)}{x^\text{u}_{i}(0)} \quad ,
\end{equation}
respectively. We assume a \emph{proportional rationing} mechanism, by setting $W_{ij}(t)$ to $W_{ij}(0)\cdot h^\text{d}_i(t)$ and $W_{li}(t)$ to $W_{li}(0)\cdot h^\text{u}_i(t)$. This implies that if a firm received a downstream shock it will forward it to it's customers {\em pro rata} according to their initial percentage shares. Similarly, if a firm  received an upstream shock it will forward it to it's customers {\em pro rata} according to their initial percentage shares. Note that this assumption also implies a static production network $W$. For an overview of the effects of rationing mechanisms on the sector level, see \cite{pichler2021modeling}.
The amount that firm $i$ can produce at $t+1$, given the suppliers' production levels, $h^\text{d}_j(t)$ at $t$ is
\begin{eqnarray}\label{eq_downstream}
	x_i^{\text{d}}(t+1) &  =  &   f_i\Big(\sum_{j=1}^{n} W_{ji} h_j^\text{d}(t) \delta_{p_j,1}, \sum_{j=1}^{n} W_{ji}h_j^\text{d}(t)\delta_{p_j,2}, \dots,  \sum_{j=1}^{n} W_{ji}h_j^\text{d}(t) \delta_{p_j,m}\Big) \quad . 
\end{eqnarray}
The amount firm $i$ produces at $t+1$, given the current production level of its buyers, $h^\text{u}_l(t)$ at $t$ is
\begin{eqnarray}\label{eq_upstream}
	x_i^{\text{u}}(t+1) & = &    \sum_{l=1}^{n} W_{il}h_l^\text{u}(t) \quad. 
\end{eqnarray}
Note that we assume here that companies do not produce simply on stock if there is no demand. The failure of firm $j$, $h^\text{u}_j(1)=h^\text{d}_j(1)=0$ is propagated through the network by iterating the Eq. (\ref{eq_downstream}) and (\ref{eq_upstream}) until the network reaches a stable state at time $T$. The stable state is reached at time
\begin{equation}
	T \equiv \min_t\{t \in \mathbb{N}|\max\big(  h^d(t)-h^d(t+1) ,   h^u(t)-h^u(t+1)   \big) \leq \epsilon \}  \quad ,
\end{equation}
where $\epsilon=10^{-2}$ is chosen as a convergence threshold. Thus, we assume that once all shocks are smaller than $\epsilon$, the propagation stops ---the production level of the firm is no longer adapted, when shocks are becoming small--- and the corresponding time point of convergence is $T$. The final production level of every firm $i$ is set to $h_i(T)=\min(h_i^d(T),h^u_i(T))$. The ESRI of firm $j$ is now calculated by weighting the relative losses in production of each firm $i$ by its out-strength, $s^\text{out}_i$, 
\begin{equation} \label{ESRI_i}
	{\rm ESRI}_j = \sum_{i=1}^{n} \frac{s^\text{out}_i }{\sum_{l=1}^n s^\text{out}_l }\big(1-h_i(T) \big) \quad .
\end{equation}
The calibration of the recursion algorithm to the specific production functions used in the empirical analysis is shown in  Appendix \ref{SI:propagation}; the derivation of the recursion equations is shown in the following Appendix \ref{SI:derivation}.

\section{Details on the derivation of the recursion} \label{SI:derivation}

We explicitly determine the relations for Eqs. (\ref{eq_downstream}-\ref{eq_upstream}) for the generalized Leontief,  the Leontief, and the linear production functions. We use the term update equation and recursion interchangeably. \\

\textbf{Downstream recursion.} 
We derive the downstream update equation for the generalized Leontief production function from Eq. (\ref{eq:def_modified_leontief_pf})
\begin{eqnarray} \label{eq:def_modified_leontief_pf_dyn}
	x_i^{\text{d}}(t+1) &  = &  \min\Bigg[
	\min_{k \in \mathcal{I}_i^\text{es}} \left( \frac{1}{\alpha_{ik}}\sum_{j=1}^{n} W_{ji} h_j^\text{d}(t) \delta_{p_j,k}\right), \: 
	\beta_{i} + \frac{1}{\alpha_i} \sum_{k \in  \mathcal{I}_i^\text{ne}}  \sum_{j=1}^{n}   W_{ji} h_j^\text{d}(t) \delta_{p_j,k}
	\Bigg] \quad . 
\end{eqnarray}
It is convenient to work with a direct recursion with $h^\text{d}_i(t+1)$ on the left hand side and $h^\text{d}_j(t)$ on the right hand side. Thus, we divide Eq. (\ref{eq:def_modified_leontief_pf_dyn})  on both sides by $x_i(0)=\sum_{l=1}^n W_{il}$,  
\begin{eqnarray} \label{eq:def_modified_leontief_pf_dyn2}
	h_i^{\text{d}}(t+1) &  = &  \min\Bigg[
	\min_{k \in \mathcal{I}_i^\text{es}} \Bigg( \frac{1}{x_i(0)} \frac{1}{\alpha_{ik}}\sum_{j=1}^{n} W_{ji} h_j^\text{d}(t) \delta_{p_j,k}\Bigg), \: 
	\frac{1}{x_i(0)} \Bigg(\beta_{i} +  \frac{1}{\alpha_i} \sum_{k \in  \mathcal{I}_i^\text{ne}}  \sum_{j=1}^{n}   W_{ji} h_j^\text{d}(t) \delta_{p_j,k} \Bigg)
	\Bigg] \quad . 
\end{eqnarray}
After a few simplifications (see Eq. (\ref{eq:simplification1}-\ref{eq:simplification3})) we can write more compactly
\begin{eqnarray} \label{eq:def_modified_leontief_pf_dyn3}
	h_i^{\text{d}}(t+1) &  = &  \min\Bigg[
	\min_{k \in \mathcal{I}_i^\text{es}} \Bigg( \sum_{j=1}^{n} \Lambda^{d1}_{ji} h_j^\text{d}(t) \delta_{p_j,k}\Bigg), 
	\tilde{\beta}_{i} +   \sum_{k \in  \mathcal{I}_i^\text{ne}}  \sum_{j=1}^{n}   \Lambda^{d2}_{ji} h_j^\text{d}(t) \delta_{p_j,k} 
	\Bigg] \quad . 
\end{eqnarray}
The elements of the matrix  $\Lambda^{d1}$  are defined as  
\begin{equation} \label{def_lambda_d1}
	\Lambda^{d1}_{ji} =  \begin{cases}  
		\frac{W_{ji}}{\sum_{\iota=1}W_{\iota i}  \delta_{p_{\iota},p_{j}} } \quad \text{if} \; W_{ij} > 0 \qquad , \\
		\qquad \quad 0 \qquad \qquad \; \text{else} \quad\qquad\qquad,
	\end{cases}
\end{equation} 
the elements of $\Lambda^{d2}$  are   
\begin{equation} \label{def_lambda_d2}
	\Lambda^{d2}_{ji} =  \begin{cases}  
		\frac{W_{ji}}{\sum_{l=1}W_{li}   } \qquad \quad \text{if} \; W_{ij} > 0 \qquad , \\
		\qquad \quad 0 \qquad \qquad \; \text{else} \quad\qquad\qquad.
	\end{cases}
\end{equation} 
and $\tilde{\beta}_i$ is simply the relative fraction of production possible with only essential inputs $k \in \mathcal{I}_i^\text{es}$. Note that the Leontief production function (Eq. \ref{eq:def_leontief}) and the linear production function (Eq. \ref{eq:def_linear}) are special cases, where either only the first, or the second part of Eq. (\ref{eq:def_modified_leontief_pf_dyn3}) is present. 

The elements of the Leontief downstream impact matrix, $\Lambda^{d1}_{ji}$, capture the impact that the failure of firm $j$ has on firm $i$, given that $j$ is supplying a ``Leontief input'' (essential) to firm $i$. Note that $\Lambda^{d1}_{ji}=1$ means that firm $j$ is the only supplier of goods of type $p_j$ to firm $i$. Similarly, the elements of the linear downstream impact matrix, $\Lambda^{d2}_{ji}$, capture the impact the failure of firm $j$ has on firm $i$ if $j$ is supplying a ``linear input'' (non-essential) to firm $i$. Note that the case $\Lambda^{d2}_{ji}=1$ is only possible if firm $j$ is the only supplier of firm $i$. This difference gives already a glimpse on the fact that the Leontief production function leads to much higher contagion levels than the linear one. Further, the case of  $\Lambda^{d1}_{ji}=1$ becomes more frequent if the product types (industry classifications) become more fine grained. This leads to the intuitive behaviour that companies producing scarce and essential (to many firms) resources, are expected to have a high systemic risk index. 

As will become visible in Appendix \ref{SI:propagation}, for the ease of numerical implementation, we split the updating of Eq. (\ref{eq:def_modified_leontief_pf_dyn3}) into two parts. First, we define the unified downstream impact matrix 
\begin{equation} 
	\Lambda_{ij}^\text{d} = \begin{cases}
		\Lambda^{d1}_{ji} \qquad \text{if} \; p_j \in \mathcal{I}_i^\text{es} \\
		\Lambda^{d2}_{ji} \qquad \text{if} \; p_j \in \mathcal{I}_i^\text{ne}
	\end{cases}
\end{equation}
where $k\in \mathcal{I}_i^\text{es}$ contains all inputs $k$ that are essential to firm $i$ and $k\in \mathcal{I}_i^\text{ne}$ contains all inputs $k$ that are non-essential to firm $i$. Again, firms having a pure Leontief or linear production function are special cases. Second, we define the relative share of input $k$, available to firm $i$ at time $t$ as
\begin{equation}
	\tilde{\Pi}_{ik}(t) =  \sum_{j=1}^n   \Lambda^{d}_{ji} h_j^\text{d}(t) \delta_{p_j,k} \quad . \notag
\end{equation}
Consequently, the $n \times m$ matrix, $\tilde{\Pi}(t)$, contains in row $i$ the relative input vector at $t$ for firm $i$. Relative refers to the initial state $t=0$. Hence, $\tilde{\Pi}(t)$ can be updated by a matrix multiplication, $(\Lambda^\text{d})^{\top} P$, where $P \in \{0,1\}^{n \times m}$ is defined as
\begin{equation} 
	P_{ik} = \begin{cases} \notag
		1 \qquad \text{if} \; \; \: p_i=k \quad, \\
		0 \qquad \text{otherwise} \quad. 
	\end{cases}
\end{equation}
Note that ---due to the definition of $\Lambda^\text{d}$---  for the inputs $k \in \mathcal{I}_i^\text{es}$ the relative input,  $ \tilde{\Pi}_{ik}(t)$, is equal to $1$ if no supplier defaulted, while for $k \in \mathcal{I}_i^\text{ne}$ it is equal to the share of input $k$'s value out of the value of all inputs, i.e., its only exactly $1$, if $i$  buys only one single non-essential input, $k$. Third, we can update the variable $h^\text{d}_i$ by evaluating  the relative production function for each firm $i$
\begin{equation} \label{eq:def_modified_leontief_pf_dyn4}
	h_i^{\text{d}}(t+1)   =   \min\Big[
	\min_{k \in \mathcal{I}_i^\text{es}} \Big(\tilde{\Pi}_{ik}(t)   \Big), \;
	\tilde{\beta}_{i} +   \sum_{k \in  \mathcal{I}_i^\text{ne}} \tilde{\Pi}_{ik}(t)  
	\Big] \quad . 
\end{equation}
Note that the quantity $ \sum_{k \in  \mathcal{I}_i^\text{ne}} \tilde{\Pi}_{ik}(t)  $ is $0$ if all inputs $k \in  \mathcal{I}_i^\text{ne}$ are not available. In the case of a pure Leontief producer this is always zero and $\tilde{\beta}_i=1$, while for the case of a pure linear producer this quantity is one at time $t=0$ and $\tilde{\beta}_i=0$. \\

\textbf{Upstream recursion.}
We follow the same procedure for the upstream update equation based on Eq. (\ref{eq_upstream}). The production of company $i$ in response to demand reductions from its customers is 
\begin{eqnarray} \label{def_h_u_update}
	x^\text{u}_{i}(t+1) &  = &  \sum_{j=1}^n W_{ij}h^u_j(t)   \notag
\end{eqnarray}
and we divide both sides by $x_i(0)$ to get
\begin{eqnarray} \label{eq:upstream_dynamics}
	h^u_{i}(t+1) & = & \sum_{j=1}^n \frac{W_{ij}}{ \sum_{l=1}^n W_{il} }h^\text{u}_j(t)  \notag  \\
	h^u_{i}(t+1) & = &  \sum_{j=1}^n \Lambda^\text{u}_{ji}h^\text{u}_j(t)   \quad.  
\end{eqnarray}
The upstream impact matrix is defined as
\begin{eqnarray} \label{eq:lambda_upstream}
	\Lambda^\text{u}_{ji} = 
	\begin{cases} \frac{W_{ij}}{ \sum_{l=1}^n W_{il} } \qquad \text{if} \quad  W_{ij}>0 \quad,  \\
		\qquad  0 \qquad \quad \text{else} \qquad \qquad \; \;  , \end{cases}
\end{eqnarray}
and its elements, $\Lambda^\text{u}_{ji}$, determine the impact of the failure of buyer $j$ on supplier $i$. $\Lambda^\text{u}_{ji}=1$ only occurs if $j$ is the only buyer of $i$. Note that in comparison to the definition of the downstream update equations, we assume that upstream shocks are independent on the production functions.  We assume implicitly that firms keep the proportion of their inputs fixed when output is reduced or increased. Therefore, we ignore  that some inputs are ``fixed costs'' and will or cannot be affected by upstream contagion. Further, we can not take into account contractual obligations, which enforce a supply transaction also in case the demand for the customers product is reduced. \\

\textbf{Incorporating exogenous shocks.}
For analysing an exogenous shock we have to make the initial exogenous shock to every firm explicit in the updating equations Eq. (\ref{eq:upstream_dynamics}) and Eq. (\ref{eq:def_modified_leontief_pf_dyn4}). Simply iterating these equations after the initial  failure of firm $j$, $x^\text{d}_j(1)=x^\text{u}_j(1)=0 \implies h^\text{d}_j(1)=h^\text{u}_j(1)=0$, leads to an instant recovery of the initially failed firm because all its suppliers and customers are not affected (yet). \footnote{Note that if there is a loop (e.g. $W_{ij}>0$ and $W_{ji}>0$) there will be a shock fed back to the initially defaulting firm.} In our model the initial failure is an abstract exogenous shock to firm $j$ that could constitute everything from the destruction of the business premises of a firm, by for example a fire, a government mandated closure in a pandemic, to a strike. Technically, such exogenous shocks affect either capital, labour, or both in the production function. As mentioned in the main text, we don't model capital and labour explicitly but simply define an exogenous constraint, $\psi_i \in [0,1]$, and set it to the fraction of production that is \emph{still possible} after the initial shock. For example, $\psi_i = 0.8$ means that 80\% of production is still possible after the initial shock. Both shocks to labour or capital can be implemented with the variable $\psi_i$. Equations (\ref{eq:upstream_dynamics}) and  (\ref{eq:def_modified_leontief_pf_dyn4}) change in the following ways
\begin{equation} \label{eq:def_modified_leontief_pf_dyn_psi}
	h_i^{\text{d}}(t+1)   =   \min\Big[
	\min_{k \in \mathcal{I}_i^\text{es}} \Big(\tilde{\Pi}_{ik}(t)   \Big), \;
	\tilde{\beta}_{i} +   \sum_{k \in  \mathcal{I}_i^\text{ne}} \tilde{\Pi}_{ik}(t)  
	, \psi_i \Big] \quad ,
\end{equation}
\begin{equation} \label{eq:upstream_dynamics_psi}
	h^u_{i}(t+1) = \min \Big[ \sum_{j=1}^n \Lambda^\text{u}_{ji}h^\text{u}_j(t), \psi_i  \Big] \quad.  
\end{equation}
Note that instead of introducing the parameter, $\psi_i$, one could also work with incremental updates, $h^\text{d}_i(t)-h^\text{d}_i(t-1)$, as e.g. in \cite{bardoscia2015debtrank}. When firms receive an exogenous shock, $(1-\psi_i)<1$ (not a 100\% failure) upstream and downstream shocks are propagated on top of the initial shock $1-\psi_i$. However, when including $\psi_i$ directly into upstream and downstream updates Eqs. (\ref{eq:def_modified_leontief_pf_dyn_psi}-\ref{eq:upstream_dynamics_psi}), the additionally received upstream and downstream shocks need to be larger than the initial shock, $1-\psi_i$, so that firm $i$ propagates them further. Note that a positive shock can be considered if $\psi_i>1$. As in \cite{inoue2019firm} the initial shock could be made time dependent $\psi_i(t)$ to also model potential recovery from the initial shock. \\

\textbf{Modelling supplier replaceability.}
So far the replaceability of suppliers was ignored. It is conceivable that in practice it is totally unrealistic for many firms (suppliers) to be not replaceable at all, even on the short-term; see also  Appendix \ref{SI:supplier_replaceability}.  A straight forward proxy for a firm's replaceability --- that can be inferred from available data --- is its market share. Intuitively, a company with a small market share (within a fine grained industry classification) will be easier to replace than one with a large market share. For example, the lack of inputs caused by a (temporary) failure of a firm with a 5\% market share can be most likely compensated to a large degree by its competitors, as long as the produced goods are reasonably standardized. This degree of standardisation varies significantly and is ignored in our approach. On the other hand a temporary failure of a firm with a 50\% market share is much more difficult to compensate by competitors, given that there are inventory and production capacity constraints. 

Modelling the replaceability of suppliers explicitly would require that a firm, whose supplier defaulted creates a new link with one or more firms of the same industry classification. This link creation depends on the available inventory level, geographical proximity, and available production capacity. To be meaningful, such a network rewiring model needs to be calibrated to appropriate data that is hard to access. Moreover, it becomes computationally significantly more expensive and involved. We therefore choose a simpler approach for calculating the systemic risk index for every firm. We model the replaceability based on firms' market shares with a simple linear factor. In particular, we calculate the replaceability factor at time $t$
\begin{equation} \label{eq:replaceability_factor}
	\sigma_i(t) =  \min \Big(\frac{s^\text{out}_i(0)}{\sum_{j=1}^{n} s^\text{out}_j(0)h_j^d(t) \delta_{p_j,p_i}},1 \Big) \quad. 
\end{equation}
Over time, $\sigma_i(t)$ can only increase and $\sigma_j(t)\in [0,1]$. It is bounded from below by firms with zero market share. When other suppliers of the same product have reduced outputs too, the supplier becomes harder to replace. Note that since the supply of firm $i$ itself is also part of the total market (denominator), a market share higher than 50\% can not be replaced anymore by competitors. This would imply a short-term doubling of their capacities. Note that $\Lambda^{d}_{ji} \big(1-h_j^\text{d}(t) \big) $ is the shock that $i$ receives from $j$ and since $\sigma_j(t)\in [0,1]$, small market shares dampen this shock substantially. 

To integrate this factor into the recursion we have to adapt the update equation for the availability of relative inputs $\tilde{\Pi}_{ik}(t)$. For $k\in \mathcal{I}_i^\text{es}$ we now update 
\begin{equation} \label{eq:relative_input_sigma1}
	\tilde{\Pi}_{ik}(t) = 1 - \sum_{j=1}^n \sigma_j(t) \Lambda^{d}_{ji} \big(1-h_j^\text{d}(t) \big) \: \delta_{p_j,k} \quad ,
\end{equation}
and for $k\in \mathcal{I}_i^\text{ne}$ we define an artificial product category $k'$ that is updated according to
\begin{equation} \label{eq:relative_input_sigma2}
	\tilde{\Pi}_{ik'}(t) = 1 - \sum_{k\in \mathcal{I}_i^\text{ne}}  \sum_{j=1}^n \sigma_j(t) \Lambda^{d}_{ji} \big(1-h_j^\text{d}(t) \big) \: \delta_{p_j,k} \quad .
\end{equation}
Note that if all $\sigma_j(t)=1$ and all $h_j^\text{d}(t)=0$ for all non-essential input suppliers ($p_j \in  \mathcal{I}_i^\text{ne}$) the right hand side of Eq. (\ref{eq:relative_input_sigma2}) is equal to $\tilde{\beta}_i$ and we can write the downstream update more compactly as
\begin{equation} \label{eq:def_modified_leontief_pf_dyn_psi_sigma}
	h_i^{\text{d}}(t+1)   =   \min\Big[
	\min_{k \in \mathcal{I}_i^\text{es}} \Big(\tilde{\Pi}_{ik}(t)   \Big), \;
	\tilde{\Pi}_{ik'}(t)
	, \psi_i \Big] \quad . 
\end{equation}
The matrix $P$ can be adjusted accordingly.

\textbf{Simplifications.}
For completeness, we specify the simplification steps for the first term in Eq. (\ref{eq:def_modified_leontief_pf_dyn2})
\begin{align} \label{eq:simplification1}
	\frac{1}{\alpha_k} \frac{1}{x_{i}(0)} \sum_{j=1}^{n} W_{ji}  h^d_j(t) \delta_{k, p_{j}} =   \notag  \\
	\frac{\sum_{l=1}^{n}  W_{il} }{\sum_{\iota=1}^{n}  W_{\iota i} \delta_{p_{\iota},k} } \frac{1}{\sum_{l=1}^{n}W_{il}} \sum_{j=1}^{n}  W_{ji}  h^d_j(t) \delta_{k, p_{j}} = \notag \\
	%\frac{1 }{\sum_{\iota=1}^{n}  W_{\iota i i} \delta_{p_{\iota},k} } \sum_{j=1}^{n} W_{ji}    h^d_j(t) \delta_{k, p_{j}} \quad\: \notag\\
	\sum_{j=1}^{n} \frac{W_{ji} }{\sum_{\iota =1}^{n}  W_{\iota i} \delta_{p_{\iota},k} }  h^d_j(t) \delta_{k, p_{j}} = \notag \\
	\sum_{j=1}^{n} \Lambda^{d1}_{ji}  h^d_j(t) \delta_{k, p_{j}} \quad .
\end{align} 
and the second term in Eq. (\ref{eq:def_modified_leontief_pf_dyn2})
\begin{align} \label{eq:simplification2}
	\frac{1}{x_{i}(0)} \frac{1}{\alpha_i} \sum_{k \in  \mathcal{I}_i^\text{ne}}  \sum_{j=1}^{n}   W_{ji} h_j^\text{d}(t) \delta_{p_j,k} = \notag \\
	\frac{1}{\sum_{l=1}^{n} W_{il}} \frac{\sum_{l=1}^{n} W_{il}}{\sum_{j=1}^{n} W_{ji}} \sum_{k \in  \mathcal{I}_i^\text{ne}}  \sum_{j=1}^{n}   W_{ji} h_j^\text{d}(t) \delta_{p_j,k}  = \notag \\
	\sum_{k \in  \mathcal{I}_i^\text{ne}}  \sum_{j=1}^{n} \frac{ W_{ji}}{\sum_{j=1}^{n} W_{ji}}  h_j^\text{d}(t) \delta_{p_j,k} =  \notag \\
	\sum_{k \in  \mathcal{I}_i^\text{ne}}  \sum_{j=1}^{n} \Lambda_{ji}^{d2}  h_j^\text{d}(t) \delta_{p_j,k}  \quad . 
\end{align}
We denote the relative fraction of production that possible with only essential inputs $k \in \mathcal{I}_i^\text{es}$ as
\begin{equation} \label{eq:simplification3}
	\tilde{\beta}_i     =   \frac{1}{x_{i}(0)} \beta_i = 
	\frac{ \sum_{k \in  \mathcal{I}_i^\text{es}}  \sum_{j=1}^{n}  W_{ji} \delta_{p_{j},k} }{  \sum_{j=1}^{n}  W_{ji} \delta_{p_{j},k} }   \quad. 
\end{equation}

\section{ESRI algorithm} \label{SI:propagation}

We finally show the necessary equations to calculate the ESRI for each firm in the production network, given the network $W$, the industry affiliation vector $p$ and the sets of essential products $ \mathcal{I}_i^\text{es}$ and the set of non-essential products $ \mathcal{I}_i^\text{ne}$. The downstream impact matrix $\Lambda^\text{d}$ is computed as 
\begin{equation}  \label{eq:algorithm_lambda_d}
	\Lambda_{ij}^\text{d} = \begin{cases}
		\Lambda^{d1}_{ji} \qquad \text{if} \; p_j \in \mathcal{I}_i^\text{es} \\
		\Lambda^{d2}_{ji} \qquad \text{if} \; p_j \in \mathcal{I}_i^\text{ne}
	\end{cases} \, .
\end{equation}
The elements of $\Lambda^{d1}$  and $\Lambda^{d2}$  are defined as  
\begin{align} \label{eq:algorithm_lambda_d1_d2} 
	\Lambda^{d1}_{ji} =  \begin{cases}  
		\frac{W_{ji}}{\sum_{\iota=1}W_{\iota i}  \delta_{p_{\iota},p_{j}} } \quad \text{if} \; W_{ij} > 0 \qquad , \\
		\qquad \quad 0 \quad \qquad \; \text{else} \quad\qquad\qquad,
	\end{cases} \\
	\Lambda^{d2}_{ji} =  \begin{cases}  
		\frac{W_{ji}}{\sum_{l=1}W_{li}   } \qquad \quad \text{if} \; W_{ij} > 0 \qquad , \\
		\qquad \quad 0 \quad \qquad \; \text{else} \quad\qquad\qquad.
	\end{cases}
\end{align} 
The upstream impact matrix is calculated as
\begin{eqnarray} \label{eq:lambda_u}
	\Lambda^\text{u}_{ji} = 
	\begin{cases} \frac{W_{ij}}{ \sum_{l=1}^n W_{il} } \qquad \text{if} \quad  W_{ij}>0 \quad,  \\
		\qquad  0 \qquad \quad \text{else} \qquad \qquad \; \;  , \end{cases}
\end{eqnarray}
Next, the initial exogenous shock parameter is set to $\psi_i=0$ for the failing firm $i$ (for which we compute $\text{ESRI}_i$) and $\psi_i=1$ for all other firms $i$. 

Then, we iterate the following set of equations. First, for each firm the market share at time $t$ is calculated as
\begin{equation} \label{eq:algorithm_sigma}
	\sigma_j(t) =  \min \Big(\frac{s^\text{out}_j(0)}{\sum_{l=1}^{n} s^\text{out}_l(0)h_l^d(t) \delta_{p_l,p_j}},1 \Big) \quad. 
\end{equation} 
Next we calculate the relative amount of essential and non-essential inputs available for each firm $i$. For essential inputs $k\in \mathcal{I}_i^\text{es}$ we update the relative share available  according to
\begin{equation} \label{eq:algorithm_rel_input_essential}
	\tilde{\Pi}_{ik}(t) = 1 - \sum_{j=1}^n \sigma_j(t) \Lambda^{d}_{ji} \big(1-h_j^\text{d}(t) \big) \: \delta_{p_j,k} \quad .
\end{equation}
For the non-essential inputs, $k\in \mathcal{I}_i^\text{ne}$, we update the relative share available of the auxiliary product category $k'$ according to
\begin{equation} \label{eq:algorithm_rel_input_non_essential}
	\tilde{\Pi}_{ik'}(t) = 1 - \sum_{k\in \mathcal{I}_i^\text{ne}}  \sum_{j=1}^n \sigma_j(t) \Lambda^{d}_{ji} \big(1-h_j^\text{d}(t) \big) \: \delta_{p_j,k} \quad .
\end{equation}
Finally, we update for each firm $i$ its relative production level determined by the received downstream shocks
\begin{equation} \label{eq:algorithm_downstream_update}
	h_i^{\text{d}}(t+1)   =   \min\Big[
	\min_{k \in \mathcal{I}_i^\text{es}} \Big(\tilde{\Pi}_{ik}(t)   \Big), \;
	\tilde{\Pi}_{ik'}(t)
	, \psi_i \Big] \quad ,
\end{equation}
and its relative production level, determined by the received upstream shocks
\begin{equation} \label{eq:upstream_update}
	h^u_{i}(t+1) = \min \Big[ \sum_{j=1}^n \Lambda^\text{u}_{ji}h^\text{u}_j(t), \psi_i  \Big] \quad.  
\end{equation}

The update equations are iterated until the algorithm converges at time 
\begin{equation}
	T \equiv \min_t\{t \in \mathbb{N}|\max\big(  h^d(t)-h^d(t+1) ,   h^u(t)-h^u(t+1)   \big) \leq \epsilon \} + 1 \quad ,
\end{equation}
where $\epsilon=10^{-2}$ is chosen as the convergence threshold. Once all shocks are smaller than $\epsilon$ the propagation stops and the corresponding time point of convergence is $T$. 

The ESRI of firm $i$ is now calculated as the weighted sum of lost output in the production network in response to the (temporary) failure of firm $i$
\begin{equation} \label{eq:algorithm_ESRI_i}
	{\rm ESRI}_i = \sum_{j=1}^{n} \frac{s^\text{out}_j }{\sum_{l=1}^n s^\text{out}_l }\big(1-h_j(T) \big) \quad .
\end{equation}
Note that the interpretation is conditional on the assumption that neither the supply of inputs nor the demand of the failing firm are replaced. Thus, the ESRI should be interpreted more as an economically motivated metric to rank firms with respect to their systemic importance than as a precise estimate of the actually lost total output. 

Note that the Hungarian production network  does not compromise all sales and purchases of firms based in Hungary, due to the reporting threshold and international transactions. When calculating the up- and downstream impact matrices, $\Lambda^\text{d}$ and $\Lambda^\text{u}$, we over-estimate the impacts single links have. The upstream impact from $j$ to $i$, $\Lambda_{ji}^\text{u}$, associated to link $W_{ij}$ needs to be adjusted for the sales transactions that are not included in the observed network, $W$. These residual sales transactions are included in the revenue of firm $i$. Thus, for the empirical application of the algorithm we re-weight the element $\Lambda^\text{u}_{ij}$ by the factor $s_j^\text{out} / r_j(0)$, where the numerator is the output recorded in the observed production network and $r_j(0)$ is the revenue of firm $j$ for the respective year. 
Similarly, we re-weight $\Lambda_{ji}^\text{d}$ by $s_j^\text{in} / c_j(0)$, where the numerator is the volume of purchased inputs within the observed production network and $c_j(0)$ denotes the material costs in the respective year. We use income statement data of the years 2017 and 2016 of the firms, wherever available to make this adjustment. For those  firms where this is not available we apply a factor of 1. We report that $(\sum_{i=1}^{n}s_i^\text{out}) / (\sum_{i=1}^{n}r_i(0))=0.60$ and 
$(\sum_{i=1}^{n}s_i^\text{in} )/ (\sum_{i=1}^{n}c_i(0))=0.76$. Omitting this re-weighting would lead to a significant overestimation of ESRI. A potential error arises by  re-weighting for firm $i$ $\Lambda_{ji}^\text{d}$ by the same factor for all input suppliers $j$, since we do not know which input types the unobserved quantity $ c_j(0)-s_j^\text{in}$ compromises. 

We apply the algorithm to the four production scenarios described in the main text. 
First, in the linear-only scenario (LIN) we assume  all firms have only non-essential inputs, i.e. $ \mathcal{I}_i^\text{ne} = \{01,\dots, 99\}$. Second, in the Leontief-only scenario (LEO) we assume  all firms only have essential inputs, i.e. $ \mathcal{I}_i^\text{es} = \{01,\dots, 99\}$. LIN (LEO) is the lower (upper) bound for shock propagation in our model. Third, in the mixed scenario (MIX) we assume all firms within NACE classes 01-45 (physical production) have only essential inputs $ \mathcal{I}_i^\text{es} = \{01,\dots, 99\} $ and all firms within NACE classes 46-99 (non-physical production) have only non-essential inputs $\mathcal{I}_i^\text{ne} = \{01,\dots, 99\} $. Fourth, in our baseline scenario (GL), we assume that for firms within NACE 01-45 the set of essential inputs consists of $\mathcal{I}_i^\text{es} = \{01,\dots, 45\}$ only physical producers and the set of non-essential inputs is $\mathcal{I}_i^\text{ne} = \{46,\dots, 99\} $ all inputs from non physical producers, while for firms with NACE 46-99 we assume they have only non-essential inputs  $ \mathcal{I}_i^\text{ne} = \{01,\dots, 99\}$. We assume that firms within NACE 01-45 have a physical production process, while firms within NACE 46-99 trade or provide services.

\section{Description of Hungarian VAT data }\label{SI:data}

To calibrate the production functions and to model the spreading of shocks, we use the  supply network, $W$, consisting of  all relevant supplier buyer transactions of firms within Hungary. For the current study we obtained data for the years 2017 and 2016. The firm level supplier-buyer connection data is collected by the National Tax and Customs Administration of Hungary as a part of the VAT reporting of firms. We got access to this dataset in an anonymized format through the Central Bank of Hungary. The data contains trade links among Hungarian companies between 2014 and 2017, when the tax content of the transactions between two firms exceeds HUF 1 million (EUR 3,000) in the given year. The cleaning and filtering of the data was based on \cite{borsos2020unfolding}. The two most important corrections are described below.

Many of the links in the network disappear between the observed periods and new links emerge to a similar extent. This happens mainly because of the presence of many one-off, incidental transactions, which relationships are not particularly relevant from the point of view of supply chain contagion. As these links increase the noise in the data, we filtered the network to contain only long-term supplier connections. We consider a link long-term if there were at least two trade events between the parties and if there is at least 90 days time difference between the first and the last transaction. Under these mild requirements, only 54\% of the links are long-term, however, these cover 93\% of the aggregate trade volume in the network (in 2017).

In the Hungarian VAT regulation there is no general rule that states if firms belonging to the same ownership based group should report individually or at the group level. To handle the potential distortions arising from this inconsistency, we collapsed the supplier network to the group level in every case based on the ownership data obtained from the OPTEN database. The details of this procedure are described in \cite{borsos2020unfolding}.

To obtain not only the transactions between nodes, but also information about the nodes the VAT data set was enriched with anonymized firm level information from the Hungarian corporate tax dataset. This contains the NACE classification on the 4 digit level\footnote{By imputing NACE classifications from 2016 we can increase the number of firms to 69,863 or 76\%.} for 64,053 firms (in 2017) as well as revenue and material cost information. For those firms where no NACE classification is available, we merge it into an artificial 569$^\text{th}$ category. In the future, classification methods could be used to predict NACE membership based on the input and output links of a company.

\begin{figure*}[t]
	\centering
	\includegraphics[width=0.3\textwidth]{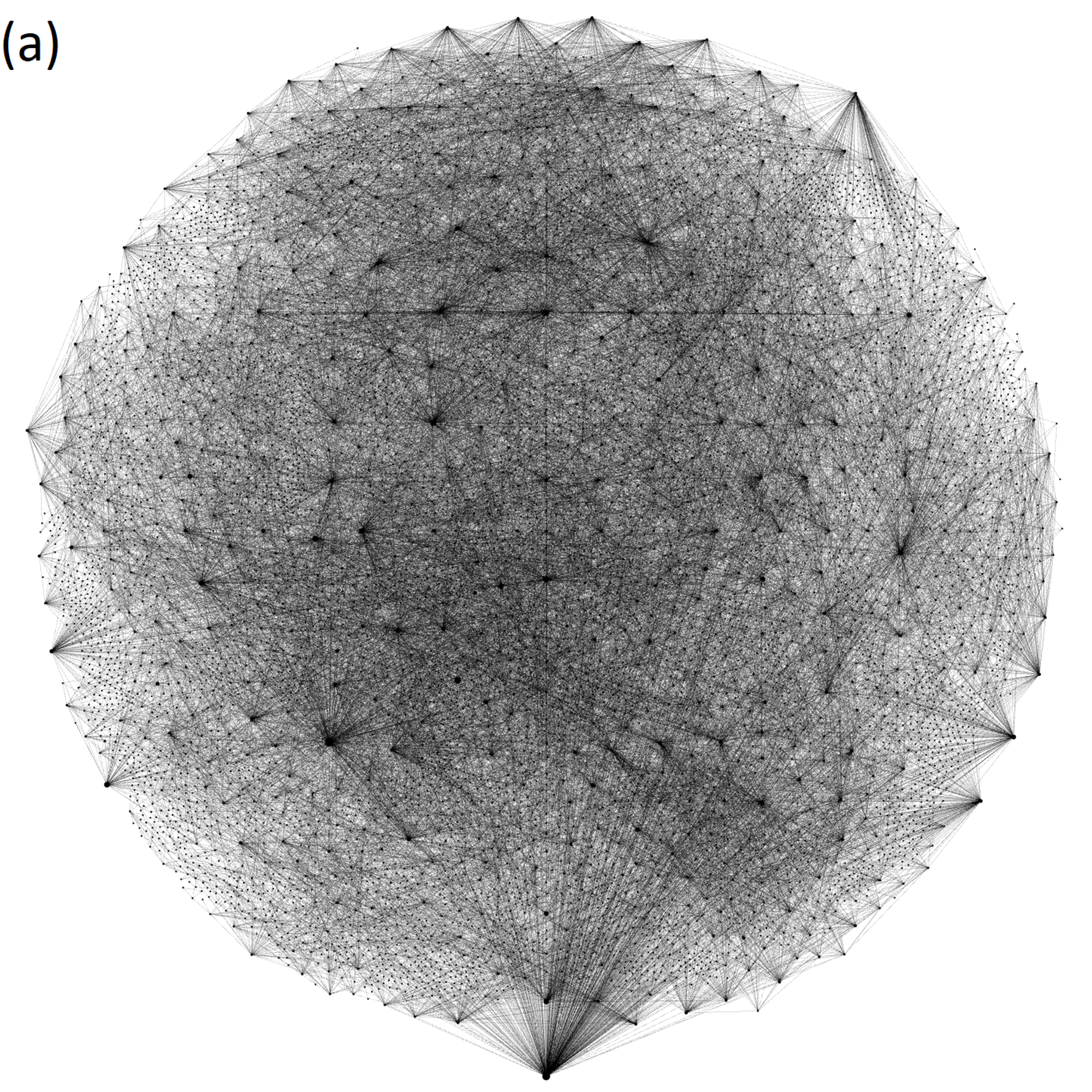}
	\includegraphics[width=0.3\textwidth]{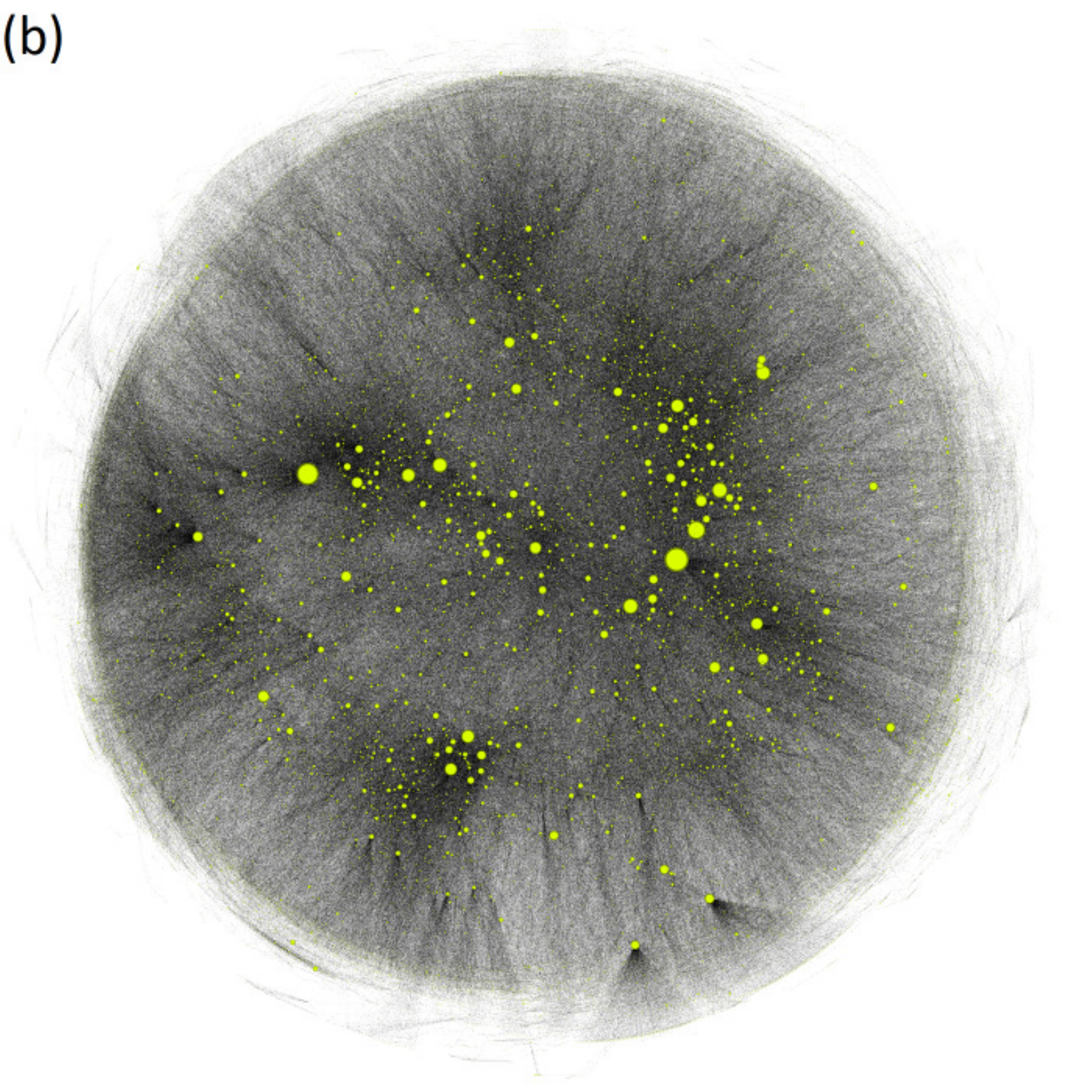}
	\includegraphics[width=0.367\textwidth]{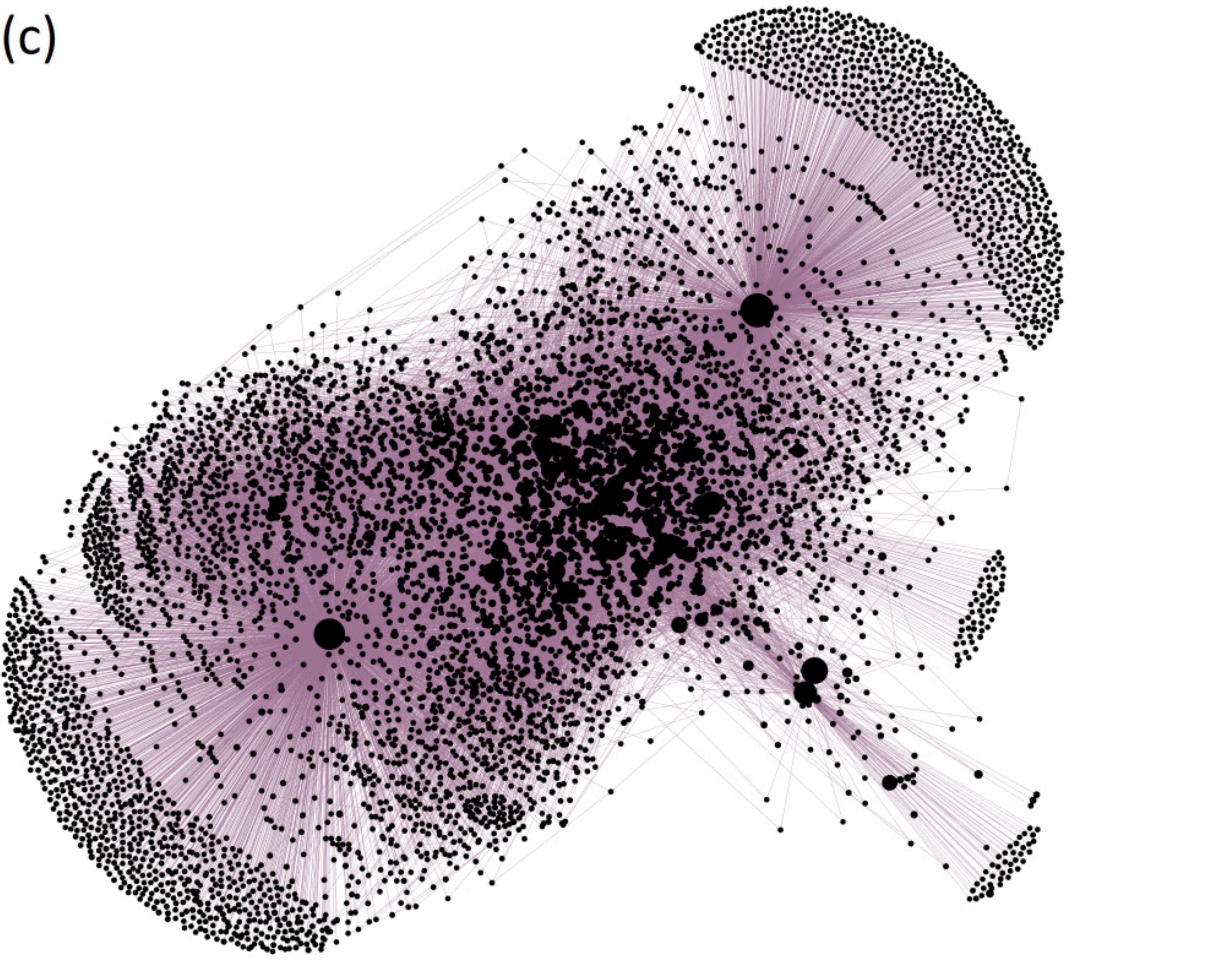} 
	\caption{Visualizing different aspects of the production network.
		(a) Section of the Hungarian production network with 5,541 nodes and 27,793 links. The network is obtained from considering the \emph{same} 1,500 randomly sampled nodes as in FIG.	\ref{fig:transaction} (c) in the main text, considering all their suppliers, yielding 6,113 nodes. Here we consider \emph{all links} between these 6,113 nodes and take the giant component from this. This shows that the true network is actually much denser. 
		(b) The supply links between all 91,595 firms in the Hungarian production network in 2017. Node size corresponds to strength. Large firms cluster in the center of the network. Darker areas correspond to denser parts. 
		(c) The induced subgraph of all neighbours from the largest three companies containing 5,945 firms and 26,783 links. It is clear that a large fraction the largest three nodes' buyers and suppliers are densely connected, while on the bottom and top right firms that are only direct neighbours of one out of the largest two firms are grouped together.  
	}
	\label{fig:full_network_plots}
\end{figure*}

\section{Network visualisation} \label{SI:network_vis}

To give a more detailed view on the network structure, we show three additional aspects of it in FIG.	\ref{fig:full_network_plots}. (a) shows a section of the network with 5,541 nodes and all 27,793 links between them. It is based on the same randomly sampled 1,500 nodes used for the visualization in FIG. \ref{fig:transaction} (c) in the main text. In FIG. \ref{fig:transaction} (c) we sample 1,500 random nodes and consider all their \emph{in-}links, yielding 6,113 nodes, and visualizing the giant component consisting of 4070 nodes. Here we consider \emph{all} links between the 6,113 nodes ---not just the in-links of the 1,500 nodes--- and visualize the resulting giant component consisting of 5,541 nodes.
The giant component becomes substantially larger, and needless to say, denser. This makes the structure more difficult to visualize.  The center of the network is denser than the periphery. However, by considering all links,  some nodes on the periphery have higher degrees (nodes with high degree on the outer part of the circle). This is most likely due to the nature of considering a random sample. In the overall network they are most likely not in the periphery, but more central.  
In FIG. \ref{fig:full_network_plots} (b) we show the giant component of 86,470 nodes and 227,355 links. The core-periphery structure is still visible with some denser regions (darker areas in the center and around); the network becomes substantially sparser towards the periphery. In the center large nodes, highlighed in yellow, are present. Size is proportional to the square root of node strength. This hints to the fact that denser regions are agglomerations around large firms and clusters of large firms. This should be verified with community detection methods in future work. In FIG. \ref{fig:full_network_plots} (c) we visualize the graph of all suppliers of the largest 3 nodes (5,945 nodes) and all 26,783  links between them. It is obvious that many suppliers of the largest firms are densely connected.

\begin{figure*}[t] 
	\centering
	\includegraphics[width=0.32 \columnwidth]{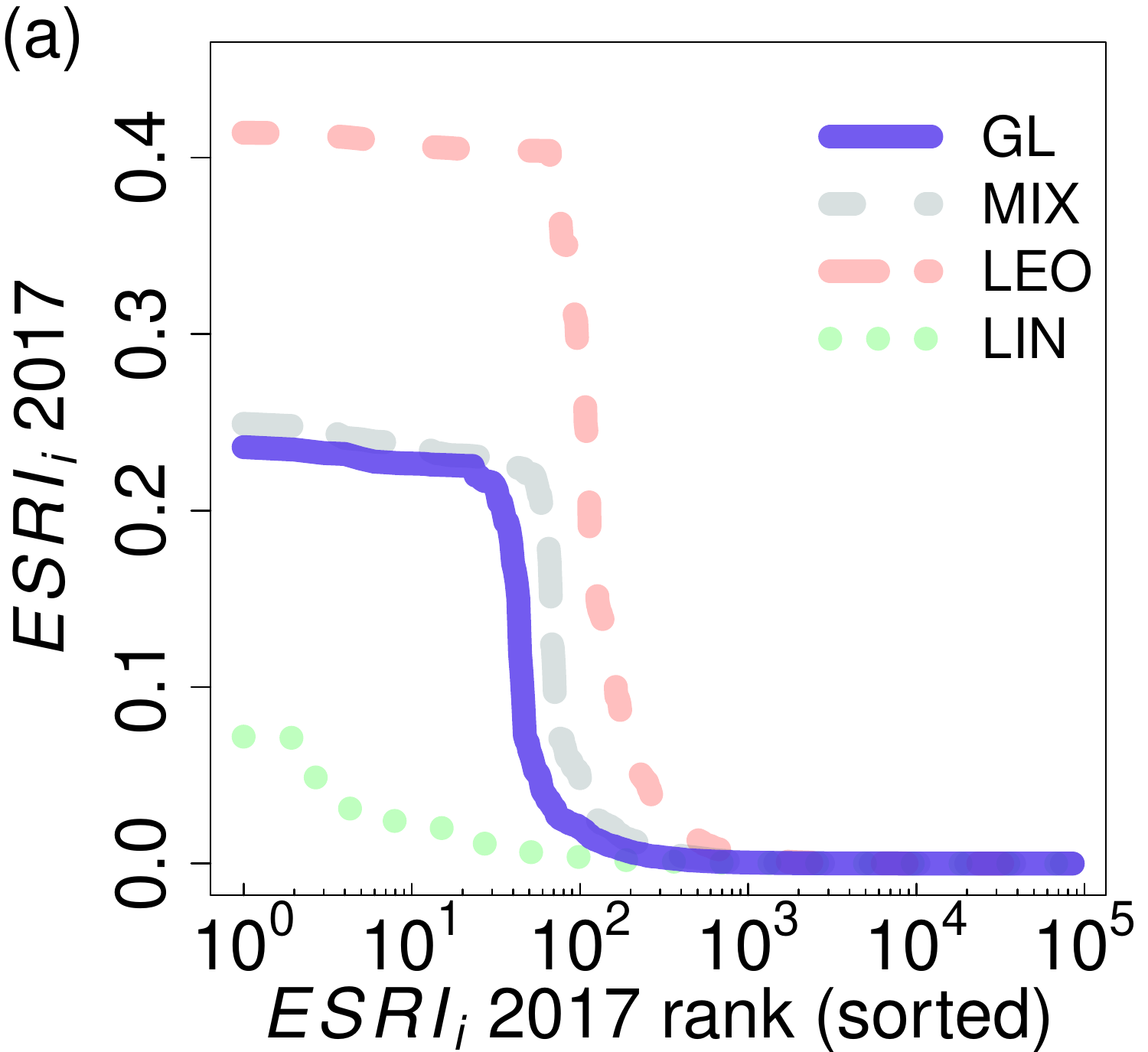} 
	\includegraphics[width=0.32 \columnwidth]{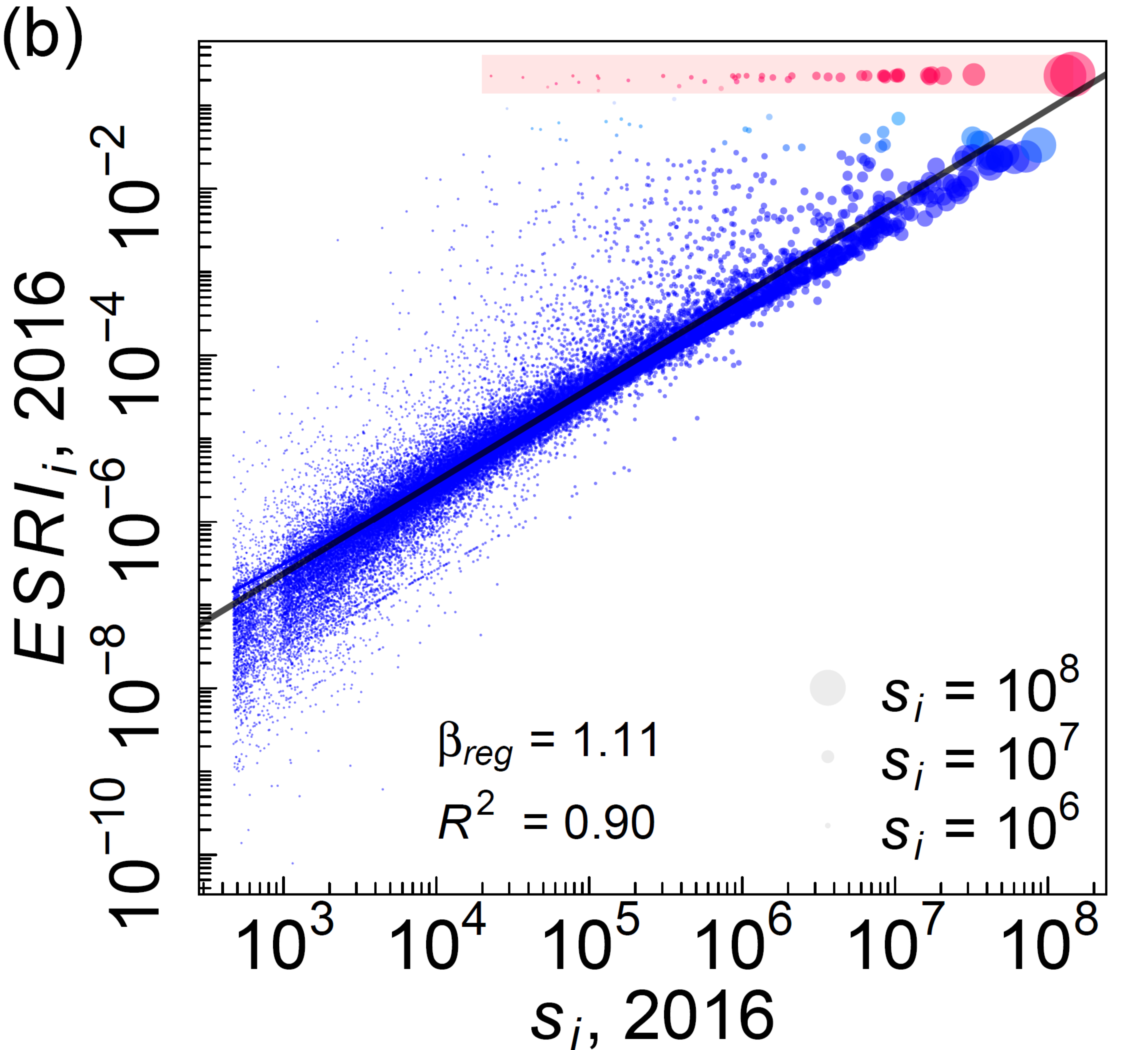} 
	\includegraphics[width=0.32 \columnwidth]{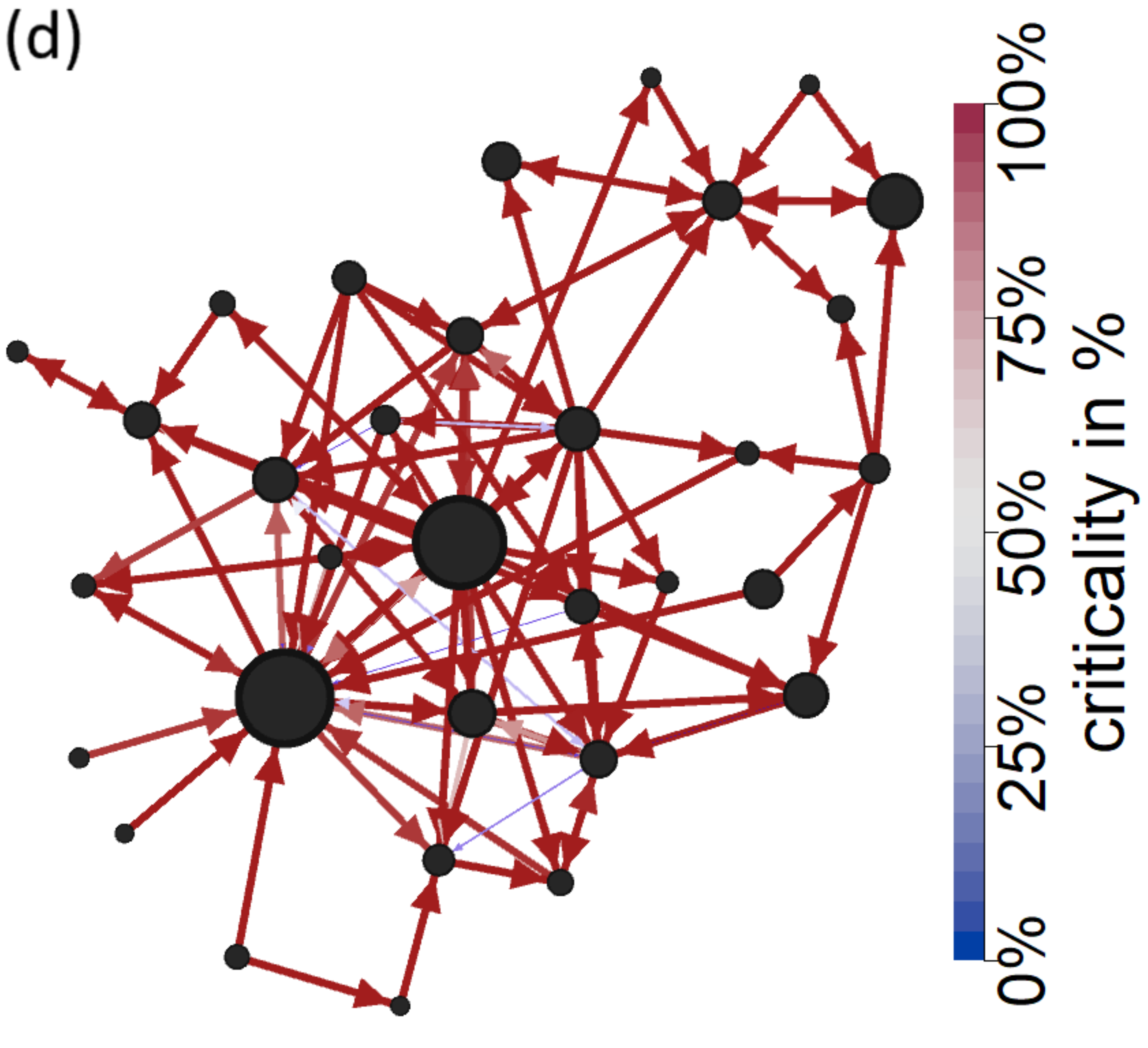} 
	\caption{Economic systemic risk of companies. (a) Economic systemic risk profile (distribution) ${\rm ESRI}_i$ for all $n=85,131$ companies in linear-log scale for the year 2016. The distributions are rank-ordered, meaning the most risky company is to the very left. The blue line shows the result for the GL scenario (production functions according to industry classification and classifying produced goods as essential or not). A plateau exists around an ESRI of 0.22, containing 33 firms for GL. There is a steep decline to ${\rm ESRI}_i \sim 0.05$ from rank 33 to 57. 154 firms have an ESRI larger than 0.01. 
		For comparison, the MIX scenario (light blue) is shown (production functions according to industry classification only). As limiting cases we show the LIN scenario (green), where all firms have linear production functions (lower bound) and for the upper bound, the situation where all firms are assumed to have Leontief production function (red). The tail of the ESRI profile decays as an approximate power-law; for the exponents see FIG. \ref{SI_fig:powerlaw}.
		(b) ESRI plotted against firm strength (representing firm-size) in log-log scale. Symbol-size corresponds to the strength $s_i$, red symbols belong to the plateau, emphasised by the shaded area. We find large and small companies in the plateau, suggesting that firm-size is not a good predictor for very high ESRI. Even though we find that for the bulk of the companies strength explains parts of the variations in ESRI ($R^2=0.90$ and slope $\beta_{reg}=1.11$ for regressing log-ESRI on log-strength, $p=2\cdot 10^{-16}$), for individual companies strength does not serve as a reliable predictor of ESRI.
		(c) Network of the 32 most systemically risky firms (plateau) for 2016. Node size is proportional to the square root of strength, $s_i$. Link (from $i$ to $j$) colors correspond to the downstream ``criticality'' i.e. the percentage of $j$'s production should $i$ stop producing, $\Lambda_{ij}^d$. Red thick (blue, thin) links indicate very large (small) losses of production. Small companies predominately supply to large high-risk companies, thereby inheriting systemic risk. Between the companies in the plateau, almost all supply relations are highly critical (red thick). In this sub-network the default of one firm's production will lead to the default of many others.}
	\label{fig:sr_profile2016}
\end{figure*}

\section{2016 results for ESRI}\label{SI:2016results}

We show the same results as in main text FIG. \ref{fig:sr_profile} (a), (b) and (c) also for the year 2016. The rank-ordered distribution of the ESRI is shown in FIG. \ref{fig:sr_profile2016} (a) in log-linear scale for 2016. For the realistic baseline scenario GL (blue), we find that 32 companies show extremely high levels of systemic risk, all being at a value of about $0.22$, meaning that about 22\% of the entire economy is affected should one of these companies fail and its supply and demand is not replaced. 57 and 154 firms have an ESRI larger than 0.05 and 0.01, respectively. For respective numbers for the other scenarios, see Tab. \ref{SI:table_large_obs2016} in Appendix \ref{SI:powerlaw}. The situation is similar for the LL scenario, where 51 companies belong to the plateau with values around $0.22$. For the (unrealistic) reference case, LEO, where {\em all} companies are of Leontief type (red), we find much higher systemic risk levels for 66 firms, with an ESRI of about $0.41$.

For ranks larger than a characteristic rank of 32, the pronounced plateau in the distribution is followed first by a steep decline and then by a slow decay of the ESRI values. The tail  (without the plateau and steep decline) of the distribution can be fitted to a power-law with an exponent of roughly  $\alpha^\text{GL}=0.68$ for GL, and $\alpha^\text{MIX}=0.63$ for MIX. For details of the power-laws fit, see FIG.  \ref{SI_fig:powerlaw} (b) in Appendix \ref{SI:powerlaw}. The shape of the rank-ordered ESRI distribution (plateau and power-law tail) is similar to what was found in FIG. 4. in \cite{moran2019may}. As expected, the reference case LIN (green) where {\em all} companies have linear production functions, generates substantially lower systemic risk than GL and MIX. The LIN case does neither show a plateau nor a power-law decay. 

To better understand which companies are forming the plateau region of extremely risky companies in Appendix FIG. \ref{fig:sr_profile2016} (b) we show ${\rm ESRI}_i$ as a function of the strength, $s_i$ (firm size within the network). Red color indicates the plateau companies; Symbol size represents the strength, $s_i$. Clearly, as for the year 2017, also in 2016 the ESRI of plateau firms (located in the shaded area) is not changing with size; in the plateau we find large and small companies (note the range of strength of 4 orders of magnitude), suggesting that firm-size is not able to predict extreme ESRI values at all also for the year 2016. For the bulk of the companies (blue), we find an overall strong correlation of log-ESRI and log-strength ($R^2=0.90$ and slope $\beta_{reg}=1.11$ for regressing log-ESRI on log-strength, $p=2\cdot 10^{-16}$). For individual companies, strength does not serve as a reliable predictor of the ESRI, since the spread of the ESRI extends over 4 orders of magnitude. Note that all firms in the plateau are physical producers (NACE 01-45). 

Figure \ref{fig:sr_profile2016} (c) shows the network between the plateau nodes. The pattern is similar to 2017. There is a strongly connected component of core nodes that are connected with each other via highly critical supply relationships. In the periphery there are mostly small nodes having only one or two critical supply relations to nodes in the strongly connected component.  

\begin{figure*}[t]
	\centering
	\includegraphics[width=.32 \columnwidth]{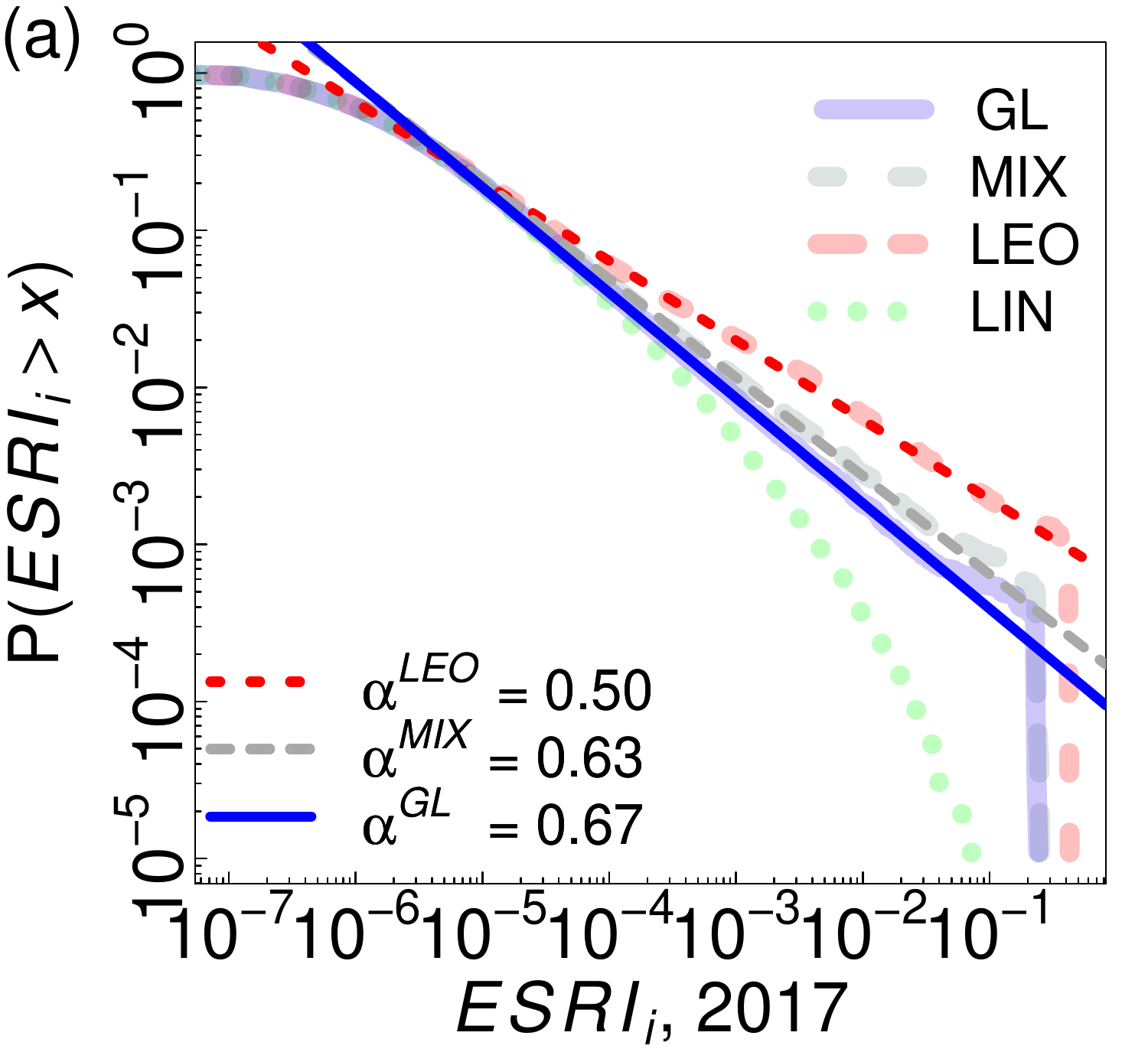} \includegraphics[width=.32 \columnwidth]{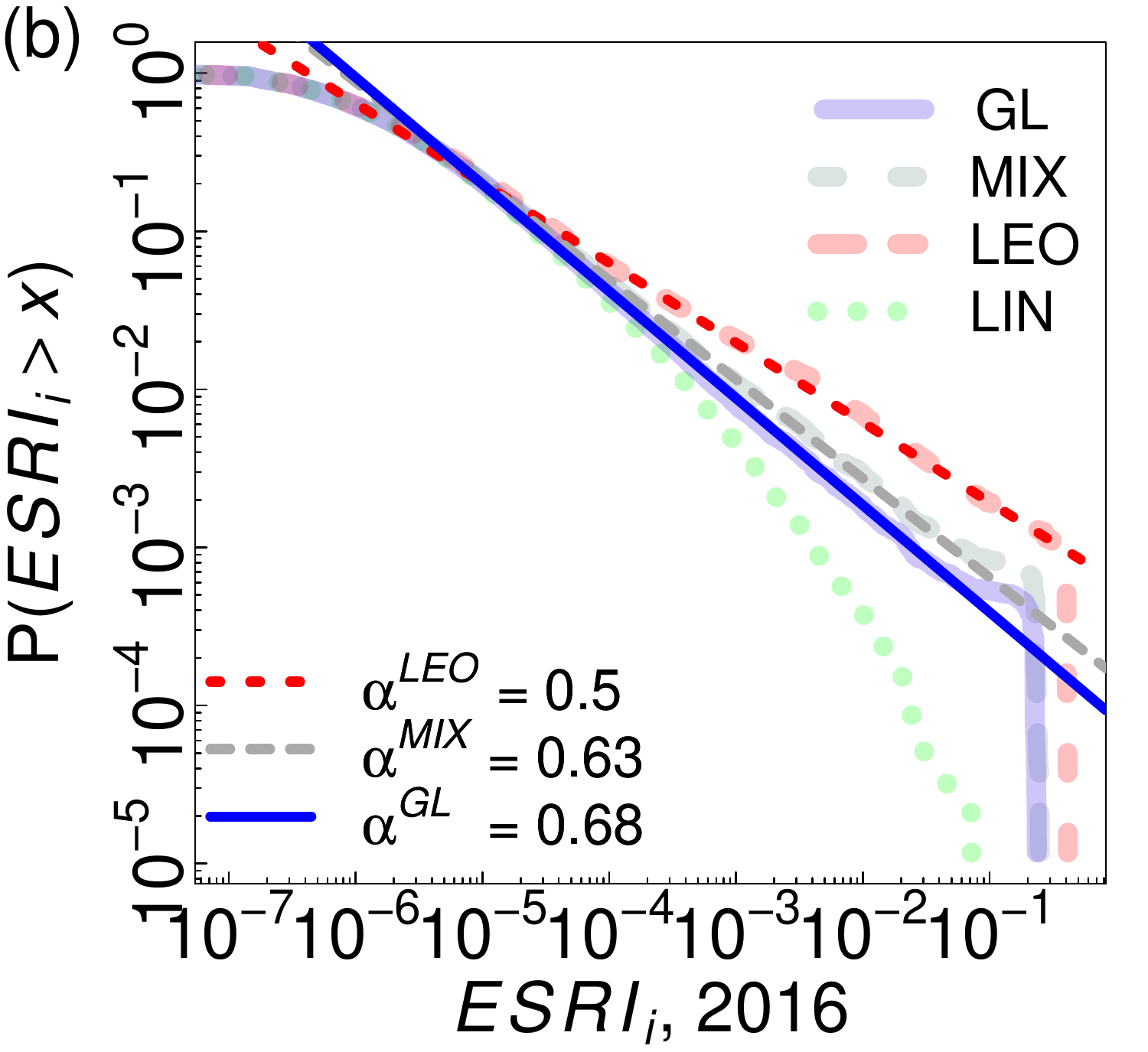} 
	\caption{Cumulative distribution of ESRI indicates a power law decay in the tail of the three scenarios: generalized Leontief (GL), mixed Leontief and linear (MIX), linear-only (LIN), and  Leontief-only (LEO). 
		(a) values for the year 2017. The estimated exponents are  0.67 for GL, 0.63 for MIX, and 0.5 for LEO.  
		(b) shows the values for the year 2016 with the exponents 0.68 for GL, 0.63 for MIX, and 0.5 for LEO.}
	\label{SI_fig:powerlaw}
\end{figure*}

\section{Estimation of power-law exponents}\label{SI:powerlaw}

We investigate the empirical distribution of the ESRI for the years 2017 and 2016 in more detail. In FIG. \ref{SI_fig:powerlaw} (a) we plot the empirical cumulative distribution function --- indicating the probability $ \text{P} ( \text{ESRI}_i > x)$ of observing an ESRI value of larger than $x$ --- in log-log scale, for the four scenarios, GL, MIX, LIN, and LEO. The three scenarios including shares of Leontief production functions behave differently than the LIN scenario. These three scenarios decay over a wide range of values as a power law (linearly in log-log), followed by a sharp cut-off (the plateau). The linear-only (LIN) scenario does not show this behaviour. We estimate the power-law exponents for the three scenarios and indicate the corresponding slopes with a red dashed line for LEO, a grey long-dashed line for MIX and a blue solid line for the GL. For the estimation we use the 
maximum likelihood estimator (MLE) 
\begin{equation} \label{eq:powerlaw_mle}
	\hat{\alpha}= 1+n \Big[ \sum_{i=1}^{n} \ln\big(\frac{x_i}{x_{\min}} \big) \Big]^{-1} \quad,
\end{equation}
see \cite{hanel2017fitting} or Eq. (3.1) in \cite{clauset2009power}. 
Since, the linear decay does not extent over the whole range of the ESRI values we restrict the estimation intervals. For LEO the estimate gives a slope $\hat{\alpha}^\text{LEO}=0.5$ for the values $\text{ESRI}_i \in [1.5\cdot 10^{-6} ,10^{-1}]$ covering 53\% of the observations. For the MIX the estimate gives a slope $\hat{\alpha}^\text{MIX}=0.63$ for the values $\text{ESRI}_i \in [5\cdot 10^{-6} ,3\cdot 10^{-2}]$, covering 29\% of the observations. 
For the GL the estimate gives a slope of $\hat{\alpha}^\text{GL}=0.67$ for the values  $\text{ESRI}_i \in [7\cdot 10^{-6} ,3\cdot 10^{-2}]$, covering 24\% of the observations.
The MLE in Eq. (\ref{eq:powerlaw_mle}) corresponds to the probability density function, while the plotted empirical ESRI distributions and the estimated slopes $\hat{\alpha}^\text{GL},\hat{\alpha}^\text{MIX},\hat{\alpha}^\text{LEO}$  in the figure corresponds to the cumulative distribution function that has an exponent $\alpha -1$. 
Figure \ref{SI_fig:powerlaw} (b) shows the results for the year 2016. The patterns are practically identical. For GL the estimate of the slope changes slightly from  $0.67$ to $0.68$. 

Table \ref{SI:table_large_obs2017} contains the number of observations above several threshold values of ESRI  in 2017. The value $ 0.41$  is the threshold for the plateau in the  LEO scenario (66 firms) and $0.22$ is the threshold for the plateau in GL (32 firms), and MIX (47 firms). In all scenarios only a few firms have a high ESRI of more than one percent.  
Same for 2016 in Table \ref{SI:table_large_obs2016}.

\begin{table}[ht]
	\caption{\centering Number of Large Observations 2017} 
	\begin{tabular}{rrrrrrrr} 
		\hline
		ESRI Levels & 0.41 & 0.22 & 0.1 & 0.05 & $10^-2$ & $10^{-3}$ & $10^{-4}$ \\ 
		\hline
		LIN & 0   & 0  & 0    & 2  &  32   & 431  & 3167  \\ 
		GL & 0   & 32  & 50  & 63  & 165  & 784  & 3740  \\ 
		MIX  & 0   & 47  & 77  & 101  & 258  & 1041  & 4178  \\ 
		LEO & 66  & 128  & 176  & 241  & 611  & 1880  & 5526 \\ 
		\hline
	\end{tabular}\label{SI:table_large_obs2017}
\end{table}

\begin{table}[ht]
	\caption{\centering Number of Large Observations 2016} 
	\begin{tabular}{rrrrrrrr}
		\hline
		ESRI Levels & 0.4 & 0.21 & 0.1 & 0.05 & $10^-2$ & $10^{-3}$ & $10^{-4}$ \\ 
		\hline
		LIN & 0  & 0  & 0  & 2  & 33  & 392  & 3052  \\ 
		GL & 0 & 32 & 45 & 57 & 154 & 687& 3540 \\ 
		MIX & 0  & 56  & 70  & 99  & 246  & 926  & 3984  \\ 
		LEO & 66  & 113  & 163  & 231  & 591  & 1765  & 5210  \\ 
		\hline
	\end{tabular}\label{SI:table_large_obs2016}
\end{table}

\section{Regression Analysis}\label{SI:regression_extended}

We conduct a more extensive regression analysis to investigate if there are firm level factors that explain ESRI$_i$ better than total strength, $s_i$. We estimate 5 additional regression models with the dependent variable in \emph{log-scale}. Details of the model specification and results are shown in Table \ref{SI:table_log_regression}. Each column corresponds to one model fit. First, we regress systemic risk on log-in-strength, $\log s_i^\text{in}$, (first column) and on log-out-strength, $\log s_i^\text{out}$,  (second column) separately and jointly (third column). Then, we regress log-systemic risk  on log-total-strength (fourth column), $\log s_i$, and on log-revenue (fith column). Additionally, we control for the number of customers, the number of suppliers in the network, market shares of firms in their respective NACE 4 sector and add fixed effects for the NACE 4 industry affiliation of firms. Note that market share is important because it determines the degree by which firms can be replaced as a supplier. Note that for in-strength and out-strength not all observations can be included, because there are firms without in- or out-links in the production network dataset. 

We see that all independent variables are highly significant, which is not surprising given the large number of observations. Note that log-out-strength seems to be a better explanatory variable ($R^2=0.82$) than log-in-strength ($R^2=0.73$) for log-ESRI. The likely reason is that due to the non-linearity of the downstream contagion (for firms with essential inputs) downstream effects weigh larger than upstream effects. Interestingly, explanatory power ($R^2=0.85$) increases only slightly when adding both log-in- and log-out-strength. The difference in sample size makes a direct comparisons difficult. Interestingly, the explanatory power of log-total-strength ($R^2=0.91$) seems to be a better explanatory variable than log-in-strength and log-out-strength jointly. This can be attributed to the difference in the observations used in the model estimation. Note that the model containing log-total-strength consists of all 91,515 observations while the model containing log-in- and log-out-strength contains 35,058 observations. The ESRI of firms having both in- and out-links seems to be more difficult to explain than the one for firms having only in- or out-links. Log-revenue can explain less than either of the other four models. This is not surprising, because revenue captures also sales transactions that are not present in the Hungarian production network and this leads to a distorted picture of firm size within the network. For example, a firm with a large revenue can export almost its entire production and thus cause relatively small downstream contagion, while the firm can import most of its inputs and thus causes only little upstream contagion. The additional control variable market share and the industry fixed effects increased the $R^2$ only marginally. 

There is a clear indication that size proxies like strength seem to explain large parts in the variation of ESRI. However, as seen in main text FIG. \ref{fig:sr_profile} (b) on the individual firm level for a given level of strength there exists still an extremely large variation in ESRI.

\begin{table*}[t] \centering 
	\caption{Regression results. The dependent variable is the firms logged economic systemic risk index (ESRI). \\ \hspace*{16.5mm} For the five models, see text.} 
	\label{SI:table_log_regression} 
	\begin{tabular}{@{\extracolsep{3pt}}lD{.}{.}{-3} D{.}{.}{-3} D{.}{.}{-3} D{.}{.}{-3} D{.}{.}{-3} } 
		\\[-1.8ex]\hline 
		\hline \\[-1.8ex] 
		& \multicolumn{5}{c}{\textit{Dependent variable: Log of economic systemic risk index (ESRI) of firms}} \\ 
		\\[-1.8ex] & \multicolumn{1}{c}{(1)} & \multicolumn{1}{c}{(2)} & \multicolumn{1}{c}{(3)} & \multicolumn{1}{c}{(4)} & \multicolumn{1}{c}{(5)}\\ 
		\hline \\[-1.8ex] 
		log(in\_strength) & 1.084^{***} &  & 0.548^{***} &  &  \\ 
		& (0.003) &  & (0.003) &  &  \\ 
		& & & & & \\ 
		log(out\_strength) &  & 0.997^{***} & 0.513^{***} &  &  \\ 
		&  & (0.002) & (0.003) &  &  \\ 
		& & & & & \\ 
		log(total\_strength) &  &  &  & 1.095^{***} &  \\ 
		&  &  &  & (0.001) &  \\ 
		& & & & & \\ 
		log(revenue) &  &  &  &  & 0.943^{***} \\ 
		&  &  &  &  & (0.004) \\ 
		& & & & & \\ 
		market\_share & 2.922^{***} & 3.108^{***} & 2.293^{***} & 1.890^{***} & 2.469^{***} \\ 
		& (0.110) & (0.076) & (0.069) & (0.054) & (0.126) \\ 
		& & & & & \\ 
		number\_of\_buyers & 0.005^{***} & -0.002^{***} & -0.00004 & 0.0003^{**} & 0.007^{***} \\ 
		& (0.0003) & (0.0002) & (0.0002) & (0.0002) & (0.0003) \\ 
		& & & & & \\ 
		number\_of\_suppliers & -0.003^{***} & 0.011^{***} & 0.001^{***} & -0.0002 & 0.011^{***} \\ 
		& (0.0003) & (0.0002) & (0.0002) & (0.0002) & (0.0004) \\ 
		& & & & & \\ 
		Constant & -24.656^{***} & -23.353^{***} & -23.951^{***} & -25.311^{***} & -24.006^{***} \\ 
		& (0.043) & (0.028) & (0.035) & (0.018) & (0.053) \\ 
		& & & & & \\ 
		\hline \\[-1.8ex] 
		Observations & \multicolumn{1}{c}{56,829} & \multicolumn{1}{c}{69,824} & \multicolumn{1}{c}{35,058} & \multicolumn{1}{c}{91,595} & \multicolumn{1}{c}{68,834} \\ 
		R$^{2}$ & \multicolumn{1}{c}{0.731} & \multicolumn{1}{c}{0.819} & \multicolumn{1}{c}{0.854} & \multicolumn{1}{c}{0.908} & \multicolumn{1}{c}{0.550} \\ 
		Adjusted R$^{2}$ & \multicolumn{1}{c}{0.730} & \multicolumn{1}{c}{0.819} & \multicolumn{1}{c}{0.854} & \multicolumn{1}{c}{0.908} & \multicolumn{1}{c}{0.550} \\ 
		Residual Std. Error & \multicolumn{1}{c}{1.163} & \multicolumn{1}{c}{0.873 } & \multicolumn{1}{c}{0.710} & \multicolumn{1}{c}{0.632 } & \multicolumn{1}{c}{1.417 } \\
		& \multicolumn{1}{c}{(df = 56738)} & \multicolumn{1}{c}{ (df = 69733)} & \multicolumn{1}{c}{(df = 34966)} & \multicolumn{1}{c}{(df = 91504)} & \multicolumn{1}{c}{ (df = 68745)} \\
		F Statistic & \multicolumn{1}{c}{1,709.566$^{***}$ } & \multicolumn{1}{c}{3,500.626$^{***}$ } & \multicolumn{1}{c}{2,256.226$^{***}$ } & \multicolumn{1}{c}{10,061.870$^{***}$} & \multicolumn{1}{c}{956.684$^{***}$ } \\ 
		& \multicolumn{1}{c}{ (df = 90; 56738)} & \multicolumn{1}{c}{(df = 90; 69733)} & \multicolumn{1}{c}{ (df = 91; 34966)} & \multicolumn{1}{c}{ (df = 90; 91504)} & \multicolumn{1}{c}{(df = 88; 68745)} \\ 
		%\hline 
		\hline \\[-1.8ex] 
		\multicolumn{4}{l}{All models are estimated with NACE 4 industry fixed effects. }  & \multicolumn{1}{r}{$^{*}$p$<$0.1; $^{**}$p$<$0.05; $^{***}$p$<$0.01} \\ 
	\end{tabular} 
\end{table*} 

\begin{figure*}[t] 
	\centering
	\includegraphics[width=0.32 \columnwidth]{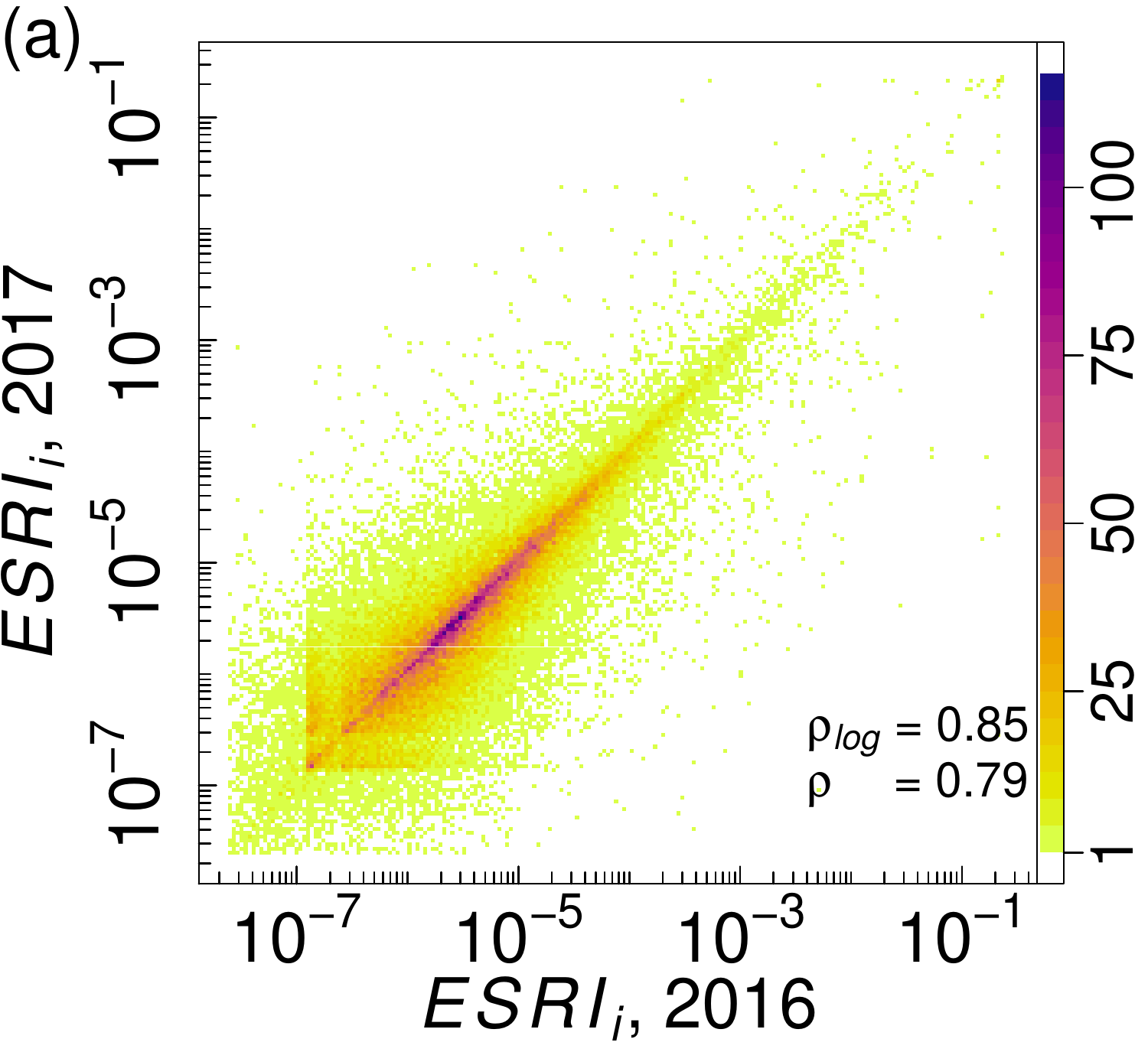}
	\includegraphics[width=0.32 \columnwidth]{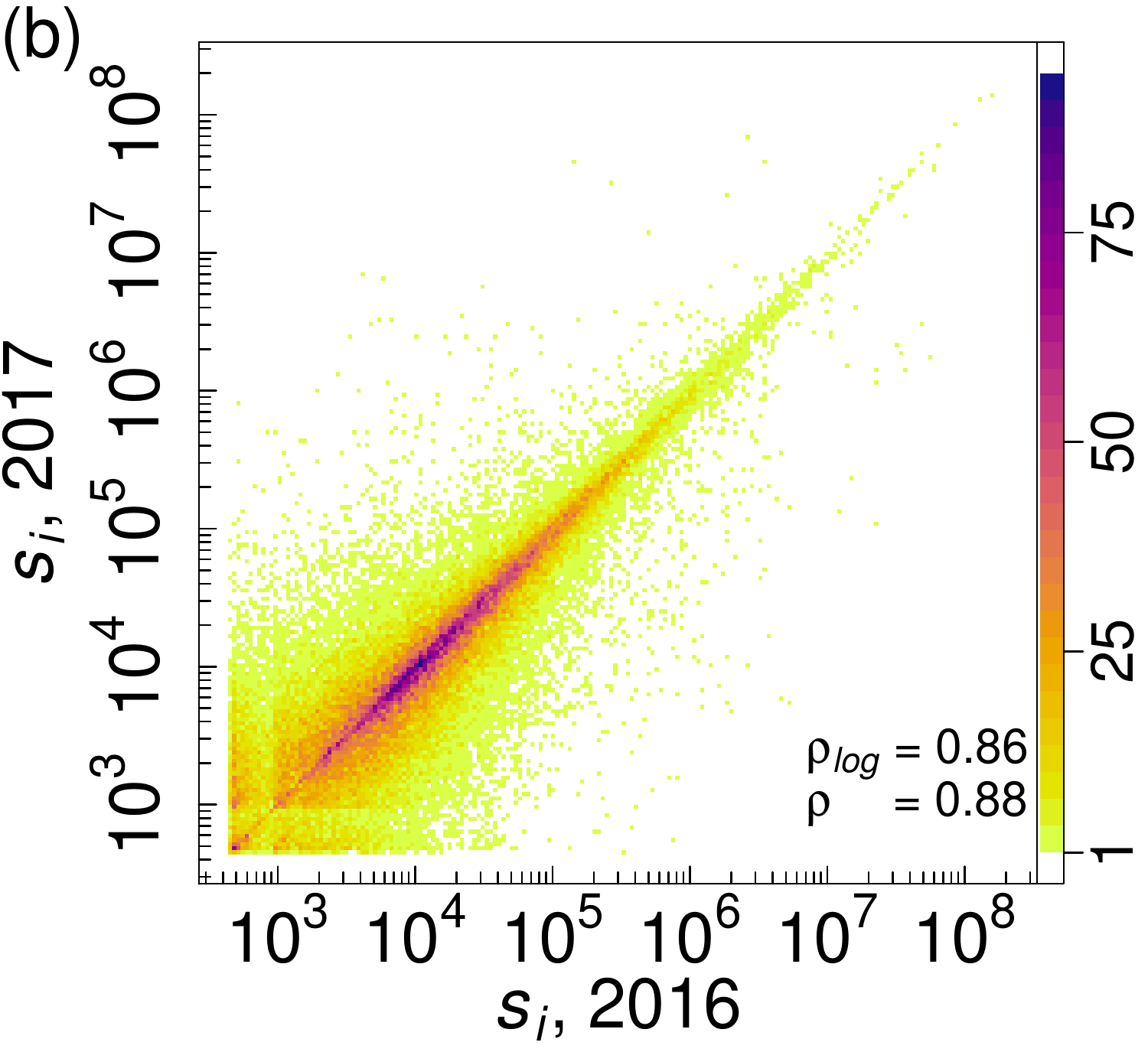}
	\includegraphics[width=0.32 \columnwidth]{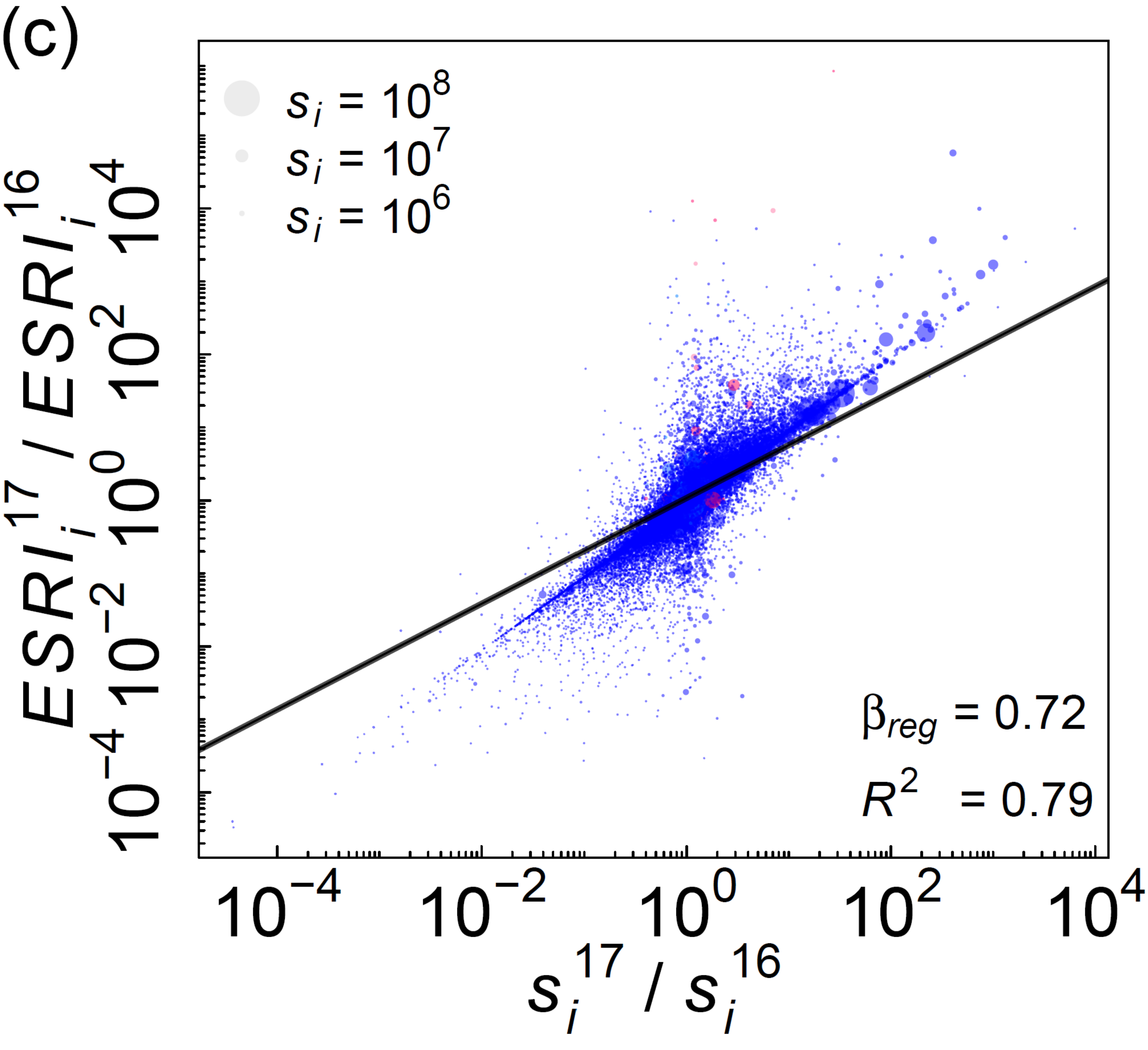} 
	\caption{Temporal changes from 2016 to 2017. 
		(a) bivariate density plot of ESRI in 2016 versus 2017 in log-log. Yellow indicates low, blue high counts. The distribution is strongly peaked along the diagonal (red and blue).  Note that the 1\% of firms with the smallest ESRI values was omitted for the figure.
		(b) bivariate density of firm strength $s_i$ 2016 versus $s_i$ 2017. The strength of smaller firms shows much stronger fluctuations than larger ones. The smallest 0.3\% of firms was omitted in the figure. 
		(c) relative changes of ESRI plotted against relative changes of firm strength in log-log scale from 2016 to 2017. There are large relative changes in both strength, $s_i$ and ESRI$_i$. The regression for the logged variables gives  $R^2=0.79$ and a slope of $\beta_\text{reg}=0.72$ for a p-value of $p=2\cdot 10^{-16}$. Thus, the variation in firm strength fails to explain a large fraction of variation in firms' systemic risk, ESRI. Note the very large fluctuations in ESRI for the region where almost no change in strength occurred. On a broader scale, two variables are positively related left-bottom and right-top. For the bulk of values the fluctuations are relatively small. 
	}
	\label{fig:sri_16_17_density}
\end{figure*}

\section{Change of ESRI from 2016 to 2017} \label{SI_annualchange}

We investigate the changes of the systemic risk index, ESRI$_i$, and firm strength, $s_i$ from 2016 to 2017. For a comprehensive view we present respective bi-variate density plots in log-log scale in FIG. \ref{fig:sri_16_17_density} (a) and (b). It is clear that most values lie on the diagonal indicated by red and dark blue colors. There are also large outliers visible (yellow). Note that these are single counts. The most notable outliers are constituted by the fluctuations in the plateau, i.e. nodes entering and leaving the plateau from 2016 to 2017. In general, large fluctuations are more likely for smaller values of ESRI while for larger values the spread away from the diagonal becomes smaller. For completeness we checked these relations by computing the correlation of the ESRI values of 2016 vs. 2017 in linear scale $\rho \big(ESRI^{16},ESRI^{17}\big)=0.78$ (p-value $2 \: 10^{-16}$), and in log-log scale $\rho \big(\log(ESRI^{16}),\log(ESRI^{17})\big)=0.85$ (p-value $2 \: 10^{-16}$). The high correlation values confirm the results seen by visual inspection. 

The relation of the most important firm-size proxy, firm strength $s_i$, from 2016 to 2017 is illustrated in FIG. \ref{fig:sri_16_17_density} (b). Again most values are found on the diagonal indicated by red and dark blue colors. Outliers are visible (yellow). Strength of smaller firms shows much larger fluctuations than the strength of of larger firms, seen by a diminishing variance around the diagonal for large values. The correlation of firm strength of 2016 and 2017 in linear scale is $\rho \big(ESRI^{16},ESRI^{17}\big)=0.86$ (p-value $2 \:10^{-16}$) and correlation in log scale is $\rho \big(\log(ESRI^{16}),\log(ESRI^{17})\big)=0.88$ (p-value $2 \: 10^{-16}$). This confirms the intuition obtained from the visual inspection. 

Figure \ref{fig:sri_16_17_density} (c) shows the relative changes in firm level systemic risk $\big(\text{ESRI}_i^{17}/\text{ESRI}_i^{16} \big)$ against relative changes in strength $\big(\text{s}_i^{17}/\text{s}_i^{16} \big)$ in log-log scale. There are three main observations. First, the bulk of values is clustered at the center with small changes in both variables. Second, there is a group of firms exhibiting a positive relation between changes in strength and ESRI. Third, there is a group of firms exhibiting large variation in ESRI, but almost no variation in strength. Overall this indicates again that changes in strength seems to have  some influence on changes in ESRI, but for many firms the changes in strength are not predictive for changes in ESRI. It seems that it matters more with whom firms form buyer-supplier relations than how large the sum of these relations is. For completeness we estimate the regression in log-log scale and receive an $R^2=0.79$ indicating that variations in relative change of strength explain 79\% of the variations in relative change of ESRI. The slope is $\beta_\text{reg}=0.72$. It is obvious from FIG. \ref{fig:sri_16_17_density} (c) that the regression line (black line) is not a good model for the data.

\section{On the necessity of working at the firm level} \label{SI:firm_level_advantages}

Here we explain deeper why sector level aggregation of the firm level production network leads to distortions in the picture of shock propagation. If a sector contains at least two firms, A and B, with linearly independent input sector vectors or customer sector vectors, it is possible to construct three firm level shock scenarios that have the same size on the sector level, but are in fact heterogeneous shocks on the firm level (single companies are  affect differently). Consequently, these three shock scenarios affect the direct input- and customer-sectors in different ways and hence actual shocks will propagate very differently for each of the scenarios. Note that shocks only propagate differently when analyzed on the firm level, but when the production network is aggregated to the sector level the cascades triggered by these three initial shocks are the same. 

\begin{figure}[t]
	\centering
	\includegraphics[width=0.32 \columnwidth]{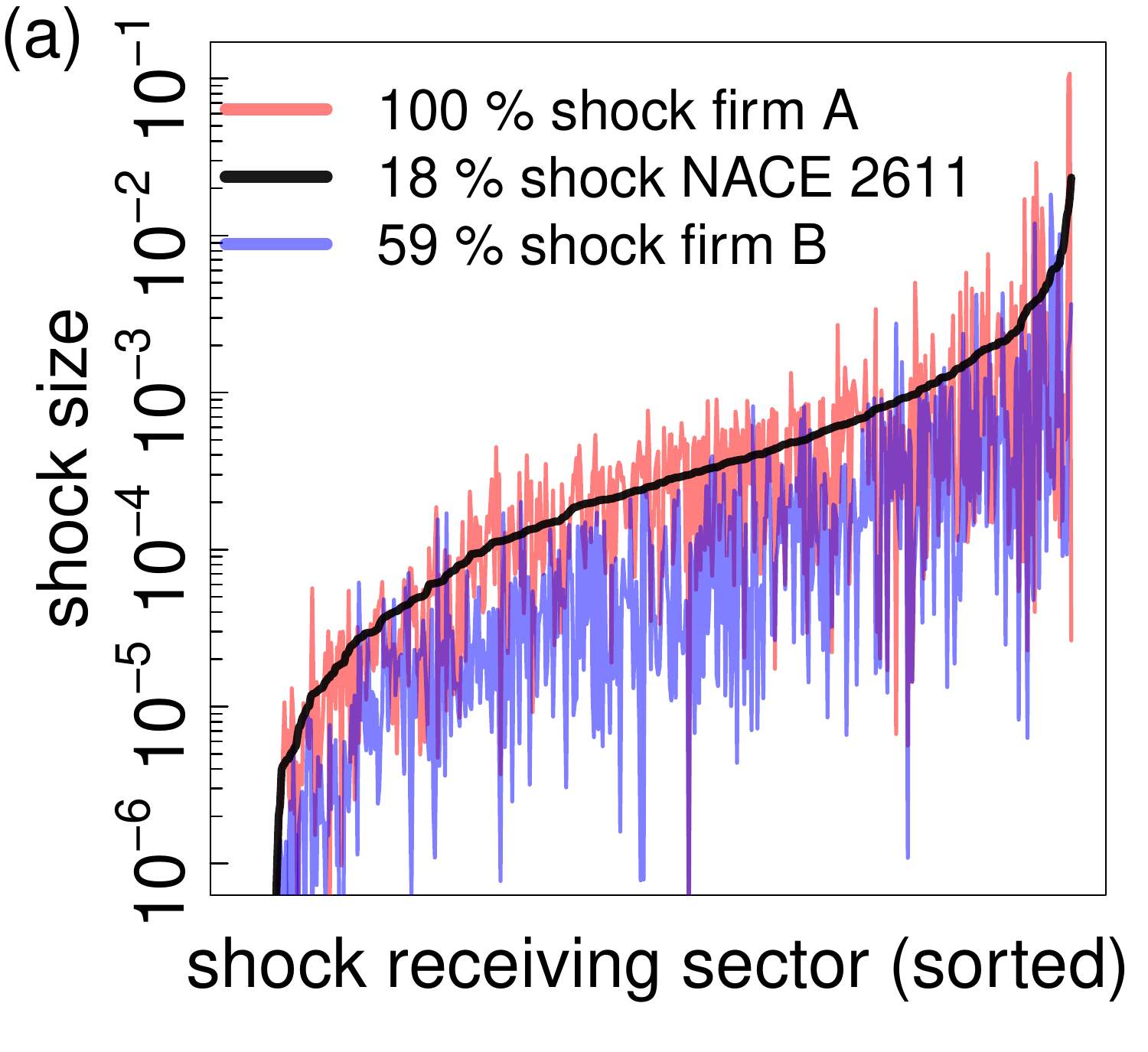}
	\includegraphics[width=0.32 \columnwidth]{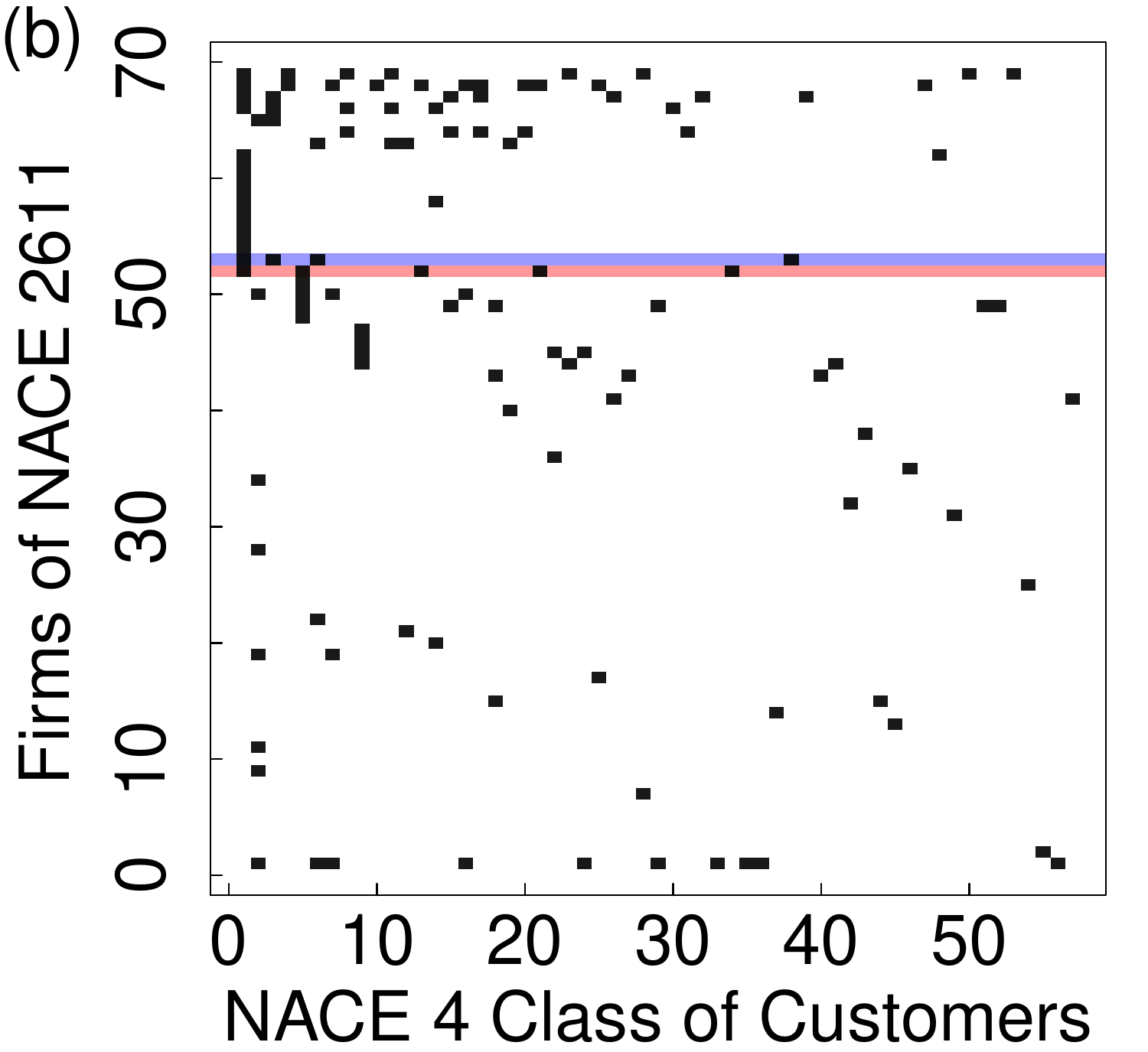}
	\caption{ 
		(a) Effects of different shock scenarios to NACE sector 2611 (Manufacture of electronic components). Shocks are of the same size but affect different companies in the other 567 NACE 4 sectors. The black line shows the baseline shock scenario where a 18\% shock is applied to all companies in sector 2611. Sectors are sorted in ascending order w.r.t. the baseline scenario.  
		The red curve corresponds to a 100\% shock to firm A (corresponds to an 18\% shock on sector). The blue curve represents a 59\% shock to firm B (corresponds to an 18\% shock on sector). so that the three shock scenarios are all of the same size on the sector level.
		The choice of the initially shocked companies have a drastically different effect on how all the other sectors are affected.
		(b) Customer sector vectors (at NACE 4 level) for 69 firms in sector 2611. Even though all companies belong to the same sector, their customer sector vectors are substantially different. We highlight the customer sectors of firm A (red, 5 distinct customer sectors) and firm B (blue highlight, 4 customer sectors). They differ substantially and have only 1 common customer sector. 
	}
	\label{fig:firm-level_SI}
\end{figure}

We present a concrete example where three shock scenarios are indistinguishable in terms of sector size affected, but have very different effects on how other sectors are affected by them. The implied assumption of a classical sector shock of size $x$\% is that all companies $i$ within the sector are affected by the same percentage shock of $x_i \%  = x\%$. In our example the sector shock is the baseline scenario. We simulate a 18\% shock to each company in NACE sector 2611 (Manufacturing of electronic components). The percentage shock is measured in percent of  the sector's strength, $s^\text{2611}=\sum_{j \: : \: p_j=2611} s_j$, that is initially affected. Then, we construct two firm level shocks that affect also 18\% of the strength $s^\text{2611}$ of sector 2611, but are not uniformly distributed across the sector's firms. In the first scenario we apply a shock to a single company ---called A for anonymity reasons--- of 100\% (de facto a temporary failure). Note that $s_A / s^\text{2611} = 0.18$. In the same way we apply a shock of 59\% to firm B (we assume that 59\% of the firms inputs and outputs are not bought and supplied). Note that $0.59 s_A / s^\text{2611} = 0.18$. Thus all three shocks are the same on the sector level in the classical sense, but affect different companies within the sector to a different degree. We know from FIG.  \ref{fig:firm-level_SI} (b) and main text FIG. \ref{fig:firm-level} (b) that firm A and B have different customer- and input-sector vectors, consequently, first order shocks spread to different sectors. 
This leads to a distinct spreading of shocks within the whole network and we expect other sectors to receive different indirect shocks for the three scenarios.  

In  FIG. \ref{fig:firm-level_SI} (a) we show how all the other 567 NACE 4 sectors are affected, measured in percent of output lost, i.e., for each NACE sector (on the x-axis)  the y-axis shows the percentage of the shock this NACE sector received in response to the initial shocks from the three scenarios uniform (black line) shock to firm A (red) and shock to firm B (blue). We sorted the NACE 4 sectors on the x-axis in ascending order with respect to the baseline scenario (the uniform sector shock), i.e., to the very left there is the NACE category that is affected least by the 18\% shock to all firms in sector 2611. We see that the effects of the shock to firm A are in general larger than the shock to firm B, but this is not the case for every single sector. The blue line is below the baseline scenario in most cases (shock to firm A affects sectors less), while in several cases the red line is above the baseline scenario  (shock to firm B affects these sectors stronger than the baseline). Note that the red (blue) line is above the black line only in 37\% (6\%) of the cases. This means that looking  at the sector level would overestimate the shocks received by the NACE 4 sectors in 63\% (94
\%) of the cases and underestimate it in the other cases if the true shock was a shock to firm B (firm A). The correlation between the percentage of output lost per sector, caused by the initial shock scenarios, is given in Table \ref{SI:table_sector_shock_cor}. We see that the two firm level shocks have a low, but significant correlation of 0.11 (p-value $p=0.0085$). Both firm level scenarios are positively correlated with the baseline shock scenario.

\begin{table}[h]
	\caption{\centering Sector shock correlations}
	\begin{tabular}{rrrr}
		\hline
		shock scenario & shock 2611 &  shock firm A & shock firm B \\ 
		\hline
		shock 2611 & 1.00 & 0.52 & 0.40 \\ 
		shock firm A  & 0.52 & 1.00 & 0.11 \\ 
		shock firm B & 0.40 & 0.11 & 1.00 \\ 
		\hline
	\end{tabular} \label{SI:table_sector_shock_cor}
\end{table}

In FIG. \ref{fig:firm-level_SI} (b)  we show for all companies in NACE sector 2611 the customer sectors at the NACE 4-digit level (one row corresponds to one firm, a column to a NACE 4 class). Companies (rows) are sorted with respect to the similarity of their customer sectors. Even though all companies belong to the same fine grained industry sector, their customer sectors vary substantially. The customer sectors of firm A (red shading) and firm B (blue shading) have a small overlap seen in a Jaccard index of 0.13. Depending on which of two companies fail, very different customer sectors are affected. Among the 51 firms (in NACE 2611 that have NACE 4 classified inputs) 85\% have no pairwise customer sector overlap (Jaccard index of zero). This means that when choosing two firms randomly within the sector the probability of having no common customer sector is 85\%. Consequently, for theses cases not only different companies, but entirely different economic sectors will be affected, depending on which of the two randomly selected firms suffers the initial shock. The most common customer sector is supplied to by 30\% of the companies in 2611 and the most common input sector is a supplier to 35\% of the companies in 2611. Note that 19 firms have no input sectors (empty rows). 

\end{document}